\documentclass[a4paper,10pt]{article}

\usepackage{jheppub} 

\usepackage[T1]{fontenc} 

\usepackage{graphicx}
\usepackage{verbatim,epsfig}
\usepackage{epstopdf}
\usepackage[utf8x]{inputenc}
\usepackage[colorlinks=true,linkcolor=blue,citecolor=blue]{hyperref}
  \usepackage{bm}
   \usepackage{amsmath}
    \usepackage{amssymb}
     \usepackage{pifont}
      \usepackage{simplewick}

\makeatletter
\newcommand\footnoteref[1]{\protected@xdef\@thefnmark{\ref{#1}}\@footnotemark}
\makeatother

\newcommand{\Ds}{\displaystyle}

\newcommand{\nn}{\nonumber}

\newcommand{\Tr}{\mathrm{Tr}}

\newcommand{\ot}{\leftarrow}

\renewcommand{\(}{\left(}
\renewcommand{\)}{\right)}
\renewcommand{\[}{\left[}
\renewcommand{\]}{\right]}

\renewcommand{\vec}[1]{\bm{#1}}
\newcommand{\fnot}[1]{\not{\! #1}}

\makeatletter
\newcommand{\pushright}[1]{\ifmeasuring@#1\else\omit\hfill$\displaystyle#1$\fi\ignorespaces}
\newcommand{\pushleft}[1]{\ifmeasuring@#1\else\omit$\displaystyle#1$\hfill\fi\ignorespaces}
\makeatother

\title{Collinear matching for Sivers function at next-to-leading order}

\author[a]{Ignazio Scimemi,}
\author[b]{Andrey Tarasov}
\author[c]{and Alexey Vladimirov}

\affiliation[a]{Departamento de F\'isica Te\'orica, Universidad Complutense de Madrid (UCM) and IPARCOS,\\E-28040 Madrid, Spain}
\affiliation[b]{Physics Department, Brookhaven National Laboratory, Upton, NY 11973, USA}
\affiliation[c]{Institut f\"ur Theoretische Physik, Universit\"at Regensburg, D-93040, Regensburg, Germany}

\emailAdd{ignazios@fis.ucm.es}
\emailAdd{atarasov@bnl.gov}
\emailAdd{alexey.vladimirov@ur.de}

\abstract{We evaluate the light-cone operator product expansion for unpolarized transverse momentum dependent (TMD) operator in the background-field technique up twist-3 inclusively. The next-to-leading order (NLO) matching coefficient for the Sivers function is derived. The method, as well as many details of the calculation are presented.}

\begin{document}
\maketitle
\flushbottom

\section{Introduction}

The exploration of the internal structure of nuclei is a fascinating task, which identifies transverse  momentum dependent (TMD) distributions as one of its most powerful tools. Transverse momentum dependent factorization theorems present a consistent description of double-inclusive processes, such as Drell-Yan/Vector/Scalar boson production(DY)\cite{Collins:2011zzd,GarciaEchevarria:2011rb} and  semi-inclusive deep inelastic scattering (SIDIS)\cite{Echevarria:2014rua,Collins:2011zzd,Bacchetta:2006tn} in the regime of small transverse momentum. Within the TMD factorization approach, the information on hadron structure is encoded in TMD parton distribution functions (TMDPDFs) and TMD fragmentation functions (TMDFFs). The presence of the transverse scale allows to resolve the internal structure of hadron with more details than collinear parton distributions. Many polarization phenomena, which are subleading in collinear factorization, are described by the leading order TMD factorization. In this work, we study the Sivers function \cite{Efremov:1981sh,Sivers:1989cc}, which describes the correlation of   an unpolarized parton  transverse momentum and a hadron polarization vector.

The Sivers function is an essential part of the single-spin asymmetry (SSA) phenomenon. Experimentally, SSA has been measured in SIDIS at Hermes \cite{Airapetian:2009ae}, COMPASS ~\cite{Alekseev:2008aa,Adolph:2012sp}, JLab~\cite{Qian:2011py} and in Drell-Yan  at RHIC \cite{Adamczyk:2015gyk,Dilks:2016ufy,Bok:2018}. Its measurement is planned also for the future Electron-Ion Collider (EIC)\cite{Accardi:2012qut}. SSA has been also an object of intensive phenomenological analysis, see e.g.~\cite{Anselmino:2005ea,Kang:2009bp,Echevarria:2014xaa,Anselmino:2016uie,Martin:2017yms,Boglione:2018dqd}. The resulting predictions differ substantially among these studies owing to  TMD evolution \cite{Aschenauer:2015ndk}, which shows the importance of a correct treatment of QCD perturbatively calculable parts. In the literature, there are several available calculations of the SSA in  perturbative QCD. The leading order (LO) (and partially the next-to-leading  order (NLO)) calculations for the SSA were performed in many works \cite{Boer:2003cm,Ji:2006ub,Ji:2006vf,Koike:2007dg,Kang:2011mr,Sun:2013hua,Dai:2014ala}. In principle, following these works it is possible to obtain the perturbative expression for Sivers function at NLO (however, different schemes are used for  different parts of the calculation, see discussion in sec.~\ref{sec:discussion}). Therefore, the SSA and the Sivers function are probably one of the most renowned and intensively studied polarized TMD quantities. 

Although the TMD distributions are genuine non-perturbative functions that should be extracted from  data, they can be evaluated in a model-independent way in terms of collinear distributions in the limit of large-$q_T$ \cite{Collins:1984kg}, or small-$b$ in the position space. This procedure is called ``matching'' and typically it serves as an initial input for the non-perturbative model of the TMD distributions, see e.g.~\cite{Echevarria:2014xaa,Scimemi:2017etj,Bacchetta:2017gcc}. The matching greatly increases the agreement with data~\cite{Scimemi:2017etj}. From the theory side, the matching procedure consists in the selection of the leading term in the light-cone operator product expansion (OPE) for the TMD operators \cite{Echevarria:2016scs,Gutierrez-Reyes:2017glx}. Alternatively, the matching can be obtained by taking the small-$q_T$ limit of  collinear factorization~\cite{Sun:2013hua,Dai:2014ala}, which, however, is not always possible~\cite{Bacchetta:2008xw}.

Only a few TMD distributions of leading-dynamical twist match the twist-2 collinear distributions. These are the unpolarized, helicity and transversity TMDPDFs and TMDFFs. The matching coefficients for these distributions are known uniformly at the next-to-leading order (NLO)~\cite{Collins:2011zzd,GarciaEchevarria:2011rb,Bacchetta:2013pqa,Echevarria:2015uaa,Gutierrez-Reyes:2017glx} and some are known at NNLO~\cite{Echevarria:2015usa,Echevarria:2016scs,Gutierrez-Reyes:2018iod}. The remaining TMD distributions match twist-3 collinear distributions (apart of the pretzelosity which is apparently of twist-4~\cite{Gutierrez-Reyes:2018iod,Chai:2018mwx}). The knowledge of the matching for these distributions is very poor: the quark TMDPDFs are all known at LO \cite{Boer:2003cm,Ji:2006ub,Kang:2011mr,Kanazawa:2015ajw,Scimemi:2018mmi} and only Sivers function is known at NLO \cite{Sun:2013hua,Dai:2014ala} (however, see discussion in sec.~\ref{sec:results}). The matching for some of quark TMDFFs, such as Collins function, is known at LO \cite{Kanazawa:2015ajw}. The matching for the majority of gluon TMD distributions is unknown.

The importance of the computation of the perturbative part of a TMD distribution in order to meet an agreement between theory and experiment has been shown already in \cite{Scimemi:2017etj} for the unpolarized case. Depending on the experimental conditions, the measured data can be sensitive to various aspects of the theory such as power corrections in the evolution \cite{Scimemi:2016ffw}, power correction \cite{Balitsky:2017gis}, small-$x$ effects in the evolution~\cite{Balitsky:2015qba} and many others. The full control of all of these sources of non-perturbative physics requires an accurate setting of the perturbative scales, as provided, for instance, by the $\zeta$-prescription of~\cite{Scimemi:2018xaf}.
 
In this work, we perform a complete NLO computation of the Sivers function starting from its operator definition and performing a light-cone OPE in  background field~\cite{DeWitt:1967ub}. To our best knowledge, this approach is used for the description of TMD operator for the first time, despite the fact that it is a standard tool in higher twist calculation, see e.g. \cite{Balitsky:1987bk,Balitsky:2016dgz}. This technique grants an unprecedented control of the operator structure and it allows a very general treatment for  twist-3 distributions. Therefore, the result obtained in this work is also interesting for a broader study. For the first time, we demonstrate how the TMD renormalization (ultraviolet and rapidity renormalization~\cite{Vladimirov:2017ksc}) is organized at the operator level. We also articulate the role of the gauge links and their direction and show (at the level of operators) the famous sing-change in-between DY and SIDIS definitions of the Sivers function~\cite{Collins:2002kn}. Motivated by these considerations, we provide a detailed and pedagogical explanation of the calculation method, which is a major target of this article. For that aim, the Sivers function represents an ideal case, because one can cross-check the calculation with other methods already used in the literature. We anticipate that our results agree with the results present in the literature only partially, however, the origin of the discrepancy is clear. 

The article is organized as following. Sec.~\ref{sec:SSA} is a general introduction to SSA in the TMD factorization approach. Here we collect the expressions for SSA structure functions and describe the role of Sivers function and its collinear matching. In sec.~\ref{sec:TMDdefs} we introduce and describe in detail the operator that defines Sivers function.  Its renormalization properties are discussed in sec.~\ref{sec:evol+ren}. Sec.~\ref{sec:evol+ren} is devoted to the detailed derivation of OPE at LO. We discuss separately the evaluation in regular (sec.~\ref{sec:LO_reg}) and light-cone (sec.~\ref{sec:LO_sing}) gauges. 
 The NLO evaluation is presented in sec.~\ref{sec:OPENLO}. We make a pedagogical introduction to the background field method in sec.~\ref{sec:intro-to-back}-\ref{sec:background}. The details on the NLO evaluation of diagrams are given in sec.~\ref{sec:eval}.
  In sec.~\ref{sec:rapidity_div}-\ref{sec:backrenormalization} we discuss the appearance of rapidity divergences and their renormalization. 
  The difference in the evaluation of DY and SIDIS operators is discussed in sec.~\ref{sec:DY-SIDIS-difference}. 
  The extra details on the calculation are given in appendices \ref{app:exampleDiag}, where we present a step-by-step calculation of a diagram and \ref{app:diag-by-diag_OP}, where we give the diagram-by-diagram expressions for OPE. 
  The collinear distributions are defined in sec.~\ref{sec:def-collinear}. Additional details of the parametrization definition are given in appendix.~\ref{app:tensor-decomposition}.
   The transition from operators to distributions is discussed in sec.~\ref{sec:TOdistr1} and the collection of diagram-by-diagram expressions can be found in appendix~\ref{app:diag-by-diag:matching}. The final result of calculation is given in sec.~\ref{sec:results}. The discussion and comparison with earlier calculations is given in~\ref{sec:discussion}.

\section{Sivers effect and  TMD factorization}
\label{sec:SSA}

TMD distributions are defined by a large set of parameters: collinear momentum fraction $x$, transverse distance $\vec b$ (or transverse momentum $\vec p_T$), polarization, parton flavor $f$, the type of hadron $h$, ultraviolet and rapidity renormalization scales ($\mu$ and $\zeta$) and the defining process (DY or DIS). An explicit designation of all these parameters would lead to a heavy notation such as
\begin{eqnarray}\nn
f^\perp_{1T,q\ot h;\text{DY}}(x,\vec b;\mu,\zeta),
\end{eqnarray}
which  should be  read as  the Sivers function for a quark $q$ with momentum faction $x$ at the transverse parameter $\vec b$ produced by hadron $h$ in the DY kinematics, measured at scales $\mu$ and $\zeta$. Most  of this information is not needed in perturbative calculations and in the following \textit{we skip the unnecessary parts of the notation}, e.g. the renormalization scales are usually dropped. We also distinguish the momentum and coordinate space TMD distributions only by their arguments. In the rest of this section we show how the Sivers function arises in SIDIS and DY cross sections.

\subsection{Sivers function in SIDIS}

The semi-inclusive deep inelastic scattering (SIDIS) is a common name for a set of processes
\begin{eqnarray}
l(l)+N(P)\to l(l')+h(P_h)+X,
\end{eqnarray}
where $l(l')$ is a lepton, $N$ is a nucleon target and $h$ is the produced hadron. The TMD factorization is applicable in the regime $|\vec P_h|\ll Q$, where  $Q^2=(l-l')^2$ is a hard scale of the scattering, $\vec P_h$ is the transverse component of the momentum $P_h$. In the following, we use the bold font notation for the transverse components of vectors.

 In the case of unpolarized lepton beam, unpolarized produced hadron $h$ and a transversely polarized target $N$, the cross-section for SIDIS contains three structures. The so-called Sivers effect  (proportional to $\sin(\phi_h-\phi_s)$), Collins effect (proportional to $\sin(\phi_h+\phi_s)$) and the $\sin(3\phi_h-\phi_s)$ asymmetry. The structure functions corresponding to these effects within TMD factorization can be found e.g. in~\cite{Bacchetta:2006tn,Vogelsang:2005cs,Anselmino:2005ea}. The structure function for the Sivers effect is denoted by $F_{UT}^{\sin(\phi_h-\phi_s)}$. Within the TMD factorization it is \cite{Bacchetta:2006tn} 
\begin{eqnarray}\label{SF:SIDIS_momentum}
F_{UT}^{\sin(\phi_h-\phi_s)}(x,z,Q,\vec P_h)&=&-xH_{\text{DIS}}(Q,\mu)\sum_f e_f^2 \int d^2\vec p d^2\vec k\delta^{(2)}\(\vec p-\vec k-\frac{\vec P_h}{z}\)
\\\nn &&\times
\frac{\vec P_h\cdot \vec p}{M|\vec P_h|}f_{1T;f\ot N;\text{DIS}}^\perp(x,\vec p;\mu,\zeta_1)D_{1;f\to h}(z,\vec k;\mu,\zeta_2)+O\(\frac{\vec P_h^2}{z^2 Q^2}\),
\end{eqnarray}
where the variables $x$ and $z$ are the momentum fractions of partons and $M$ is the hadron mass. The functions $D_1$ and $f_{1T}^\perp$ are unpolarized and Sivers TMD distributions. The factorization scale $\mu$ is typically chosen to be of order $Q$. The scales of soft exchanges (rapidity factorization) $\zeta_{1,2}$ satisfy $\zeta_1\zeta_2=Q^4$.

The TMD factorization is naturally formulated in  position space, where the Fourier convolution in eq.~(\ref{SF:SIDIS_momentum}) turns into a product of functions. In  position space the structure function reads 
\begin{eqnarray}
F_{UT}^{\sin(\phi_h-\phi_s)}(x,z,Q,\vec P_h)&=&ixMH_{\text{DIS}}(Q,\mu) \sum_f e_f^2 \int \frac{d^2\vec b}{(2\pi)^2} e^{i(\vec b \vec P_h)/z}
\\\nn &&\times
\frac{\vec P_h\cdot \vec b}{|\vec P_h|}f_{1T;f\ot N;\text{DIS}}^\perp(x,\vec b;\mu,\zeta_1)D_{1;f\to h}(z,\vec b;\mu,\zeta_2)+O\(\frac{\vec P_h^2}{z^2 Q^2}\).
\end{eqnarray}
The functions $D_1$ and $f_{1T}^\perp$ depend only on the length of the vector $\vec b$ but not on its direction and one can also simplify the angular dependence \cite{Boer:2011xd,Scimemi:2018mmi}
\begin{eqnarray}\label{SF:SIDIS_bessel}
F_{UT}^{\sin(\phi_h-\phi_s)}(x,z,Q,\vec P_h)&=&-xMH_{\text{DIS}}(Q,\mu) \sum_f e_f^2 \int_0^\infty \frac{d|\vec b|}{2\pi} |\vec b|^2 J_1\(\frac{|\vec b||\vec P_h|}{z}\)
\\\nn &&\times
f_{1T;f\ot N;\text{DIS}}^\perp(x,\vec b;\mu,\zeta_1)D_{1;f\to h}(z,\vec b;\mu,\zeta_2)+O\(\frac{\vec P_h^2}{z^2 Q^2}\),
\end{eqnarray}
where $J_1$ is the Bessel function of the first kind. The equation (\ref{SF:SIDIS_bessel}) is the usual starting point for the parametrization of the Sivers effect in  TMD factorization.

\subsection{Sivers function in DY}

The Sivers effect also appears in the Drell-Yan/vector boson production process
\begin{eqnarray}
h_a(P_a)+h_b(P_b)\to Z/\gamma^*(q)+X \to l(l)+\bar l(l')+X,
\end{eqnarray}
where one of the initial hadrons is polarized~\cite{Boer:1999mm,Anselmino:2002pd,Efremov:2004tp,Vogelsang:2005cs}. In general one refers to  structure functions $F_{UT}^1$ when the hadron $h_a$ is polarized and $F_{TU}^1$ when the hadron $h_b$ is polarized. The structure function $F_{TU}^1$ in  TMD factorization (i.e. for $q_T\ll Q$) reads \cite{Arnold:2008kf}
\begin{eqnarray}\label{SF:DY_momentum}
F_{TU}^1(Q,\vec q_T)&=&\frac{-H_{\text{DY}}(Q,\mu)}{N_c}\sum_f e_f^2 \int d^2\vec k_{a} d^2\vec k_{b}\delta^{(2)}\(\vec q_T-\vec k_{a}-\vec k_{b}\)
\\\nn &&\times
\frac{\vec q_T\cdot \vec k_{a}}{M|\vec q_T|}f_{1T;f\ot h_a;\text{DY}}^\perp(x_a,\vec k_a;\mu,\zeta_1)f_{1;\bar f\ot h_b}(x_b,\vec k_b;\mu,\zeta_2)+O\(\frac{\vec q_T^2}{Q^2}\),
\end{eqnarray}
where $Q^2=(l+l')^2$ is the hard scale of the process, $x_{a,b}$ are momentum fractions of partons, $\vec q_T$ is the transverse component of $q=l+l'$ relative to the scattering plane
and $f_1$ is the unpolarized TMD distribution. The factorization scales are defined similarly to the SIDIS case, i.e. $\mu\sim Q$ and $\zeta_1\zeta_2=Q^4$.  The transformation of the structure function under interchange of the polarized hadron ($h_a\leftrightarrow h_b$) is $F_{UT}^1=-F_{TU}^1$. 

The structure functions can be also written in the form
\begin{eqnarray}
F_{TU}^1(Q,\vec q_T)&=&
\frac{iMH_{\text{DY}}(Q,\mu)}{N_c}\sum_f e_f^2 \int \frac{d^2\vec b}{(2\pi)^2}e^{i(\vec b \vec q_T)}
\frac{\vec q_T\cdot \vec b}{|\vec q_T|}
\\\nn && \times 
f_{1T;f\ot h_a;\text{DY}}^\perp(x_a,\vec b;\mu,\zeta_1)f_{1;\bar f\ot h_b}(x_b,\vec b;\mu,\zeta_2)+O\(\frac{\vec q_T^2}{Q^2}\)\ ,
\end{eqnarray}
and
\begin{eqnarray}\label{SF:DY_bessel}
F_{TU}^1(Q,\vec q_T)&=&
\frac{-MH_{\text{DY}}(Q,\mu)}{N_c}\sum_f e_f^2 \int_0^\infty \frac{d|\vec b|}{2\pi}|\vec b|^2 J_1(|\vec b||\vec q_T|)
\\\nn && 
\times f_{1T;f\ot h_a;\text{DY}}^\perp(x_a,\vec b;\mu,\zeta_1)f_{1;\bar f\ot h_b}(x_b,\vec b;\mu,\zeta_2)+O\(\frac{\vec q_T^2}{Q^2}\),
\end{eqnarray}
where we have integrated out the angular dependence.

The Sivers functions in SIDIS, eq.~(\ref{SF:SIDIS_momentum}) and DY, eq.~(\ref{SF:DY_momentum}), have different labels that specify the processes. These functions have different operator definitions (see sec.~\ref{sec:TMDdefs}). However, de facto, the process-dependence reduces to a simple sign change \cite{Brodsky:2002cx, Brodsky:2002rv, Collins:2002kn, Boer:2003cm}
\begin{eqnarray}\label{SF:DY<->DIS}
f_{1T;f\ot h_a;\text{DY}}^\perp(x,\vec b;\mu,\zeta)=-f_{1T;f\ot h_a;\text{DIS}}^\perp(x,\vec b;\mu,\zeta).
\end{eqnarray}
In the following, we demonstrate the origin of the sign-change at the level of OPE.

\subsection{TMD evolution and operator product power expansion}

The practical application of TMD factorization relies on the concept of TMD evolution, which allows to relate structure functions at different values of $Q$. 
Here, we should stress that a TMD distribution is an involved non-perturbative function. In fact, in addition to the non-perturbative structure of TMD distribution (which involves the dependence on the variables ($x$, $\vec b$)), the TMD factorization also contains a non-perturbative part of the evolution factor (which depends only on $\vec b$). An efficient implementation of the TMD approach should be able to disentangle these non-perturbative contributions. The parametrization and extraction of three non-perturbative functions (two TMD distributions and the evolution kernel) of two variables would be a hopeless task if the TMD factorization would not allow us to separate the problem into pieces. 

First of all, the TMD evolution is regulated by two scales $(\mu,\zeta)$ and it is process independent. It factors out the non-perturbative evolution effects into an evolution factor which is strictly universal for all structure functions and for all TMD factorizable processes. Nonetheless, the TMD evolution still non-trivially affects the ($x$, $\vec b$) dependence of the distribution which should be modeled as a function of two variables. To simplify this procedure one can use any available information that restricts the functional form of the TMD. In particular, at small values of $\vec b$ a TMD distribution can be related to collinear distributions in a model-independent way in  perturbation theory. Such a relation has the general form provided by OPE
\begin{eqnarray}\label{small-$b$}
f(x,\vec b)=C_1(x,\mathbf{L}_\mu)\otimes f_1(x)+\vec b^2 C_2(x,\mathbf{L}_\mu)\otimes f_2(x)+...,
\end{eqnarray}
where $C_i$ are perturbatively calculable Wilson coefficient functions which depend on $\vec b$ only logarithmically via $\mathbf{L}_\mu$ (to be defined in eq.~(\ref{def:Lmu})), $f_i$ are collinear distributions of increasing twist and $\otimes$ is an integral convolution in the variable $x$. This expansion is valid only in a certain range of $\vec b$, say $|\vec b| < R$, where $R$ is some matching scale. For values of $\vec b$ larger than $R$ TMD distribution is completely non-perturbative. In fact,  as the value of $\vec b$ gets closer to $R$, the contribution of higher order terms in the small-$b$ expansion becomes more important. However, our knowledge of the corresponding higher-twist distributions is very limited.
 
Thus, it is of practical convenience to use only the first term of the small-$b$ expansion in eq.~(\ref{small-$b$}) and replace the rest by a generic non-perturbative function, i.e. \begin{eqnarray}\label{model}
f(x,\vec b)=C_1(x,\mathbf{L}_\mu)\otimes f_1(x)f_{NP}(x,\vec b).
\end{eqnarray}
The practical success of such an ansatz can be easily understood if we notice that the main contribution to the Fourier integrals in eqs.~(\ref{SF:SIDIS_bessel}, \ref{SF:DY_bessel}) comes from the small-$b$ region. Therefore, we can expect that the function $f_{NP}$ has a simple behavior in $x$ and $\vec b$, which is indeed confirmed by phenomenological applications of this formula. The details of the modeling procedure which is based on eq.~(\ref{model}) are different in different approaches,  but the core picture described here remains unchanged.

The small-$b$ matching is an essential part of the modern TMD phenomenology. In ref.~\cite{Scimemi:2017etj} a comparison of different orders of the matching to experimental results has been performed. It has been shown that the NLO matching is essential for the predictive power of the approach. The NNLO matching provides further improvements and it can be necessary for the description of the most precise experiments.

The achievable precision can also be affected by the choice of scales in the matching. Let us also mention that in~\cite{Scimemi:2018xaf} the authors have proved the possibility to disentangle the procedure of small-$b$ matching and TMD evolution using the $\zeta$-prescription which is not entirely possible in other formulations.  The $\zeta$-prescription allows using different perturbative orders for TMD evolution and small-$b$ matching. This means that the modeling of the TMD through eq.~(\ref{model}) is completely separated from the evolution part of the TMD (that is, the scale choice does not mix up non-perturbative pieces of different origin). This fact results to be extremely useful for phenomenology since it allows to use the highest allowed/known expression of evolution \cite{Vladimirov:2016dll} in combination with  polarized observables whose high perturbative orders are unknown. The universal non-perturbative part of evolution can be extracted from the most precise data (such as Z-boson production at LHC) \cite{Bertone:2019TOBE}.

Let us conclude this section recalling that the hard coefficient functions $H_{\text{DIS}}$ and $H_{\text{DY}}$ within  TMD factorization are given by the quark form factor evaluated in the different analytical regions. At the NLO they differ only by a $\pi^2$-term,
\begin{eqnarray}
H_{\text{DIS}}(Q,\mu)&=&|C_V(Q^2,\mu^2)|^2=1+2a_s C_F\(-\mathbf{l}_{Q^2}^2-3\mathbf{l}_{Q^2}-8+\frac{\pi^2}{6}\)+O(a_s^2),
\\
H_{\text{DY}}(Q,\mu)&=&|C_V(-Q^2,\mu^2)|^2=1+2a_s C_F\(-\mathbf{l}_{Q^2}^2-3\mathbf{l}_{Q^2}-8+\frac{7\pi^2}{6}\)+O(a_s^2),
\end{eqnarray}
where $\mathbf{l}_{Q^2}=\ln(\mu^2/Q^2)$ and $a_s=g^2/(4\pi)^2$. The NNLO and NNNLO expression can be found in \cite{Gehrmann:2010ue}.

\section{Operator  definitions for unpolarized and Sivers TMD distributions}
\label{sec:definitions}

In this section, we introduce and review the main properties of TMD distributions. 

\subsection{Definition of TMD distributions}
\label{sec:TMDdefs}

Through the article we use the standard notation for the light-cone decomposition of a vector
\begin{eqnarray}
v^\mu=v^+ \bar n^\mu+v^- n^\mu+v_T^\mu,
\end{eqnarray}
where $v^+=(nv)$, $v^-=(\bar n v)$ and $v_T$ is the transverse component $(v_Tn)=(v_T\bar n)=0$. The vectors $n$ and $\bar n$ are light-like
\begin{eqnarray}\label{def:n}
n^2=\bar n^2=0,\qquad (n\bar n)=1.
\end{eqnarray}
Their particular definition is related to the factorization frame of the scattering process. The transverse part (with respect to vectors $n$ and $\bar n$) of the metric and Levi-Civita tensors are
\begin{eqnarray}
g_T^{\mu\nu}=g^{\mu\nu}-\frac{n^\mu \bar n^\nu+\bar n^\mu n^\nu}{(n\bar n)}, \qquad \epsilon_T^{\mu\nu}=\frac{n_\alpha \bar n_\beta}{(n\bar n)}\epsilon^{\alpha\beta\mu\nu},
\end{eqnarray}
where $\epsilon^{\mu\nu\rho\sigma}$ is in the Bjorken convention ($\epsilon_{0123}=-\epsilon^{0123}=1$). In four dimensions (with $n$ and $\bar n$ localized in the plane $(0,3)$) both tensors have only two non-zero components, $g^{11}_T=g_T^{22}=-1$ and $\epsilon_T^{12}=-\epsilon_T^{21}=1$. 

Since the transverse subspace is Euclidian, the scalar product of transverse vectors is negative, $v_T^2<0$. In the following, we adopt the bold font notation to designate the Euclidian scalar product of transverse vectors, i.e. $\vec b^2=-b^2>0$, when it is convenient.

\begin{figure}[t]
\begin{center}
\includegraphics[width=0.7\textwidth]{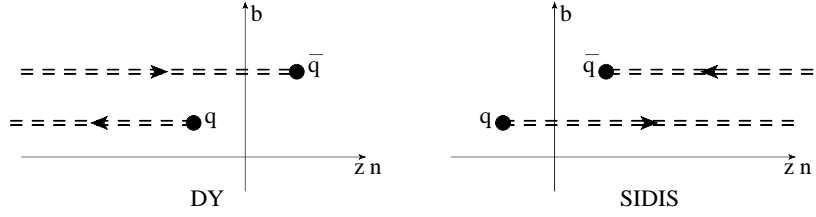}
\end{center}
\caption{\label{fig:contours} Illustration for the definition of TMD operators in DY and SIDIS. The Wilson lines (shown by dashed lines) are oriented along past (DY) or future (SIDIS) light cone direction. At light-cone infinities the Wilson lines are connected by transverse gauge links (not shown).}
\end{figure}

Using this notation,
the transverse momentum dependent parton distribution functions (TMDPDFs) for \textit{unpolarized quark} are defined by the matrix element~\cite{Tangerman:1994eh,Collins:2011zzd,GarciaEchevarria:2011rb}
\begin{eqnarray}\label{def:TMDPDF_Qop}
\Phi_{q\ot h}^{[\gamma^+]}(x,\vec b)&=&\int\!\frac{dz}{2\pi}e^{-ixzp^+}\!\langle p,S|\bar T\{\bar q\(zn+\vec b\)[z n+\vec b,\pm\infty n+\vec b]\}\gamma^+T\{
[\pm\infty n,0] q(0)\}|p,S\rangle,
\end{eqnarray}
where $[a,b]$ are Wilson lines defined in eq.~(\ref{def:WilsonLine}). The notation $\pm \infty n$ indicates different cases of TMD distributions, which appear in different processes. The TMD distributions that appear in SIDIS have Wilson lines pointing to $+\infty n$, while in Drell-Yan they point to $-\infty n$ as in  fig.~\ref{fig:contours}. The Wilson lines within the TMD operator are along the light-like direction $n$.

The matrix element in eq.~(\ref{def:TMDPDF_Qop}) for the \textit{polarized hadron} is parametrized by two independent functions  ~\cite{Boer:2011xd,Scimemi:2018mmi}
\begin{eqnarray}\label{param:TMDv}
\Phi_{q\ot h}^{[\gamma^+]}(x,\vec b)&=&f_1(x,\vec b)+i\epsilon_T^{\mu\nu} b_\mu s_{T\nu} M f_{1T}^\perp(x,\vec b),
\end{eqnarray}
where $M$ is the mass of the hadron and $s_T$ is the transverse part of the hadron spin-vector $S$, i.e. $s^\mu_T=g_T^{\mu\nu}S_\nu$. The function $f_1$ is the \textit{unpolarized} TMDPDF, which measures the unpolarized quark distribution in an unpolarized hadron. The function $f_{1T}^\perp$ is known as the \textit{Sivers function}, which measures the unpolarized quark distribution in a polarized hadron.

The parametrization of eq.~(\ref{param:TMDv}) is given in  position space. The distributions in momentum space are defined in the usual manner
\begin{eqnarray}\label{def:p<->b}
\Phi_{q\ot h}^{[\gamma^+]}(x,\vec p)=\int \frac{d^2 \vec b}{(2\pi)^2}e^{+i(\vec b\vec p)}\Phi^{[\gamma^+]}_{q\ot h,ij}(x,\vec b),
\end{eqnarray}
where the scalar product $(\vec b\vec p)$ is Euclidian. Correspondingly, the momentum space parameterization reads \cite{Goeke:2005hb,Bacchetta:2006tn} 
\begin{eqnarray}\label{def:gammaP_momentum}
\Phi_{q\ot h}^{[\gamma^+]}(x,\vec p)&=& f_1(x,\vec p)-\frac{\epsilon_T^{\mu\nu}p_{\mu}s_{T\nu}}{M}f_{1T}^\perp(x,\vec p).
\end{eqnarray}
Some explicit relations among particular TMDPDFs can be found in the appendix of ref.~\cite{Scimemi:2018mmi}. These relations are used to relate structure functions in momentum and coordinate representations in sec.~\ref{sec:SSA}.

The anti-quark TMD distribution is defined as
\begin{eqnarray}\label{def:TMDPDF_aQop}
\Phi_{\bar q\ot h}^{[\gamma^+]}(x,\vec b)&=&\int\frac{dz}{2\pi}e^{-ixz(pn)}\langle p,S|\Tr\(\gamma^+\bar T\{
[\pm\infty n,0] q_i(0)\} T\{\bar q\(zn+\vec b\)[z n+\vec b,\pm\infty n]\}\)|p,S\rangle.\nn \\
\end{eqnarray}
Using charge-conjugation, one can relate the quark and anti-quark TMD distributions~\cite{Tangerman:1994eh},
\begin{eqnarray}
\Phi_{q\ot h}^{[\gamma^+]}(x,\vec b)=-\(\Phi_{\bar q\ot h}^{[\gamma^+]}(-x,\vec b)\)^*,
\end{eqnarray}
from which it follows
\begin{eqnarray}
f_{1;q\ot h}(x,\vec b)&=&-f_{1;\bar q\ot h}(-x,\vec b),
\\
f_{1T;q\ot h}^\perp(x,\vec b)&=&f_{1T;\bar q\ot h}^\perp(-x,\vec b).
\end{eqnarray}
Therefore, in the following we associate the anti-quark distributions with the negative values of $x$ and  we define the TMD distributions in the range $-1<x<1$ as
\begin{eqnarray}\label{def:unpol_allX}
f_{1;q\ot h}(x,\vec b)&=&\theta(x)f_{1;q\ot h}(x,\vec b)-\theta(-x)f_{1;\bar q\ot h}(-x,\vec b),
\\\label{def:sivers_allX}
f_{1T;q\ot h}^\perp(x,\vec b)&=&\theta(x)f_{1T;q\ot h}^\perp(x,\vec b)+\theta(-x)f_{1T;\bar q\ot h}^\perp(-x,\vec b).
\end{eqnarray}

The small-$b$ expansion (often called small-$b$ matching or collinear matching) presents a TMD distributions as a series of collinear distributions  and Wilson coefficients in the vicinity of $\vec b=0$ as in eq.~(\ref{small-$b$}). For instance, the leading term of the small-$b$ expansion for unpolarized TMD  is expressed by the (unpolarized) collinear PDF $f_1(x)$
\begin{eqnarray}
f_{1,q\ot h}(x,\vec b;\mu,\zeta)&=&\sum_f\int_x^1 \frac{dy}{y}C_{1;q\ot f}(y,\vec b,\mu,\zeta)f_{1,f\ot h}\(\frac{x}{y},\mu\)+O(\vec b^2),
\end{eqnarray}
where the sum index $f$ indicates gluons, quarks and antiquarks of all flavors. The coefficient function $C$ is the perturbative Wilson coefficient, which depends on $\vec b$ logarithmically. Its leading term is $\delta(1-y)$ and the perturbative corrections are known up to NNLO~\cite{Echevarria:2015usa}. The power corrections (as in eq.~(\ref{small-$b$})) contain collinear distributions of twist-2 and twist-4 and they are currently unknown. 

The expression for the small-$b$ matching of the Sivers function is 
\begin{eqnarray}
f_{1T;q\ot h}^{\perp}=\sum_f C_{1T;q\ot f}^\perp(x_1,x_2,x_3,\vec b,\mu,\zeta)\otimes T_{f\to h}(x_1,x_2,x_3,\mu)+O(\vec b^2),
\end{eqnarray}
where $T$ are the collinear distributions of twist-3,  to be defined in secs.~\ref{sec:coll-quark}, \ref{sec:coll-gluon}.
 The symbol $\otimes$ denotes an integral convolution in the variables $x_{1,2,3}$. At  leading order the expression for the coefficient function is known to be $\pm \pi \delta(x_1+x_2+x_3)\delta(x_2)\delta(x-x_3)$~\cite{Boer:2003cm,Ji:2006ub,Kang:2011mr,Scimemi:2018mmi} (and we also re-derive it in the next section). The status of the NLO expressions is cumbersome. In principle, the quark-to-quark part can be found in~\cite{Sun:2013hua}, where it has been extracted from computation of the cross-section made in \cite{Ji:2006ub,Ji:2006vf,Koike:2007dg}. However, the computations made in~\cite{Ji:2006ub,Ji:2006vf,Koike:2007dg} miss certain parts and for this reason they are partially incorrect (see extended discussion in \cite{Braun:2009mi}). The quark-to-gluon part is evaluated in~\cite{Dai:2014ala}, however,  the authors use a  scheme which is different from the standard one for twist-2 computations. We return to this discussion in sec.~\ref{sec:results}.

\subsection{Evolution and renormalization}
\label{sec:evol+ren}

The renormalized TMD,  unlike usual parton distributions, depend on a pair of scales. This is a consequence  of the TMD factorization procedure, which decouples the hard scattering factorization and the factorization of the soft-gluon exchanges~\cite{Collins:2011zzd,Echevarria:2012js,Chiu:2012ir,Vladimirov:2017ksc}. As a result the evolution of TMD  is given by a pair of equations
\begin{eqnarray}\label{evol:gamma}
\mu^2 \frac{d}{d\mu^2}\Phi_{f\ot h}(x,\vec b;\mu,\zeta)&=&\frac{\gamma_F^f(\mu,\zeta)}{2}\Phi_{f\ot h}(x,\vec b;\mu,\zeta),
\\\label{evol:D}
\zeta\frac{d}{d\zeta}\Phi_{f\ot h}(x,\vec b;\mu,\zeta)&=&-\mathcal{D}^f(\mu,\vec b)\Phi_{f\ot h}(x,\vec b;\mu,\zeta),
\end{eqnarray}
where $\gamma_F$ and $\mathcal{D}$ are  respectively the ultraviolet (UV) and rapidity anomalous dimensions. Eq.~(\ref{evol:gamma}-\ref{evol:D}) are independent of  polarization and TMD structure. The double-scale nature of factorization and evolution opens  also unique possibilities for the phenomenological implementation of TMD. In particular, it allows a universal scale-independent definition of a TMD distribution~\cite{Scimemi:2018xaf}.

At the operator level the double-scale nature of evolution is reflected by the presence of two types of divergences, namely UV and rapidity divergences. Both divergences are to be renormalized. The UV renormalization factor is known as TMD-renormalization factor $Z_f^{\text{TMD}}$ and it can be extracted from the UV renormalization of quark (or gluon) vertex attached  to the (light-like) Wilson line. The rapidity renormalization is made through the rapidity renormalization factor $R_f$ (for the proof of multiplicativity of rapidity divergence renormalization,   see ref. \cite{Vladimirov:2017ksc}). It is compulsary that \textit{both renormalizations are made at the level of operator} and thus do not depend on the hadron states. 
The renormalized TMD operators $\mathcal{U}_f$  that defines the physical TMD distribution, reads
\begin{eqnarray}\label{def:renormalization}
\mathcal{U}_f(x,\vec b;\mu,\zeta)&=&Z_i^{-1}(\mu)Z_f^{TMD}\(\frac{\mu^2}{\zeta}\)R_f\(\vec b;\mu,\zeta\)\mathcal{U}_f^{bare}(x,\vec b),
\end{eqnarray}
where we explicitly write the scaling variables for each expression. In eq.~(\ref{def:renormalization}) $Z_i$ is the renormalization of the field wave functions ($Z_2$ for the quark field and $Z_3$ for the gluon field). The TMD operators $\mathcal{U}$ relevant for this work are defined later in eq.~(\ref{def:TMDop_DY},~\ref{def:TMDop_DIS}). 

Both renormalizations are scheme dependent. We use the conventional $\overline{\text{MS}}$-scheme together with the dimensional regularization for the UV divergences. 
For the rapidity renormalization we  use the conventional scheme~\cite{Becher:2010tm,Collins:2011zzd,GarciaEchevarria:2011rb,Chiu:2012ir,Vladimirov:2017ksc} that is fixed by the requirement that no remnants of the soft factor contribute to the hard scattering. Apart from this one should worry about the overlap between  collinear and soft modes in the factorization of the cross sections, which is rapidity regulator dependent.  This is resolved in the $\delta$-regulator scheme where the form of the rapidity renormalization factor is given by the inverse square root of the TMD soft factor $R=1/\sqrt{S}$, see ref. \cite{Echevarria:2015byo}.  
This regulator has been already used several times in higher order calculations, see refs. \cite{Echevarria:2015usa,Echevarria:2015byo,Echevarria:2016scs,Gutierrez-Reyes:2018iod}.

The particular expression depends on the order of application of the renormalization factors. In this work, we fix the order as in eq.~(\ref{def:renormalization}) and we use the $\delta$-regularization, whose definition is given in sec.~\ref{sec:rapidity_div}. Then the rapidity renormalization factor in $\overline{\text{MS}}$-scheme reads \cite{Vladimirov:2017ksc}
\begin{eqnarray}\label{renorm:R_1loop}
R_q(\vec b;\mu,\zeta)&=&1+2a_s C_F \mathbf{B}^\epsilon\mu^{2\epsilon}e^{-\epsilon\gamma_E}\Gamma(-\epsilon)\(\ln\(\mathbf{B} \delta^2\frac{\zeta}{(p^+)^2}\)-\psi(-\epsilon)+\gamma_E\)+O(a_s^2),
\end{eqnarray}
where $\mathbf{B}=\vec b^2/4$ and $a_s=g^2/(4\pi)^2$. The UV renormalization constant is \cite{Echevarria:2016scs}
\begin{eqnarray}\label{renorm:Z_1loop}
Z_2^{-1}Z_q^{TMD}\(\frac{\mu^2}{\zeta}\)&=&\(1-C_F\frac{a_s}{\epsilon}+\mathcal{O}(a_s^2)\)^{-1}\[1-2a_sC_F\(\frac{1}{\epsilon^2}+\frac{2+\ln(\mu^2/\zeta)}{\epsilon}\)+O(a_s^2)\]
\\\nn &=&1-a_sC_F\(\frac{2}{\epsilon^2}+\frac{3+2\ln(\mu^2/\zeta)}{\epsilon}\)+O(a_s^2).
\end{eqnarray}
Here, we list only the renormalization constants for quark operators at one-loop, since they are the only required in the present calculation. The gluon case, as well as, two-loop expressions can be found in ref. \cite{Echevarria:2016scs}. 

We emphasize that the rapidity renormalization factor depends on the boost-invariant combination of scales $\delta/p^+$ \cite{Echevarria:2012js}
(here, $\delta$ regularizes rapidity divergences in $n$-direction and thus transforms as $p^+$ under Lorentz transformations). Such a combination appears in the factorization of 
the cross section of DY and SIDIS and when splitting
the soft factor into parts with rapidity divergences associated with different TMD distributions~\cite{GarciaEchevarria:2011rb}. In the course of factorization procedure, the accompanying TMD distribution (e.g. $D_1$ in (\ref{SF:SIDIS_bessel}) or $f_1$ in (\ref{SF:DY_bessel})) gets the rapidity renormalization factor with $(\delta^-/p^-)\bar \zeta$ argument, where $\delta^-$ regularizes rapidity divergences in $\bar n$-direction. The values of $p^+$ and $p^-$ are arbitrary, however, they dictate the value of $\zeta$ and $\bar \zeta$, since $\zeta \bar \zeta=(2p^+p^-)^2$. The standard and convenient choice of scales is $\zeta \bar \zeta=Q^4$, which is the only physical hard scale appearing  in the reference processes.
This scale determines the value of $p^+$ and $p^-$ as momenta of partons that couple to test current, see also sec.\ref{sec:rapidity_div}. For an extended discussion see sec.~6.1.1 in ref. \cite{Vladimirov:2017ksc} and also refs.~\cite{Echevarria:2012js,Chiu:2012ir}.

\section{Light-cone OPE at  leading order}
\label{sec:LO}

In this section we present the  operators that enter in the definition of the Sivers function and their LO limit  for small-$b$, recovering the results of \cite{Scimemi:2018mmi}.
The notation for operators established in  this section is the one used in the NLO computation. 

\subsection{Light-cone OPE in a regular gauge}
\label{sec:LO_reg}

Let us denote the operator that defines the TMD distributions in DY case as
\begin{eqnarray}\label{def:TMDop_DY}
\mathcal{U}_{\text{DY}}^{\gamma^+}(z_1,z_2,\vec b)&=&\bar T\{\bar q(z_1 n+\vec b)[z_1 n+\vec b,-\infty n+\vec b]\}\,\gamma^+ T\{
[-\infty n-\vec b,z_2 n-\vec b] q(z_2n -\vec b)\},
\nn \\
\end{eqnarray}
where the Wilson lines are defined as
\begin{eqnarray}\label{def:WilsonLine}
[a_1 n+\vec b,a_2n+\vec b]&=& P\exp\(ig\int_{a_2}^{a_1} d\sigma n^\mu A_\mu(\sigma n+\vec b)\).
\end{eqnarray}
The operator that defines the TMD distributions in the SIDIS case reads
\begin{eqnarray}\label{def:TMDop_DIS}
\mathcal{U}_{\text{DIS}}^{\gamma^+}(z_1,z_2,\vec b)&=&\bar T\{\bar q(z_1 n+\vec b)[z_1 n+\vec b,+\infty n+\vec b]\}\,\gamma^+ 
T\{[+\infty n-\vec b,z_2 n-\vec b] q(z_2 n -\vec b)\}.\nn \\
\end{eqnarray}
Generally, the links which connect the end points of Wilson lines at  a distant transverse plane must be added in both operators (for DY and for SIDIS) \cite{Belitsky:2002sm,Idilbi:2010im}. Here, we omit them for simplicity, assuming that some regular gauge (e.g. covariant gauge) is in use. In non-singular gauges the field nullifies at infinities, $A_\mu(\pm \infty n)=0$ and the contribution of distant gauge links vanishes. The case of singular gauges is discussed in the following section.

We point out that for convenience of calculation and presentation the operators in eq.~(\ref{def:TMDop_DY},~\ref{def:TMDop_DIS}) are defined differently in comparison to original operator in eq.~(\ref{def:TMDPDF_Qop}). In particular, we double the transverse distance between fields and write it in symmetric form. Also, the operators in eq.~(\ref{def:TMDop_DY},~\ref{def:TMDop_DIS}) are defined for arbitrary light cone positions $z_1$ and $z_2$, although the definition of a TMD distribution depends only on the difference of these points. Such a generalization does not complicate the calculation, moreover, it allows to cross-check certain results. These modifications are undone on the last step of calculation, see eq.~(\ref{main-fourier}). Note, that the operators  in eq.~(\ref{def:TMDop_DY},~\ref{def:TMDop_DIS}) define the generalized transverse momentum distributions (GTMDs) and thus the obtained OPE can be applied for  generalized  TMD (GTMD) kinematics as well. 

It is straightforward to check that the spatial separations between any pair of fields in the operators defined in eq.~(\ref{def:TMDop_DY},~\ref{def:TMDop_DIS}) are space-like\footnote{There is a single exception. The fields of anti-quark operator and the attached Wilson line have light-like separations but anti-time-ordered. However, the reordering of the operator can performed in the light-cone gauge, where the gauge links vanish. The detailed discussion on the ordering properties of quasi-partonic operators can be found in ref.~\cite{Jaffe:1983hp}.}. For that reason we can replace the $T$- and $\bar T$- orderings by a single $T$-ordering. This significantly simplifies the calculation and  in the following we do not explicitly show the symbol of  T-ordering, but we suppose that each operator is T-ordered. The possibility to reorder the fields is not a general feature, e.g. TMD operators for fragmentation functions do not allow this simplification and thus, their properties are drastically different.

At LO in  perturbation theory one can treat the fields as classical fields, i.e. omit their interaction properties. In this approximation, the small-$b$ expansion is just the Taylor expansion at $\vec b=0$.  Expanding  $\mathcal{U}$   in $\vec b$ up to linear terms we obtain
\begin{eqnarray}\label{U_Taylor}
\mathcal{U}^{\gamma^+}(z_1,z_2,\vec b)=\mathcal{U}^{\gamma^+}(z_1,z_2,\vec 0)+b^\mu \frac{\partial}{\partial  b^\mu} \mathcal{U}^{\gamma^+}(z_1,z_2,\vec b)\Big|_{\vec b=0}+O(\vec b^2).
\end{eqnarray}
The leading term is the same for DY and SIDIS cases
\begin{eqnarray}
\mathcal{U}_{\text{DY}}^{\gamma^+}(z_1,z_2,\vec 0)&=&\mathcal{U}_{\text{DIS}}^{\gamma^+}(z_1,z_2,\vec 0)=\bar q(z_1 n)[z_1 n,z_2 n]\gamma^+ q(z_2 n).
\end{eqnarray}
Note that the half-infinite segments of Wilson lines compensate each other due to the unitarity of the Wilson line and the resulting operator is spatially compact. 

The derivative term in eq.~(\ref{U_Taylor}) is different for different kinematics
\begin{eqnarray}\label{U_DY_der_0}
\frac{\partial}{\partial b^\mu} \mathcal{U}_{\text{DY}}^{\gamma^+}(z_1,z_2,\vec b)\Big|_{\vec b=0}&=&
\bar q(z_1 n)[z_1 n,-\infty n](\overleftarrow{\partial_{T\mu}}-\overrightarrow{\partial_{T\mu}})\gamma^+ [-\infty n,z_2 n] q(z_2n),
\\\label{U_DIS_der_0}
\frac{\partial}{\partial b^\mu} \mathcal{U}_{\text{DIS}}^{\gamma^+}(z_1,z_2,\vec b)\Big|_{\vec b=0}&=&
\bar q(z_1 n)[z_1 n,+\infty n](\overleftarrow{\partial_{T\mu}}-\overrightarrow{\partial_{T\mu}})\gamma^+ [+\infty n,z_2 n] q(z_2n).
\end{eqnarray}
Here, the derivative prevents the compensation of infinite segments of Wilson lines. Acting by derivative explicitly we obtain 
\begin{eqnarray}\label{U_DY_der}
\frac{\partial}{\partial b^\mu} \mathcal{U}_{\text{DY}}^{\gamma^+}(z_1,z_2,\vec b)\Big|_{\vec b=0}&=&
\bar q(z_1n)\(\overleftarrow{D_\mu}[z_1n,z_2n]-[z_1n,z_2n]\overrightarrow{D_\mu}\)\gamma^+ q(z_2n)
\\\nn &&+ig\(\int_{-\infty}^{z_1} +\int_{-\infty}^{z_2}\)d\tau~
\bar q(z_1n)[z_1n,\tau n]\gamma^+ F_{\mu+}(\tau n)[\tau n,z_2n]q(z_2n),
\\
\label{U_DIS_der}
\frac{\partial}{\partial b^\mu} \mathcal{U}_{\text{DIS}}^{\gamma^+}(z_1,z_2,\vec b)\Big|_{\vec b=0}&=&
\bar q(z_1n)\(\overleftarrow{D_\mu}[z_1n,z_2n]-[z_1n,z_2n]\overrightarrow{D_\mu}\)\gamma^+ q(z_2n)
\\\nn &&-ig\(\int^{\infty}_{z_1} +\int^{\infty}_{z_2}\)d\tau~
\bar q(z_1n)[z_1n,\tau n]\gamma^+ F_{\mu+}(\tau n)[\tau n,z_2n]q(z_2n).
\end{eqnarray}
where the covariant derivative and the field-strength tensor are defined as usual
\begin{eqnarray}\label{def:DandF}
\overrightarrow{D}_\mu=\overrightarrow{\partial}_\mu-igA_\mu,\qquad
\overleftarrow{D}_\mu=\overleftarrow{\partial}_\mu+igA_\mu,\qquad F_{\mu\nu}=\partial_\mu A_\nu-\partial_\nu A_\mu-ig[A_\mu,A_\nu].
\end{eqnarray}
The operators which contribute to each order of the small-$b$ expansion have different geometrical twists \footnote{By the term geometrical twist we refer to the standard definition of the twist as ``dimension minus spin'' of the operator. This definition is formulated for a local operator, but it can be  naturally extended to the light-cone operators as a generating function for local operators.}. In particular, the first term in eq.~(\ref{U_DY_der}) is a mixture of twist-2 and twist-3 operators, while the second term is a pure twist-3 operator (the same for  eq.~(\ref{U_DIS_der})). The procedure of separation of different twist contributions is explained in details in \cite{Scimemi:2018mmi}. In the present paper, we skip this discussion because the Sivers function contains only contribution of geometrical twist-3 operator. Indeed, comparing the results for DY in eq.~(\ref{U_DY_der}) and SIDIS in eq.~(\ref{U_DIS_der}) kinematics we observe that the first terms are the same, while the last terms differ. Therefore, already at this stage it is clear that the Sivers function is made of the operators from the last terms, i.e. pure twist-3 operator.

\subsection{Light-cone OPE in the light-cone gauge}
\label{sec:LO_sing}

Before entering a detailed description of the background field method  it is convenient to formulate the derivation of the small-$b$ limit  of the  TMD functions  at LO in the light-cone gauge. This gauge will then be used in the following to describe the background fields.

The definition of TMD operators is gauge invariant. In order to demonstrate this explicitly, let us restore the formal structure of gauge links in eq.~(\ref{def:TMDop_DY},~\ref{def:TMDop_DIS}). We have
\begin{eqnarray}\label{def:TMDop_DY_full}
&&\mathcal{U}_{\text{DY}}^{\gamma^+}(z_1,z_2,\vec b)=
\\&&\nn\quad
\bar q(z_1 n+\vec b)[z_1 n+\vec b,-\infty n+\vec b][-\infty n+\vec b,
-\infty n-\vec b][-\infty n-\vec b,z_2 n-\vec b]
 \,\gamma^+ \,q(z_2n -\vec b),
\\
\label{def:TMDop_DIS_full}
&&\mathcal{U}_{\text{DIS}}^{\gamma^+}(z_1,z_2,\vec b)=
\\&&\nn\quad
\bar q(z_1 n+\vec b)[z_1 n+\vec b,+\infty n+\vec b][+\infty n+\vec b,
+\infty n-\vec b][+\infty n-\vec b,z_2 n-\vec b]
 \,\gamma^+ \,q(z_2n -\vec b).
\end{eqnarray}
Notice, that in order to write eq.~(\ref{def:TMDop_DY_full},~\ref{def:TMDop_DIS_full}) we have explicitly used the fact that the T-ordering can be removed. In the absence of such assumption the finite distance transverse link must be replaced by two half-infinite links~\cite{Belitsky:2002sm}. 

The light-cone gauge is defined by the condition
\begin{eqnarray}\label{gauge:A+=0}
n^\mu A_\mu(x)=A_+(x)=0.
\end{eqnarray}
The application of this condition removes the contribution of gauge links along vector $n$ in the TMD operator, i.e. $[z n+\vec b,\pm\infty n+\vec b]=1$ and $[\pm\infty n-\vec b,-z n-\vec b]=1$. However, the status of the transverse gauge links is unresolved. This reflects the known fact that the gauge fixing condition (\ref{gauge:A+=0}) does not fix the gauge dependence entirely but should be supplemented by an additional boundary condition. There are two convenient choices for boundary conditions in our case\footnote{The names selected here could be misleading since the limit is taken along the light cone, rather then along a time axis. Also the vector boundary condition assumption is too strong.  The quantized Yang-Mills condition $g_T^{\mu\nu}A_\nu$ could be replaced by a weaker $\partial_\mu g_T^{\mu\nu}A_\nu$ as it is shown in~\cite{Chirilli:2015fza}. Nonetheless, for our purposes the condition in eq.~(\ref{gauge:ret},~\ref{gauge:adv}) is sufficient.}
\begin{eqnarray}\label{gauge:ret}
\text{retarded:}&& g_T^{\mu\nu}A_\nu(-\infty n)=0,
\\\label{gauge:adv}
\text{advanced:}&& g_T^{\mu\nu}A_\nu(+\infty n)=0.
\end{eqnarray}
Clearly, each of these boundary conditions is advantageous in some particular kinematics. As so, we apply \textit{the retarded boundary condition for the DY operator}. That is, the transverse link at $-\infty n$ vanishes,
\begin{eqnarray}\label{def:TMDop_DY_LC}
\mathcal{U}_{\text{DY}}^{\gamma^+}(z_1,z_2,\vec b)=
\bar q(z_1 n+\vec b) \,\gamma^+ \,q(z_2n -\vec b),\qquad \text{in the retarded light-cone gauge.}
\end{eqnarray}
Whereas \textit{for the SIDIS operator we apply the advanced boundary condition}. That is, the transverse link at $+\infty n$ vanishes,
\begin{eqnarray}\label{def:TMDop_DIS_LC}
\mathcal{U}_{\text{DIS}}^{\gamma^+}(z_1,z_2,\vec b)=
\bar q(z_1 n+\vec b) \,\gamma^+ \,q(z_2n -\vec b),\qquad \text{in the advanced light-cone gauge.}
\end{eqnarray}
Thus,  the operators have the same expression in different gauges.  In order to recover the structure of gauge links (and hence to obtain the explicitly gauge-invariant operators), we can make a gauge transformation of the operator and subsequently replace each gauge-transformation factor by a Wilson line along the vector $n$ to the selected boundary. 

The OPE in the light-cone gauge has a compact form. The leading term of eq.~(\ref{U_Taylor}) is
\begin{eqnarray}
\mathcal{U}_{\text{DY}/\text{DIS}}^{\gamma^+}(z_1,z_2,\vec 0)&=&\bar q(z_1 n)\,\gamma^+\,q(z_2 n).
\end{eqnarray}
The expression for the derivative of the operator is also independent of the  underlying kinematics (compare to eq.~(\ref{U_DY_der_0},~\ref{U_DIS_der_0}))
\begin{eqnarray}\label{U_der_LC}
\frac{\partial}{\partial b^\mu} \mathcal{U}_{\text{DY}/\text{DIS}}^{\gamma^+}(z_1,z_2,\vec b)\Big|_{\vec b=0}&=&
\bar q(z_1 n)(\overleftarrow{\partial_{T\mu}}-\overrightarrow{\partial_{T\mu}})\gamma^+ q(z_2n),
\end{eqnarray}
and in fact, it already gives the final expression of the correction linear in $\vec b$ in the light-cone gauge. 

Let us show how the results for LO OPE in eq.~(\ref{U_DY_der},~\ref{U_DIS_der}) are recovered starting from eq.~(\ref{U_der_LC}).
 One starts  rewriting eq.~(\ref{U_der_LC})   explicitly  in a gauge-invariant form. With this purpose  we replace the partial derivatives in eq.~(\ref{U_der_LC}) with covariant derivatives, see eq.~(\ref{def:DandF}), by adding (and subtracting) appropriate gluon fields
\begin{eqnarray}\label{U_der_LC+}
\frac{\partial}{\partial b^\mu} \mathcal{U}_{\text{DY}/\text{DIS}}^{\gamma^+}(z_1,z_2,\vec b)\Big|_{\vec b=0}&=&
\bar q(z_1 n)(\overleftarrow{D_{\mu}}-\overrightarrow{D_{\mu}}-ig A_\mu(z_1n)-ig A_\mu(z_2n))\gamma^+ q(z_2n).
\end{eqnarray}
To proceed further,  we have to recall the used boundary condition in the form
\begin{eqnarray}\label{gauge:A=F_ret}
A^\mu(x)&=&-\int_{-\infty}^0 d\sigma~F^{\mu+}(\sigma n+x),\qquad \text{in the retarded light-cone gauge,}
\\\label{gauge:A=F_adv}
A^\mu(x)&=&\int^{\infty}_0 d\sigma~F^{\mu+}(\sigma n+x),~~~\qquad \text{in the advanced light-cone gauge,}
\end{eqnarray}
where $x$ is an arbitrary point. Substituting these expressions into eq.~(\ref{U_der_LC+}) we arrive to eq.~(\ref{U_DY_der},~\ref{U_DIS_der}).

\subsection{Light-cone OPE for the gluon TMD operator}
\label{sec:LO_gluon}

The small-b OPE at NLO contains both quark and gluon collinear operators. The gluon operators that appear in a quark TMD are those that would appear in the small-$b$ OPE for gluon TMD operator. Since this expansion for gluons has never been considered in the literature we briefly describe it here. 

We define the gluon TMD operator as (compare to eq.~(\ref{def:TMDop_DY},~\ref{def:TMDop_DIS}))
\begin{eqnarray}\label{def:TMDop_DY_gluon}
\mathcal{G}_{\text{DY}}^{\mu\nu}(z_1,z_2,\vec b)&=&F^{\mu+}(z_1 n+\vec b)[z_1 n+\vec b,-\infty n+\vec b]
[-\infty n-\vec b,z_2 n-\vec b] F^{\nu+}(z_2n -\vec b),
\\
\label{def:TMDop_DIS_gluon}
\mathcal{G}_{\text{DIS}}^{\mu\nu}(z_1,z_2,\vec b)&=&F^{\mu+}(z_1 n+\vec b)[z_1 n+\vec b,+\infty n+\vec b]
[+\infty n-\vec b,z_2 n-\vec b] F^{\nu+}(z_2n -\vec b),
\end{eqnarray}
where the Wilson lines are in the adjoint representation, i.e. the contraction of the color indices\footnote{This is the only color structure that appears in the leading power of TMD factorization. The so-called dipole TMD distributions that couples to opposite directed Wilson lines in the fundamental representation do not appear in the factorization of SIDIS or DY processes.} is $F^A(z_1)[..]^{AB}F^B(z_2)$.  The parametrization of the corresponding TMD matrix elements can be found e.g. in~\cite{Echevarria:2015uaa}.

The evaluation of the light-cone OPE for gluon operators is totally analogous to the one made in sec.~\ref{sec:LO_reg}. The only difference is that the quark fields are replaced by $F^{+\mu}$ and the covariant derivatives act in the adjoint representation. We obtain the following analog of eq.~(\ref{U_DY_der},~\ref{U_DIS_der})
\begin{eqnarray}\label{G_DY_der}
\frac{\partial}{\partial b^\rho} \mathcal{G}_{\text{DY}}^{\mu\nu}(z_1,z_2,\vec b)\Big|_{\vec b=0}&=&
F^{\mu+}(z_1n)\(\overleftarrow{D_\rho}[z_1n,z_2n]-[z_1n,z_2n]\overrightarrow{D_\rho}\)F^{\nu+}(z_2n)
\\\nn &&+ig\(\int_{-\infty}^{z_1} +\int_{-\infty}^{z_2}\)d\tau~
F^{\mu+}(z_1n)[z_1n,\tau n]F_{\rho+}(\tau n)[\tau n,z_2n]F^{\nu+}(z_2n),
\\
\label{G_DIS_der}
\frac{\partial}{\partial b^\rho} \mathcal{G}_{\text{DIS}}^{\mu\nu}(z_1,z_2,\vec b)\Big|_{\vec b=0}&=&
F^{\mu+}(z_1n)\(\overleftarrow{D_\rho}[z_1n,z_2n]-[z_1n,z_2n]\overrightarrow{D_\rho}\)F^{\nu+}(z_2n)
\\\nn &&-ig\(\int^{\infty}_{z_1} +\int^{\infty}_{z_2}\)d\tau~
F^{\mu+}(z_1n)[z_1n,\tau n] F_{\rho+}(\tau n)[\tau n,z_2n]F^{\nu+}(z_2n),
\end{eqnarray}
where the covariant derivatives are in the adjoint representation. Alike the quark case, the only operators which contribute to the Sivers function are given in the second lines of these equations.

\section{Light-cone OPE at  next-to-leading order}
\label{sec:OPENLO}

The object of this section is to introduce the calculation of OPE for $\mathcal{U}$ up to terms linear in $\vec b$  at NLO in perturbation theory. 
The OPE is realized when $\vec b^2\ll \Lambda^{-2}$  and it  looks like
\begin{eqnarray}\label{OPE_gen4}
\mathcal{U}(z,\vec b)&=&\sum_n C^{\text{tw-2}}_n(z,\mathbf{L}_\mu,a_s(\mu))\otimes \mathcal{O}^{\text{tw2}}_{n}(z;\mu)
\\\nn &&+
b_\nu \sum_n C^{\text{tw-3}}_n(z,\mathbf{L}_\mu,a_s(\mu))\otimes \mathcal{O}^{\nu,\text{tw3}}_{n}(z;\mu)+O(\vec b^2),
\end{eqnarray}
where $C$ are the coefficient functions which depend on $\vec b^2$ logarithmically, $n$ enumerates all available operators at this order and $\otimes$ is some integral convolution in variables $z$. Here, we also introduce the notation for the coupling constant $a_s=g^2/(4\pi)^2$ and for the logarithm combination that typically enters in perturbative calculations
\begin{eqnarray}\label{def:Lmu}
\mathbf{L}_\mu=\ln\(\frac{\mu^2 \vec b^2}{4 e^{-2\gamma_E}}\).
\end{eqnarray}
The variable $\mu$ represents the scale of OPE. 

The complexity of the computations for OPE increases drastically passing from LO to NLO in perturbative QCD. In the latter case one  cannot omit the field interactions, as it happens in ordinary Taylor expansion as in eq.~(\ref{U_Taylor}). The propagation of fields between different points is responsible of the fact that eq.~(\ref{U_Taylor}) is to be modified in the presence of interactions which can pick up additional fields from the vacuum. Moreover, the  OPE with interacting fields  contains all possible operators with correct (as  prescribed by the theory) quantum numbers.

An additional difficulty in the present calculation is that only a few computing methods have been tested on higher twist operators. For the twist-2 TMD operators the matching procedure is simple because in the OPE a TMD is in a one-to-one correspondence with the on-shell matrix elements over collinear-parton states. 
In the case of higher twist operators the only matrix elements of collinear partons are not suitable for obtaining the matching coefficients, since a transverse component of momentum is needed to carry the operator indices. It can also happen that a matrix element over collinear partons is not infrared-safe and it requires an additional regularization with a (specific) separation of pole contributions, see e.g.~\cite{Ji:2006vf,Chen:2017lvx}. These problems are solved using off-shell matrix elements, which is significantly more complicated, due to the fact that the higher-twist operators mix with each other via QCD equations of motion and that off-shell colored states are not generally gauge invariant.  The best method to evaluate the coefficient functions at higher twist results to be the \textit{background-field method}.  At the diagram level, the method is equivalent to the evaluation of a generic matrix elements, with the main difference that the result of the calculation is given explicitly in  operator form.  The method  allows to keep track of gauge properties and significantly simplifies the processing of equations of motion. Altogether, these properties make the background-field method very effective for  higher twist calculations.  In the following we concentrate on this method, for which  we provide a brief general introduction in sec.~\ref{sec:intro-to-back}. The details of the calculation are given in sec.~\ref{sec:background}-\ref{sec:eval}. The treatment of rapidity divergences and renormalization needs a special discussion which is provided in sec.~\ref{sec:rapidity_div}-\ref{sec:backrenormalization}. All the computation is done for the DY case, but the passage to the SIDIS case does not present particular difficulties and the comparison of the two cases is provided in sec.~\ref{sec:DY-SIDIS-difference}.

\subsection{OPE in  background field method}
\label{sec:intro-to-back}

The background-field method is founded on the idea of mode separation. The operator matrix element between states $S_1$ and $S_2$ is defined as
\begin{eqnarray}\label{gen:func_int1}
\langle S_1|\mathcal{U}|S_2\rangle = \int \mathcal D \Phi ~\Psi^*_{S_1}[\Phi] \,\mathcal{U}[\Phi] \,\Psi_{S_2}[\Phi]\,e^{i \mathcal{S}[\Phi]},
\end{eqnarray}
where the  letter $\Phi$ represents  any QCD field $\{\bar q, q,A_\mu\}$, $\Psi_S$ is the wave function of the state $S$ and $\mathcal{S}$ is the action of QCD. Let us split the fields into the ``fast'' and ``slow'' (or ``short-correlated'' and ``long-correlated'' in position space terminology)  components, as 
\begin{eqnarray}
\Phi(x) =  \varphi(x;\mu)+\phi(x;\mu).
\end{eqnarray}
Here, the ``fast'' modes $\phi$ have momentum $p>\mu$, while ``slow'' modes have momentum $p<\mu$. The (factorization) scale $\mu$  is not explicitly defined but it is large enough to guarantee the convergence of the perturbative series. In the following we omit the argument $\mu$ for the fields.  We postulate that physical states (hadrons) are built from the ``slow'' components, i.e. $\Psi_S[\Phi]=\Psi_S(\varphi)$ so that eq.~(\ref{gen:func_int1}) turns into
\begin{eqnarray}
\langle S_1|\mathcal{U}|S_2\rangle = \int \mathcal D \varphi\, \mathcal D \phi ~\Psi^*_{S_1}[\varphi] \,\mathcal{U}[\varphi+\phi](x) \,\Psi_{S_2}[\varphi]\,e^{i \mathcal{S}[\varphi+\phi]}.
\end{eqnarray}
In this expression the integral over ``fast'' components can be evaluated and the expression for observables has the following effective form
\begin{eqnarray}\label{gen:slow_op}
\langle S_1|\mathcal{U}|S_2\rangle = \int \mathcal D \varphi\, \Psi^*_{S_1}[\varphi] \,\widetilde{\mathcal{U}}[\varphi](x) \,\Psi_{S_2}[\varphi]\,e^{i \mathcal{S}[\varphi]},
\end{eqnarray}
where
\begin{eqnarray}
\widetilde{\mathcal{U}}[\varphi](x)=\int \mathcal D \phi ~\mathcal{U}[\varphi+\phi](x) ~e^{i \mathcal{S}[\varphi+\phi]-i \mathcal{S}[\varphi]}.
\end{eqnarray}
The mode separation then assumes that  the ``slow'' fields can be treated as free-fields on distances $~x^2$. This hypothesis is typical for effective field theories (see for instance \cite{Beneke:2002ph,Bauer:2000yr,Bauer:2001yt} for the application of similar concepts in soft collinear effective theory (SCET) or \cite{Balitsky:2016dgz} for TMD factorization at small-x). 

One can interpret the construction in eq.~(\ref{gen:slow_op}) as an evaluation of the perturbative QCD fields in a general parton background, which gives the method its name. After the integration of the ``fast'' fields in eq.~(\ref{gen:slow_op}), the resulting effective operator is then expanded using  free-theory twist expansion,  as it was done in sec.~\ref{sec:LO}. It is important to realize that in  background calculation the result is gauge-invariant and satisfies QCD equations of motion at each step of the evaluation (even for each diagram). The result then is also universal, that is, it is valid for all states (we do not even specify them) and thus, we can operate only with fields $\varphi$.  Essentially, the background field methods is concentrated in a single definition, eq.~(\ref{gen:slow_op}). 

The background field method is an essential tool of the modern small-x calculations. In this case the separation of kinematic modes is based on the strong ordering in rapidity, which is a distinctive feature of the small-x kinematics. To define different modes one has to introduces a rapidity cutoff parameter $\sigma$, which separates ``fast'' ($p^+<\sigma$) and  ``slow'' ($p^+>\sigma$) fields based on the value of the longitudinal component of the momenta $p^+$. Instead of the twist expansion the calculation of the functional integral over ``fast'' fields (\ref{gen:slow_op}) is now performed in the so-called shock-wave approximation. Since the procedure of separation of modes is quite general, the method can incorporate different kinematic regimes, which has been recently employed in ~\cite{Balitsky:2015qba, Balitsky:2016dgz}.

\subsection{QCD in  background field}
\label{sec:background}

The QCD Lagrangian reads
\begin{eqnarray}
\mathcal{L}=\bar q(i\fnot \!D)q+\frac{1}{4}F_{\mu\nu}^a F^{\mu\nu}_a+\text{gauge fix},
\end{eqnarray}
where the covariant derivative and $F_{\mu\nu}$ are defined in eq.~(\ref{def:DandF}). Following the mode separation we split the fields as $A_\mu\to A_\mu+B_\mu$ and $q\to q+\psi$, where $\psi$ and $B_\mu$ are ``fast'' fields and $q$ and $A_\mu$ are ``slow'' (background) fields. The separation of modes in the main body of the Lagrangian is straightforward, but the gauge fixing term should be considered with caution. The ultimately convenient point of the background field method is the possibility to choose different classes of gauge fixing for different modes. The detailed discussion on gauge fixing in QCD with background method is given in \cite{Abbott:1980hw,Abbott:1981ke}. 

We choose the most convenient combination of gauges for our task. For ``fast'' components we use the background-field gauge,
\begin{eqnarray}\label{gauge:back}
(\partial_\mu \delta^{AC}+g f^{ABC}A_\mu^B)B^{\mu,C}=D_\mu[A]B^\mu=0\ ,
\end{eqnarray}
which is the analog of covariant gauge fixing in the usual QCD perturbation theory. 
In particular, the propagator has the familiar form
\begin{eqnarray}\label{gauga:alpha}
\contraction{}{B}{\,(x)~}{B}
B_\mu^A(x)B_\nu^B(0)&=&\int \frac{d^dk}{(2\pi)^d}e^{-ikx}\frac{-i \delta^{AB}}{k^2+i0}\(g^{\mu\nu}-(1-\alpha) \frac{k^\mu k^\nu}{k^2+i0}\),
\end{eqnarray}
where $\alpha$ is a free parameter. For background fields we use  light-cone gauge eq.~(\ref{gauge:A+=0}) with retarded boundary condition eq.~(\ref{gauge:A=F_ret}) for DY operators and advanced boundary condition eq.~(\ref{gauge:adv}) for SIDIS operators.

In  background field formulation, the Lagrangian of QCD splits into three parts
\begin{eqnarray}
\mathcal{L}=\mathcal{L}[q,A]+\mathcal{L}[\psi,B]+\delta \mathcal{L},
\end{eqnarray}
where the first two terms are usual QCD Lagrangians built for particular modes and the last term is the ``fast-slow'' modes interaction,
\begin{eqnarray}\label{QCD:deltaL}
\delta \mathcal{L}=g\(\bar q \fnot B\psi+\bar \psi \fnot B q+\bar \psi \fnot A \psi\)+\delta \mathcal{L}_{ABB}+\delta \mathcal{L}_{AABB}+\delta \mathcal{L}_{ABBB},
\end{eqnarray}
where $\delta \mathcal{L}_{ABB}$ ($\delta \mathcal{L}_{ABBB}$) is the interaction of a single field $A_\mu$ with two (three) fields $B_\mu$ and $\delta \mathcal{L}_{AABB}$ is the interaction of two fields $A_\mu$ with two fields $B_\mu$. 
These terms depend on the gauge fixing condition. For our calculation we need only the $\delta \mathcal{L}_{ABB}$ interaction. It reads
\begin{eqnarray}
\delta\mathcal{L}_{ABB}&=&-gf^{ABC} A_\mu^A (\partial_\alpha B^B_\beta)B_\gamma^C\(2g^{\mu\beta} g^{\alpha \gamma}-g^{\mu \alpha}g^{\beta \gamma}-\frac{1+\alpha}{\alpha} g^{\mu\gamma}g^{\alpha \beta}\).
\end{eqnarray}
The rest of the terms can be found in \cite{Abbott:1980hw}. In the following, we consider the case $\alpha=1$, which corresponds to the ``Feynman gauge version'' of the background gauge.

\subsection{Evaluation of diagrams}
\label{sec:eval}

\begin{figure}[t]
\begin{center}
\includegraphics[width=0.99\textwidth]{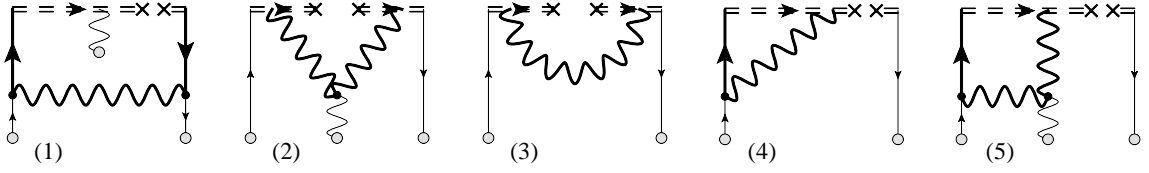}
\caption{\label{fig:0} Example of diagrams that vanish in our scheme of calculation. Diagrams (1) and (2) vanish due to $A_+=0$. Diagram (3) is proportional to $1-\alpha$  and vanish at $\alpha=1$. Diagrams (4) and (5) vanish since the dimensionally regularized loop integral does not have a scale. The bold lines denote the propagators of quantum fields. The thin lines with bubbles are background fields. The double dashed lines are Wilson lines and crosses show that they are pointing to light cone infinity.}
\end{center}
\end{figure}

We would like to evaluate the effective operator in eq.~(\ref{gen:slow_op}) up to twist-3 corrections, at $a_s$ order. The computation proceeds  expanding the  interaction part of the exponent in eq.~(\ref{gen:slow_op}) and integrating the ``fast'' modes by the Gaussian integration formula. i.e we obtain the Feynman diagrams with background fields as the external sources.  The divergences of loop-integrals are regularized 
by dimensional regularization and $\delta$-regulator as in \cite{Echevarria:2016scs,Echevarria:2015usa,Echevarria:2015byo}, which allows us to use renormalization factors of eq.~(\ref{renorm:R_1loop},~\ref{renorm:Z_1loop}). 

In summary,  the calculation follows this path:
\begin{itemize}
\item The dynamical fields are in background gauge, eq.~(\ref{gauge:back}) with the parameter $\alpha =1$, eq.~(\ref{gauga:alpha}).
\item The background fields are in  light-cone gauge, eq.~(\ref{gauge:A+=0}) with the retarded eq.~(\ref{gauge:A=F_ret}) (advanced eq.~(\ref{gauge:A=F_adv})) boundary condition for DY (SIDIS) operator.
\item The UV and collinear divergences are regularized by the dimensional regularization with $d=4-2\epsilon$. We use the conventional $\overline{\text{MS}}$ scheme with $(e^{-\gamma_E}/4\pi)^\epsilon$ factor for each $a_s=g^2/(4\pi)^2$.
\item The rapidity divergences are regularized by $\delta$-regularization, defined in \cite{Echevarria:2016scs}. See detailed discussion in sec.~\ref{sec:rapidity_div}.
\end{itemize}
Within this scheme many diagrams vanish. Some examples of null diagrams are shownin fig.~\ref{fig:0}. \textit{(i)} and more specifically
we have the following cases of vanishing diagrams:
\textit{(i)} The diagrams with the background field coupled directly, or through a sub-graph, to the Wilson lines, such as diagrams diagrams (1) and (2) in fig.~\ref{fig:0}. They vanish due to light-cone gauge fixing, $A_+=0$. \textit{(ii)} The diagrams with a ``Wilson-lines reducible subgraph'', such as the diagram (3) in fig.~\ref{fig:0}. They are proportional to $1-\alpha$ and thus vanish at $\alpha=1$. \textit{(iii)} The diagrams without interaction of fields at different transverse positions (i.e. with $\vec b$ and $-\vec b$), such diagrams are diagrams (4) and (5) in fig.~\ref{fig:0}. They are zero in  dimensional regularization, since loop-integrals in such diagrams are scaleless.

\begin{figure}[t]
\begin{center}
\includegraphics[width=0.6\textwidth]{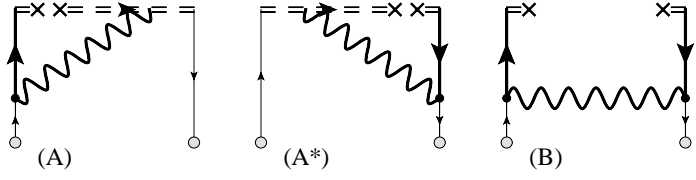}
\caption{\label{fig:2point} The non-vanishing diagrams with two insertions of background fields. The bold lines denote the propagators of quantum fields. The thin lines with bubbles are background fields. The double dashed lines are Wilson lines and crosses show that they are pointing to light-cone infinity.}
\end{center}
\end{figure}

\begin{figure}[t]
\begin{center}
\includegraphics[width=0.8\textwidth]{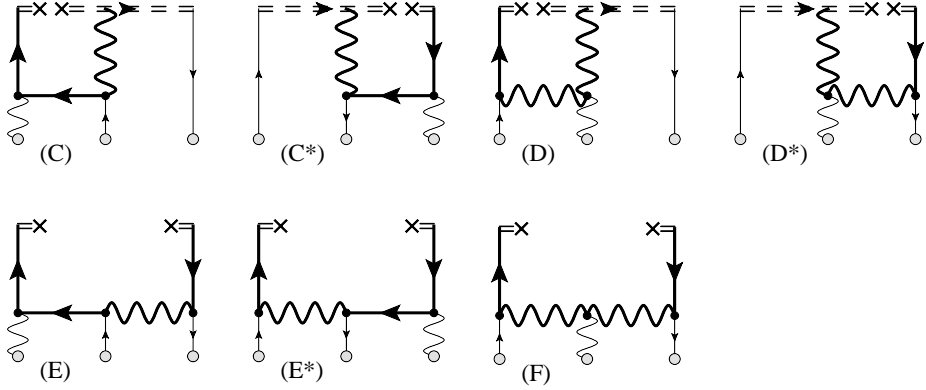}
\caption{\label{fig:3point} The non-vanishing diagrams with three insertions of background fields. The bold lines denote the propagators of quantum fields. The thin lines with bubbles are background fields. The double dashed lines are Wilson lines and crosses show that they are pointing to light-cone infinity.}
\end{center}
\end{figure}

\begin{figure}[t]
\begin{center}
\includegraphics[width=0.4\textwidth]{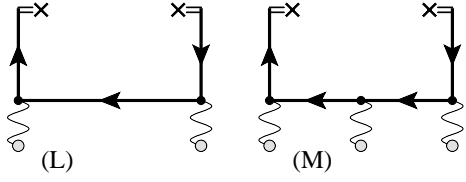}
\caption{\label{fig:QG} The non-vanishing diagrams that mix quark and gluon operators. The bold lines denote the propagators of quantum fields. The thin lines with bubbles are background fields. The double dashed lines are Wilson lines and crosses show that they are pointing to light-cone infinity.}
\end{center}
\end{figure}

The rest of contributions  are conveniently ordered with respect to the number of background fields. Since the number of fields in the operator is less or equal to the twist of the operator, only the diagrams with two or three background fields contribute at a specific power of OPE. There are 6 non-vanishing diagrams at this order (4 of them have charge conjugated diagrams). The diagrams with two quark fields are shown in fig.~\ref{fig:2point}. The diagrams with two quark and gluon fields are shown in fig.~\ref{fig:3point}. There are also diagrams (with two and three field) that mix the quark operator with the gluon operator, as  in fig.~\ref{fig:QG}. In principle, there  could be also diagrams with more gluon insertions, which are to be combined with a single gluon insertion into a gauge invariant combination $F_{\mu\nu}$ (with both transverse indices). However, we recall that only $F_{\mu+}$ contributes to operators of twist-3 and in the light-cone gauge $F_{\mu+}=-\partial_+A_\mu$. Thus, such diagrams should not be considered at twist-3 accuracy.

The process of diagrams computation is almost elementary. Let us show here the evaluation of the simplest diagram, diagram A. A similar evaluation (with the only difference in the path of Wilson lines) is presented in~\cite{Balitsky:1987bk}, which allows an instructive comparison. Also, in ref.~\cite{Vladimirov:2014aja} the diagram A (and the diagram B) has been calculated in  momentum space for all values of $\vec b$, which allows to match the scheme factors. Importantly, the diagram A plays a special role in TMD physics, since it is the only diagram which has rapidity divergences as discussed in the next section. In appendix~\ref{app:exampleDiag} we also present a detailed  explanation of the computation technique for one of the most difficult diagrams (diagram E).  

The diagram A  comes from the following contraction of fields in eq.~(\ref{gen:slow_op})
\begin{eqnarray}\label{diagA:0}
&&\widetilde{\mathcal{U}}_A=
\\\nn &&\qquad
\contraction[6pt]{
\Big\{\bar q(z_1 n+\vec b)\Big[ig \int_{-\infty}^{z_1} d\sigma n^\mu t^A B^A_\mu(n \sigma+\vec z_1)\Big]\gamma^+}{\psi)}{(z_2 n-\vec b)\Big)\Big(ig\int d^d y}{ \bar \psi}
\contraction[8pt]{\Big(\bar q(z_1 n+\vec b)\Big[ig \int_{-\infty}^{z_1} d\sigma n^\mu t^A }{B^A_\mu}{(n \sigma+\vec z_1)\Big]\gamma^+\psi(z_2 n-\vec b)\Big\}\Big(ig\int d^d y \bar \psi(y)}{\fnot B}
\Big\{\bar q(z_1 n+\vec b)\Big[ig \int_{-\infty}^{z_1} d\sigma n^\mu t^A B^A_\mu(n \sigma+\vec b)\Big]\gamma^+\psi(z_2 n-\vec b)\Big\}\Big(ig\int d^d y \bar \psi(y)\fnot B(y)q(y)\Big),
\end{eqnarray}
where the factor in the square brackets is  part of the Wilson line and the factor in the round brackets is  part of $\delta \mathcal{L}$ (see eq.~(\ref{QCD:deltaL})). Note, that here we consider the DY operator, which dictates the integration limits over $\sigma$. The propagators in  dimensional regularization (with $d=4-2\epsilon$) are
\begin{eqnarray}
\contraction{}{\psi_i}{(x)}{\bar \psi}\psi_i(x)\bar \psi_j(0)&=&\frac{\Gamma(2-\epsilon)}{2\pi^{d/2}}\frac{i\fnot x_{ij}}{(-x^2+i0)^{2-\epsilon}}
\\
\contraction{}{B}{\,(x)~}{B}
B_\mu^a(x)B_\nu^b(0)&=&\frac{\Gamma(1-\epsilon)}{4\pi^{d/2}}\frac{-g_{\mu\nu}\delta^{ab}}{(-x^2+i0)^{1-\epsilon}},
\end{eqnarray}
where the gluon propagator is taken with $\alpha=1$. Explicitly, the diagram reads
\begin{eqnarray}
\widetilde{\mathcal{U}}_A&=&
-ig^2C_F\frac{\Gamma(2-\epsilon)\Gamma(1-\epsilon)}{8 \pi^d}
\\\nn &&\int_{-\infty}^{z_1} d\sigma  \int d^d y\,\bar q(z_1 n+\vec b)\frac{2\gamma^+ y^+}{(-(y-nz_2+\vec b)^2+i0)^{2-\epsilon}(-(y-n \sigma-\vec b)^2+i0)^{1-\epsilon}} q(y),
\end{eqnarray}
where we have simplified gamma- and color-algebra.

To proceed further we join the propagator with a usual Feynman trick, introducing a single Feynman parameter $\alpha$. The resulting propagators is $(-y^2+2y^+(\sigma \alpha +(1-\alpha)z_2)+2(yb)(1-2\alpha)+\vec b^2).$ We diagonalize it by a shift $y^\mu \to y^\mu+n^\mu(\alpha \sigma+(1-\alpha)z_2)-(1-2\alpha)b^\mu$ and obtain
\begin{eqnarray}\label{diagA:2}
\widetilde{\mathcal{U}}_A&=&-ig^2C_F\frac{\Gamma(3-2\epsilon)}{4 \pi^d}
\\\nn &&
\int_{-\infty}^{z_1} d\sigma  
\int d^d y\int_0^1 d\alpha \bar q(z_1n+\vec b)\frac{\gamma^+ y^+ \alpha^{-\epsilon}\bar \alpha^{1-\epsilon}}{(-y^2+4 \alpha \bar \alpha \vec b^2+i0)^{3-2\epsilon}} q(y+nz_{2\sigma}^{\alpha}-(1-2\alpha)\vec b),
\end{eqnarray}
where $\vec b^2=-b^2>0$, $\bar \alpha=1-\alpha$ and $z_{2\sigma}^\alpha=z_2 \bar \alpha+\sigma \alpha $.  Starting from here we use the following notation
\begin{eqnarray}
z_{ij}^\alpha=z_i \bar \alpha+z_j \alpha,\qquad \bar \alpha =1-\alpha.
\end{eqnarray}
If the  indices $i$ ($j$) are replaced by $\sigma$, the $z_i$ ($z_j$) is replaced by $\sigma$.

In order to evaluate the integral over $y$, we recall that the background field is a classical field and the expressions of the form eq.~(\ref{diagA:2}) should be understood as a generating function for the whole tower of twist-operators. Therefore, \textit{we are allowed to make the twist-expansion under the loop-integral sign.} In the considered case, we make the Taylor expansion at $y^\mu=0$, $q(y+x)=(1+y^\mu\partial_\mu+y^\mu y^\nu/2\, \partial_\mu\partial_\nu+...) q(x)$. The loop-integration can be taken for each term in the series. The necessary loop-integral reads 
\begin{eqnarray}
\int \frac{y^{\mu_1}...y^{\mu_{2n}}}{(-y^2+X+i0)^{3-2\epsilon}}&=&-i\pi^{d/2}\frac{\Gamma(1-\epsilon-n)}{\Gamma(3-2\epsilon)}\frac{(-1)^n g_s^{\mu_1...\mu_{2n}}}{2^n X^{1-\epsilon-n}},
\end{eqnarray}
where $g_s$ is a completely symmetric composition of metric tensors. For an odd number of indices the loop-integral is zero.

Metric tensors produced by loop-integration can contract derivatives, vectors $b^\mu$ and $n^\mu$. Each term in the series should be sorted with respect to its twist. The thumb rule is that each transverse derivative increases the twist of an operator, but the light-cone derivative does not. Thus, the higher derivative term could be dropped. Alternatively, one can count the power of the vector $\vec b$. In our current calculation, we evaluate up to terms linear in $\vec b$. Note, that strictly speaking we should also expand fields in the powers of $\vec b$, but it does not affect the diagram evaluation and can be postponed until later stage.

The expression in eq.~(\ref{diagA:2}) has a very simple numerator, which is linear in $y$. So, only odd terms of Taylor series contribute. Moreover, already the second term in the expansion, the one with three derivatives $\sim y^\mu y^\nu y^\rho \partial_\mu \partial_\nu\partial_\rho q/3!$, vanishes after contraction. Indeed, it generates $\partial_+\partial^2 q$, that is at least twist-4 (on top, this contributions is proportional to $\vec b^2$). Therefore, we consider only the single-derivative term of the series and obtain
\begin{eqnarray}\label{diagA:3}
\widetilde{\mathcal{U}}_A&=&
2a_sC_F\Gamma(-\epsilon) \vec b^{2\epsilon}
 \int_{-\infty}^{z_1} d\sigma  
\int_0^1 d\alpha \,\bar \alpha~\bar q(nz_1+\vec b)\gamma^+ \overrightarrow{\partial_+}q(n z^\alpha_{2\sigma}-(1-2\alpha)\vec b)+O(\vec b^2\partial^2q).~~
\end{eqnarray}
Charge-conjugated diagrams can be evaluated independently, or obtained from the direct diagrams by reversing the order of field arguments and with the replacement $z_1\leftrightarrow z_2$. I.e. the diagram A$^*$ reads
\begin{eqnarray}
\widetilde{\mathcal{U}}_{A^*}&=&2a_sC_F\vec b^{2\epsilon} \Gamma(-\epsilon)
 \int_{-\infty}^{z_2} d\sigma  
\int_0^1 d\alpha\,\bar\alpha ~\bar q(z_{1\sigma}^\alpha n+(1-2\alpha)\vec b)\overleftarrow{\partial_+}\gamma^+ q(z_2n-\vec b)+O(\vec b^2\partial^2\bar q).~~
\end{eqnarray}
These expressions contain rapidity divergences, which are discussed in the next section. All other diagrams are evaluated similarly. 

The expression for the diagram $A$ in SIDIS kinematics is almost identical to DY case. The only modification is the lower limit for integration over $\sigma$ iin eq.~(\ref{diagA:0}), which must be changed to $(+\infty)$ for the SIDIS case. Such a replacement does not affect the evaluation of the diagram and thus the analog of eq.~(\ref{diagA:3}) in the SIDIS kinematics is obtained replacing $(-\infty)$ by $(+\infty)$.

\subsection{Treatment of rapidity divergences}
\label{sec:rapidity_div}

The rapidity divergences appear due to the localization of a gluon field in the transverse plane at the light-cone infinity \cite{Vladimirov:2017ksc}. 
There are three diagrams that have interactions with a Wilson line and thus, that are potentially rapidity divergent. 
These are diagrams A, C and D. However, according to the general counting rule \cite{Vladimirov:2017ksc}, only the diagram A is rapidity divergent. In this section, we demonstrate how rapidity divergences arise in background field calculation.

The fact that diagram A is rapidity divergent is well-known. It has been calculated in numerous works, see e.g. the discussions in ref.~\cite{Echevarria:2015usa,Collins:2011zzd,GarciaEchevarria:2011rb,Gutierrez-Reyes:2017glx,Vladimirov:2014aja}. In all these works, the diagrams have been calculated in  momentum space, where the loop-integral 
is explicitly divergent. In our case the loop-integral in the diagram A has been evaluated without any problems,   however, as we demonstrate shortly, the result of the integral in eq.~(\ref{diagA:3}) is ambiguous and the resolution of this ambiguity gives rise to the rapidity divergence.

The ambiguity in diagram A is hidden in the argument of the quark field. Indeed, its value at  point $(\alpha,\sigma)=(0,-\infty)$ depends on the path  used to approach this point. In particular, we find
\begin{eqnarray}
\lim_{\alpha\to 0}\lim_{\sigma\to -\infty} q(n z^\alpha_{2\sigma})&=&
q(-\infty)=0,
\\
\lim_{\sigma\to -\infty}\lim_{\alpha\to 0}q(n z^\alpha_{2\sigma})&=&
q(z_2),
\end{eqnarray}
and  the integration over $\sigma$ and $\alpha$ does not commute in the vicinity of $(0,-\infty)$. 

In order to resolve the ambiguity, the dependence on $\alpha$ and $\sigma$ should be separated. Let us rewrite eq.~(\ref{diagA:3}) as
\begin{eqnarray}\label{diagA:4}
\widetilde{\mathcal{U}}_A&=&
2a_sC_F\Gamma(-\epsilon) \vec b^{2\epsilon}
 \int_{-\infty}^{z_1} d\sigma  
\int_0^1 d\alpha \,\frac{\bar \alpha}{\alpha}~\bar q(nz_1)\gamma^+ \frac{\partial}{\partial \sigma} q(n z^\alpha_{2\sigma}),
\end{eqnarray}
where we set $\vec b$ in the arguments of the fields to $\vec 0$, for demonstration purposes (the presence of $\vec b$ in the argument does not change the procedure of rapidity divergence elaboration and we restore it at the end of the section). In eq.~(\ref{diagA:4}) the ambiguity at $(0,-\infty)$ is enforced by the divergence of the integrand at $\alpha\to 0$. We isolate the ambiguous part of the diagram splitting the integration into two parts
\begin{eqnarray}
\widetilde{\mathcal{U}}_A=\widetilde{\mathcal{U}}^{\text{reg}}_A+\widetilde{\mathcal{U}}^{\text{sing}}_A,
\end{eqnarray}
where
\begin{eqnarray}\label{diagA:reg1}
\widetilde{\mathcal{U}}^{\text{reg}}_A&=&2a_sC_F\Gamma(-\epsilon) \vec b^{2\epsilon}
 \int_{z_2}^{z_1} d\sigma  
\int_0^1 d\alpha \,\frac{\bar \alpha}{\alpha}~\bar q(nz_1)\gamma^+ \frac{\partial}{\partial \sigma} q(n z^\alpha_{2\sigma}),
\\\label{diagA:sing1}
\widetilde{\mathcal{U}}^{\text{sing}}_A&=&2a_sC_F\Gamma(-\epsilon) \vec b^{2\epsilon}
 \int_{-\infty}^{z_2} d\sigma  
\int_0^1 d\alpha \,\frac{\bar \alpha}{\alpha}~\bar q(nz_1)\gamma^+ \frac{\partial}{\partial \sigma} q(n z^\alpha_{2\sigma}).
\end{eqnarray}
The regular part does not contain the problematic point and thus the order of integration is irrelevant. Taking the integral over $\sigma$ by parts, we obtain
\begin{eqnarray}
\widetilde{\mathcal{U}}^{\text{reg}}_A&=&2a_sC_F\Gamma(-\epsilon) \vec b^{2\epsilon}
\int_0^1 d\alpha \,\frac{\bar \alpha}{\alpha}~\Big[\bar q(nz_1)\gamma^+ q(n z^\alpha_{21})-\bar q(nz_1)\gamma^+ q(n z_{2})\Big].
\end{eqnarray}
This expression is regular at $\alpha\to 0$ since $z_{21}^{\alpha=0}=z_2$ and it  is a position representation form of the well-known ``plus''-distribution.

To evaluate the singular part we introduce a regulator. Here, we use the $\delta$-regularization, which consists in the following modification of the Wilson line
\begin{eqnarray}
P\exp\(ig \int_{-\infty}^z d\sigma A_+(n\sigma+x)\) \to P\exp\(ig \int_{-\infty}^z d\sigma A_+(n\sigma+x)e^{-\delta |\sigma|}\) ,
\end{eqnarray}
where $\delta >0$. Such modification breaks gauge invariance by power corrections and therefore, only the limit $\delta \to 0$ is gauge invariant. For the detailed discussion of this issue we refer to \cite{Echevarria:2015byo}. In  $\delta$-regularization the interaction vertex with Wilson line as in eq.~(\ref{diagA:0}) receives a factor  $e^{\sigma \delta}$, which passes through all calculation untouched and appears in the integrand of eq.~(\ref{diagA:sing1}). With such a factor the ambiguity is resolved because the integrand is zero at $\sigma\to-\infty$ irrespectively of the value of $\alpha$. In order to evaluate it, we make the change of variable $\tau=\alpha(\sigma-z_2)$ and  we obtain
\begin{eqnarray}\label{diagA:sing2}
\widetilde{\mathcal{U}}^{\text{sing}}_A&=&2a_sC_F\Gamma(-\epsilon) \vec b^{2\epsilon}
 \int_{-\infty}^{0} d\tau
\int_0^1 d\alpha \,e^{\delta \frac{\tau}{\alpha}}\frac{\bar \alpha}{\alpha}~\bar q(nz_1)\gamma^+ \frac{\partial}{\partial \tau} q(n (z_2+\tau)).
\end{eqnarray}
The integral over $\alpha$ is singular in the limit  $\delta\to 0$
\begin{eqnarray}
\int_0^1 d\alpha \,e^{\delta \frac{\tau}{\alpha}}\frac{\bar \alpha}{\alpha}\sim \ln\delta.
\end{eqnarray}
The logarithm of $\delta$ represents the rapidity singularity. In order to evaluate the construction (\ref{diagA:sing2}) explicitly we rewrite
\begin{eqnarray}
q(n (z_2+\tau))=e^{i\tau (n\cdot\hat{p}_q)}q(n z_2),
\end{eqnarray}
where $(\hat{p}_q)_\mu=-i\overrightarrow{\partial_\mu}$ is the momentum operator acting on the quark field. Then the integral (\ref{diagA:sing2}) can be taken formally
\begin{eqnarray}
\int_{-\infty}^{0} d\tau\int_0^1 d\alpha e^{\delta \frac{\tau}{\alpha}}\frac{\bar \alpha}{\alpha}\frac{\partial}{\partial \tau}e^{i\tau (n\cdot\hat{p}_q)}
&=&-1+\(1-\frac{i\delta}{(n\cdot\hat{p}_q)}\)\ln\(\frac{\delta+i(n\cdot\hat{p}_q)}{\delta}\)
\\\nn &=&-1-\ln\(\frac{\delta}{i(n\cdot\hat{p}_q)}\)+O(\delta).
\end{eqnarray}
The singular part of the diagram A is
\begin{eqnarray}\label{diagA:sing3}
\widetilde{\mathcal{U}}^{\text{sing}}_A&=&2a_sC_F\Gamma(-\epsilon) \vec b^{2\epsilon}\(-1-\ln\(\frac{\delta}{i(n\cdot\hat{p}_q)}\)\)\bar q(nz_1)\gamma^+ q(n z_2).
\end{eqnarray}
This expression literally (including the complex part) coincides with the calculation of the rapidity divergent part in $\delta$-regularization in the momentum space \cite{Vladimirov:2014aja,Echevarria:2016scs}.

The same method can be used when the position of fields is shifted by $\vec b$. The result for the diagrams A can be written in the form
\begin{eqnarray}\label{diagA:final}
\widetilde{\mathcal{U}}_A&=&2a_sC_F\Gamma(-\epsilon) \vec b^{2\epsilon}\Bigg\{\int_0^1 d\alpha \frac{\bar \alpha}{\alpha} \Big[\mathcal{U}^{\gamma^+}(z_1,z_{21}^\alpha;\bar \alpha \vec b)-\mathcal{U}^{\gamma^+}(z_1,z_{2};\vec b)\Big]
\\\nn && \qquad\qquad\qquad\qquad\qquad\qquad-\(1+\ln\(\frac{\delta}{i(n\cdot\hat{p}_q)}\)\)\mathcal{U}^{\gamma^+}(z_1,z_{2};\vec b)\Bigg\}+O(\vec b^2\partial^2 q),
\\\label{diagA:final*}
\widetilde{\mathcal{U}}_{A^*}&=&2a_sC_F\Gamma(-\epsilon) \vec b^{2\epsilon}\Bigg\{\int_0^1 d\alpha \frac{\bar \alpha}{\alpha} \Big[\mathcal{U}^{\gamma^+}(z_{12}^\alpha,z_{2};\bar \alpha \vec b)-\mathcal{U}^{\gamma^+}(z_1,z_{2};\vec b)\Big]
\\\nn &&\qquad\qquad\qquad\qquad\qquad\qquad -\(1+\ln\(\frac{\delta}{i(n\cdot\hat{p}_{\bar q})}\)\)\mathcal{U}^{\gamma^+}(z_1,z_{2};\vec b)\Bigg\}+O(\vec b^2\partial^2 q),
\end{eqnarray}
where $\hat{p}_{\bar q}=-i\overleftarrow{\partial_\mu}$ is the momentum operator acting on the anti-quark field. Note, that we have added a total shift $\sim \alpha \vec b$ to the first operators, to make the expression more compact. Including  such a shift does not affect the expression for the TMD distribution, since it is proportional to the difference between the momenta of initial and final states. Notice that while in TMD distributions this difference is null, it is not the case for generalized TMD distributions (GTMD).

\subsection{Renomalization}
\label{sec:backrenormalization}

Performing the evaluation of all the other diagrams in a similar manner (see an explicit example for diagram E in the appendix \ref{app:exampleDiag}), we get the OPE for the bare TMD operator, which schematically can be written as  
\begin{eqnarray}\label{bare_OPE}
\widetilde{\mathcal{U}}(z_1,z_2;\vec b)&=&\sum_{i}\Big[1_i+a_s\Gamma(-\epsilon)\vec b^{2\epsilon} \tilde C_i^{\text{tw2}}+O(a_s^2)\Big]\otimes \mathcal{O}_{i,\text{tw2}}(z_1,z_2)
\\\nn &&+b_\mu\sum_{i}\bigg[
1_i+a_s\Gamma(-\epsilon)\vec b^{2\epsilon} \tilde C_i^{\text{tw3}}+O(a_s^2)\bigg]\otimes \mathcal{O}^\mu_{i,\text{tw3}}(z_1,z_2)
+O(\vec b^2),
\end{eqnarray}
where the indices $i$ enumerate all operators that enter the expression, $\otimes$ is some integral convolution in the light cone positions variables $z$,  and $1_i=1 (0)$ for the operators that contribute at LO (otherwise). Here, the coefficients $\tilde C$ depend on $\epsilon$, $\delta$ and light-cone positions $z_{1,2}$, the dependence $\vec b$ is concentrated entirely in the factors $\vec b^{2\epsilon}$. The explicit form of  each term in eq.~(\ref{bare_OPE}) is rather lengthy. We  present it diagram-by-diagram (since there is practically no simplification in the diagram sum) in appendix \ref{app:diag-by-diag}.

The bare OPE eq.~(\ref{bare_OPE}) requires renormalization as in eq.~(\ref{def:renormalization}), i.e. both sides of eq.~(\ref{bare_OPE}) are to be multiplied by $Z^{-1}_2Z_q^{TMD}R_q$, whose LO expressions are given in eqs.~(\ref{renorm:R_1loop}) and ~(\ref{renorm:Z_1loop}). We recall that this renormalization is  universal, in the sense that, it is common for all terms of the small-$b$ expansion and for various Lorentz structures of TMD operator. An example of this universality is already provided  by the diagram A, discussed in the previous section. Indeed, according to eqs.~(\ref{diagA:final},~\ref{diagA:final*}) the rapidity divergence enters the expression multiplying the bare TMD operator $\mathcal{U}(z_1,z_2;\vec b)$. In other words, we can extract the rapidity divergent terms from eq.~(\ref{bare_OPE}) and write it as
\begin{eqnarray}
\widetilde{\mathcal{U}}(z_1,z_2;\vec b)&=&\Big[
1-2 a_s C_F \Gamma(-\epsilon)\vec b^{2\epsilon} \ln\(\frac{\delta^2}{(p^+)^2}\)\Big]\mathcal{U}(z_1,z_2;\vec b)+a_s(\text{rapidity finite terms}),
\end{eqnarray}
where $p^+$ is the momentum of the parton\footnote{In GTMD case, initial and final partons have different momenta. We cannot specify which momentum appears in the soft factor in the absence of the process and factorizaton theorem which would fix the kinematic scales. Nonetheless, in any case, the rapidity divergences are renomalized by factor $R_q$, but possibly leave extra terms of the form $\ln(p_{q}^+/p_{\bar q}^+)$.}. Multiplying it by $R_q$, given in eq.~(\ref{renorm:R_1loop}), \textit{the logarithm of $\delta$ cancels for all terms of the small-$b$ expansion to all orders of $\epsilon$}. To our best knowledge this is the first explicit demonstration of rapidity divergences renormalization of TMD at higher twists.

The renormalization of eq.~(\ref{bare_OPE}) makes this expression finite. However, coefficients $\tilde C$ contain singularities in $\epsilon$. These singularities are collinear singularities and are compensated by UV behavior of light-cone operators. To remove them explicitly we replace the bare operators on r.h.s. by the renormalized operators $\mathcal{O}^{bare}=Z^{-1}\otimes \mathcal{O}^R(\mu)$. The factor $Z^{-1}$ being convoluted with coefficient function removes the remaining poles in $\epsilon$. 

Concluding, the renormalized expression for small-$b$ OPE has the form
\begin{eqnarray}\label{renorm:final}
\widetilde{\mathcal{U}}(z_1,z_2;\vec b;\mu,\zeta)&=&\sum_{i}\big[1_i+a_s(\mu)C_i^{\text{tw2}}(\mu,\zeta)+O(a_s^2)\Big]\otimes \mathcal{O}_{i,\text{tw2}}(z_1,z_2;\mu)
\\\nn &&\qquad +b_\mu\sum_i\big[
1_i+a_s(\mu)C_i^{\text{tw3}}(\mu,\zeta)+O(a_s^2)\big]\otimes \mathcal{O}^\mu_{i,\text{tw3}}(z_1,z_2;\mu)
+O(\vec b^2),
\end{eqnarray}
where the operators are renormalized at scales $\mu$ and $\zeta$ and we have set the scale of renormalization for light-cone operators to be the same as for TMD operator for simplicity. The expression for the  coefficient functions at NLO for any twist can be written as
\begin{eqnarray}\label{renorm:final-final}
C_i^{\text{tw-n}}(\mu,\zeta)&=&\Bigg\{\Gamma(-\epsilon)\vec b^{2\epsilon}\mu^{2\epsilon}e^{-\epsilon \gamma_E} \Big[\tilde C_i^{\text{tw-n}}+2 C_F\(\ln\(\vec b^2\delta^2\frac{\zeta}{(p^+)^2}\)-\psi(-\epsilon)+\gamma_E\)\Big]
\\\nn &&\qquad\qquad\qquad\qquad\qquad\qquad\qquad
-C_F\(\frac{2}{\epsilon^2}+\frac{3+2\ln(\mu^2/\zeta)}{\epsilon}\)\Bigg\}_{\epsilon-\text{finite}},
\end{eqnarray}
where  the rapidity divergences  in $\tilde C_i^{\text{tw-n}}$ are explicitly canceled and we have expressed   the renormalization factors in $\overline{\text{MS}}$-scheme, see eq.~(\ref{renorm:R_1loop}, \ref{renorm:Z_1loop}). With this formula it is simple enough to obtain the coefficient functions for the small-$b$ OPE in coordinate space. 
However, they are of little use, since in practice, one operates in terms of momentum fractions $x$ and the corresponding collinear distributions. The transition to the distribution and the corresponding expressions are discussed in sec.~\ref{sec:TOdistr}.

\subsection{Difference in the evaluation of DY and SIDIS operators}
\label{sec:DY-SIDIS-difference}

The operators for the DY and SIDIS initiated TMD distributions differ by the geometry of Wilson lines. This dependence influences the calculation in two aspects. The first one is the explicit expression for diagrams that have interaction with Wilson line, such as diagrams A, C and E. The second one is the preferred boundary conditions for the gauge fixing for the background field, the retarded for DY-type operators, eq.~(\ref{gauge:ret}) and advanced one for SIDIS-type operators, eq.~(\ref{gauge:adv}). Let us note, that boundary conditions do not influence the process of diagram evaluation, but rather the procedure of recompilation of the expressions in terms of gauge-invariant operators, see eq.~(\ref{gauge:A=F_ret},~\ref{gauge:A=F_adv}).

In both cases the only difference between expressions for DY and SIDIS kinematic is the sign of infinity in the integration limits. I.e. a term contributing to OPE for DY operator has the form
\begin{eqnarray}
\text{DY}:\qquad \int_{-\infty}^{z_i}d\sigma~ ...~ F^{\mu +}(\sigma),
\end{eqnarray}
whereas the same term in the OPE for SIDIS operator is
\begin{eqnarray}
\text{SIDIS}:\qquad \int_{+\infty}^{z_i}d\sigma~ ...~ F^{\mu +}(\sigma).
\end{eqnarray}
Here, dots indicate various compositions of fields, functions and integrals that do not change. Such a structure is already evident at the tree level order, as one finds comparing eq.~(\ref{U_DY_der}) and eq.~(\ref{U_DIS_der}). As we will see, in  terms of distributions this difference will result into a different global sign of the coefficient functions.

\section{Definition of collinear distributions}
\label{sec:def-collinear}

In order to proceed further we need to evaluate the hadronic matrix element of OPE. This procedure is scheme dependent in the following sense: We recall that our computation is made in  dimensional regularization and after the renormalization procedure the expressions are finite for $\epsilon\to 0$.   
Nonetheless, the finite part of the results  depends on $\epsilon$ and  moreover the expressions so obtained have a tensor structure which  also depends on the number of dimensions. Thus, in order to completely define the scheme, we should specify the order of operations with respect to the limit $\epsilon\to 0$. 

There are two major options. The first one consists in setting $\epsilon\to 0$ before the evaluation of matrix elements (i.e. at the level of operators) and defining the distributions in 4-dimensions. The second one is to \textit{define the distributions in $d$-dimensions and to perform the limit $\epsilon\to0$ after the evaluation of matrix elements}. Both schemes have positive and negative aspects. In fact, this problem has not been accurately addressed in the TMD-related literature. Checking the traditional calculations of TMD matching at twist-2 \cite{Collins:2011zzd,Echevarria:2016scs,Gutierrez-Reyes:2017glx,Aybat:2011zv}, we conclude that  the second scheme is used in all these cases. Therefore, to be consistent with earlier calculations, \textit{we use the second scheme.}  
Nonetheless, we have also performed the calculation in the first scheme and we have found that for the Sivers function  some differences appear only in the quark-gluon mixing diagrams. These differences are $\epsilon$-suppressed and thus the expression for the NLO matching coefficient is the same in both schemes.   In appendix \ref{app:diag-by-diag:matching} we present the expressions for diagrams with an explicit designation of the origin of $\epsilon$ which  allows to re-derive the complete result.

In the rest of this section we define the twist-2 and twist-3 matrix collinear distributions and evaluate the TMD matrix element over the small-$b$ OPE obtained in the previous section.

\subsection{Quark distributions}
\label{sec:coll-quark}

The forward matrix elements of the light-cone operators are parametrized by collinear distributions, or parton distribution functions (PDFs). For this work we need the forward matrix element of twist-2 and twist-3 operators only. We start  discussing  the required quark distributions, while the gluon distributions are treated in the next section.

There are three quark operators contributing to the OPE of the Sivers function,
\begin{eqnarray}\label{distr:O2}
\mathcal{O}_{\gamma^+}(z_1,z_2)&=&\bar q(z_1 n)[z_1 n,z_2 n]\gamma^+ q(z_2 n),
\\\label{distr:T1}
\mathcal{T}^{\mu}_{\gamma^+}(z_1,z_2,z_3)&=&g\bar q(z_1n)[z_1n,z_2n]\gamma^+ F^{\mu+}(z_2n)[z_2n,z_3n]q(z_3n),
\\\label{distr:T2}
\mathcal{T}^{\nu}_{\gamma^+\gamma_T^{\nu\mu}}(z_1,z_2,z_3)&=&g\bar q(z_1n)[z_1n,z_2n]\gamma^+\gamma^{\nu\mu}_T F^{\nu+}(z_2n)[z_2n,z_3n]q(z_3n),
\end{eqnarray}
where
\begin{eqnarray}
\gamma_T^{\mu\nu}=g_T^{\mu\mu'}g_T^{\nu\nu'}\frac{\gamma_{\mu'}\gamma_{\nu'}-\gamma_{\nu'} \gamma_{\mu'}}{2}.
\end{eqnarray}
The operator in eq.~(\ref{distr:O2}) is  twist-2, whereas the operators  in eq.~(\ref{distr:T1},~\ref{distr:T2}) are twist-3.  We emphasize that all indices appearing in  eq.~(\ref{distr:T1},~\ref{distr:T2})  are transverse.

The forward matrix element depends only on the distance between fields, but not on the absolute position. A shift of the common position  can be written as a total derivative of the operator, which is a momentum transfer between initial and final states. It is the consequence of the quantum-mechanical definition of the momentum operator:
\begin{eqnarray}\label{dO=0}
\langle p_1|\partial_\mu\{O\}|p_2\rangle= i(p_2-p_1)_\mu \langle p_1|O|p_2\rangle,
\end{eqnarray}
where $O$ is any operator.  It allows to move each term of OPE to a convenient position and to drop terms with total derivatives. Altogether it significantly simplifies the evaluation. To resolve the total derivative terms one should consider a non-forward kinematics, that defines GTMD distributions and generalized parton distributions. In the following, we consider each operator in a convenient point.

The  standard unpolarized PDF comes from the forward matrix element of $\mathcal{O}_{\gamma^+}$,
\begin{eqnarray}\label{def:f1}
\langle p,S|O_{\gamma^+}(z_1,z_2)|p,S\rangle=2p^+ \int dx e^{ix(z_1-z_2)p_+} f_1(x).
\end{eqnarray}
The PDF is non-zero for $-1<x<1$ and
\begin{eqnarray}
f_1(x)=\theta(x)q(x)-\theta(-x)\bar q(x),
\end{eqnarray}
where $q(x)$ and $\bar q (x)$ are the quark and anti-quark parton densities in the infinite momentum frame. 

\begin{figure}[t]
\begin{center}
\includegraphics[width=0.6\textwidth]{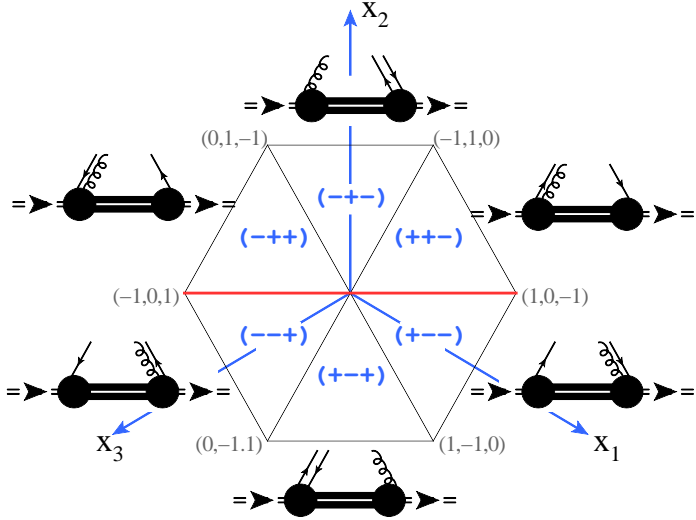}
\end{center}
\caption{\label{fig:domain} The support of the twist-3 functions, drawn in the barycentric coordinates, $x_1+x_2+x_3=0$. The diagrams demonstrate the interpretation of distribution in the terms of emission-absorption of partons by a hadron. Red dashed line is the line on which the Qui-Sterman distribution is defined.}
\end{figure}

The definition of twist-3 PDFs is more cumbersome since they depend on two momentum fractions $x_i$ and they have a different interpretation relative to a domain of variables. The notation simplifies considerably if one writes the twist-3 distributions as a functions of three momentum factions $x_{1,2,3}$. 
Each momentum fraction is the Fourier conjugate of the corresponding coordinate $z_{1,2,3}$. We define
\begin{eqnarray}\label{def:T}
\langle p,S|\mathcal{T}^{\mu}_{\gamma^+}(z_1,z_2,z_3)|p,S\rangle
&=& 2\tilde s^\mu (p^+)^2 M\int [dx]e^{-ip^+(x_1z_1+x_2z_2+x_3z_3)}T(x_1,x_2,x_3),
\\\label{def:DeltaT}
\langle p,S|\mathcal{T}^{\nu}_{\gamma^+\gamma_T^{\nu\mu}}(z_1,z_2,z_3)|p,S\rangle
&=& -2\tilde s^\mu (p^+)^2 M\int [dx]e^{-ip^+(x_1z_1+x_2z_2+x_3z_3)}\Delta T(x_1,x_2,x_3),
\end{eqnarray}
where $M$ is the mass of the hadron and the integral measure is defined as
\begin{eqnarray}\label{def:[dx]}
\int [dx]f(x_1,x_2,x_3)=\int_{-1}^1 dx_1dx_2dx_3 \delta(x_1+x_2+x_3)f(x_1,x_2,x_3).
\end{eqnarray}
Such an integral measure automatically takes into account the independence of forward matrix element on the total shift, eq.~(\ref{dO=0}).

The functions of three variables $T(x_1,x_2,x_3)$ have several symmetry properties. It is natural to consider them as functions defined on the hyperplane $x_1+x_2+x_3=0$, since only this domain contributes to forward matrix element.  The domain can be split into six regions, corresponding to different signs of the variables $x_i$, see fig.~\ref{fig:domain}. Each of these regions has a different interpretation in parton language: depending on the sign of $x_i$ the corresponding parton is either emitted ($x_i>0$) or absorbed by a hadron \cite{Jaffe:1983hp}, as it is shown schematically in fig.~\ref{fig:domain}.

The functions $T$ and $\Delta T$ are not independent and mix under the evolution. In ref.~\cite{Braun:2009mi} it is shown that there exist a combination of $T$ and $\Delta T$ which evolve autonomously, but we do not use it in this work.

The definitions in eq.~(\ref{def:T},~\ref{def:DeltaT}) are understood in $d$-dimensions. That is,  the vector $\tilde s^\mu$ is some vector that turns into $\tilde s^\mu=\epsilon_T^{\mu\nu}s_\nu$ when $\epsilon\to 0$. The definition of the non-perturbative functions $T$ and $\Delta T$ coincides\footnote{To compare the definitions that we have used, consider the 4-dimensional relation $\gamma^+\gamma^{\mu\nu}_T=-i\epsilon^{\mu\nu}_T\gamma^+\gamma^5$.} with the one made in \cite{Scimemi:2018mmi}. Also it is coincides (up to a factor $M$) with the definition given in \cite{Braun:2009mi}. The articles \cite{Ji:2006vf,Koike:2007dg,Kang:2008ey,Kang:2011mr,Sun:2013hua} use a less convenient two-variable definition, which is related to the definition with three variables by (here we compare to \cite{Kang:2008ey})
\begin{eqnarray}\label{def:T(3)->T(2)}
\tilde{\mathcal{T}}_{q,F}(x,x+x_2)&=&MT(-x-x_2,x_2,x),\qquad
\tilde{\mathcal{T}}_{\Delta q,F}(x,x+x_2)=M\Delta T(-x-x_2,x_2,x).
\end{eqnarray}

\begin{figure}[t]
\begin{center}
\includegraphics[width=0.6\textwidth]{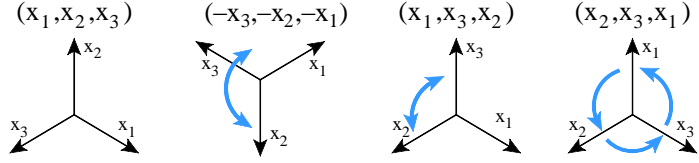}
\end{center}
\caption{\label{fig:transformation} The illustration for the transformation of the barycentric coordinates. From left to right: original, time-inversion, permutation of variables, cyclic permutation of variables.}
\end{figure}

Using  time-reversal and hermiticity, one can show that the functions $T$ and $\Delta T$ are real and obey the property
\begin{eqnarray}\label{quark:symT}
T(x_1,x_2,x_3)&=&T(-x_3,-x_2,-x_1),
\\\label{quark:symdT}
\Delta T(x_1,x_2,x_3)&=&-\Delta T(-x_3,-x_2,-x_1).
\end{eqnarray}
These properties are central in the following calculation. They represent the simple statement that gluon is a neutral particle. 
In barycentric coordinates the time-reversal transformation turns the picture upside down as shown in fig.~\ref{fig:transformation}. 
Therefore, the function $T$ ($\Delta T$) is (anti)symmetric with respect to the horizontal line $x_2=0$ (given by red dashed line in fig.~\ref{fig:domain}). 
PDFs defined on these lines are known as Qui-Sterman distribution. 
They play a special role in TMD physics, since they provide the  LO  matching, as it is shown in the next sections.

\subsection{Gluon distributions}
\label{sec:coll-gluon}

The gluon operators of twist-2 and twist-3 are
\begin{eqnarray}\label{distr:Omunu}
\mathcal{O}^{\mu\nu}(z_1,z_2)&=&F^{\mu+}(z_1n)[z_1n,z_2n]F^{\nu+}(z_2n),
\\\label{distr:T+}
\mathcal{T}^{\mu\nu\rho}_+(z_1,z_2,z_3)&=&igf^{ABC}F^{A;\mu+}(z_1n)F^{B;\nu+}(z_2n)F^{C;\rho+}(z_3n),
\\\label{distr:T-}
\mathcal{T}^{\mu\nu\rho}_-(z_1,z_2,z_3)&=&gd^{ABC}F^{A;\mu+}(z_1n)F^{B;\nu+}(z_2n)F^{C;\rho+}(z_3n),
\end{eqnarray}
where $f^{ABC}$ and $d^{ABC}$ are symmetric and anti-symmetric structure constants of the gauge-group. In the definitions (\ref{distr:T+}) we have dropped the Wilson lines for simplicity\footnote{ The complete expression with Wilson lines is like
\begin{eqnarray}\nn
\mathcal{T}^{\mu\nu\rho}_+(z_1,z_2,z_3)=gF^{A';\mu+}(z_1n)F^{B';\nu+}(z_2n)F^{C';\rho+}(z_3n)[z_1n,rn]^{A'A}[z_2n,rn]^{B'B}[z_3n,rn]^{C'C}if^{ABC},
\end{eqnarray}
and analogous for $\mathcal{T}^{\mu\nu\rho}_-$. The expression is independent on $r$, thanks to Jacobi identity.}.

The forward matrix element is parametrized by
\begin{eqnarray}\label{distr:tensor-decomposition}
\langle p,S|\mathcal{O}^{\mu\nu}(z_1,z_2)|p,S\rangle&=& (p^+)^2\int dx e^{i(z_1-z_2)xp^+}\,x\,\Big(\frac{g_T^{\mu\nu}}{2(1-\epsilon)}g(x)+\lambda\frac{a^{\mu\nu}}{2}\Delta g(x)\Big),
\end{eqnarray}
where $\lambda$ is a hadron helicity and $a^{\mu\nu}$ is an antisymmetric tensor such that 
\begin{eqnarray}
\lim_{\epsilon\to 0}a^{\mu\nu}=\epsilon_T^{\mu\nu}.
\end{eqnarray}
Generally, the decomposition (\ref{distr:tensor-decomposition}) should additionally contain a symmetric-traceless component. The corresponding distribution is however zero in  forward kinematics. The distributions $g(x)$ and $\Delta g$ are conventional unpolarized and polarized gluon distributions.

There is no standard parametrization for the twist-3 gluon operator. Here we introduce the parameterization that is convenient for our calculation. It is different (but equivalent) to other parameterizations used e.g. in \cite{Braun:2009mi,Kang:2008ey,Beppu:2010qn,Dai:2014ala,Chen:2016dnp,Chen:2017lvx}.   The main difference is that we use two distributions with different properties, instead of a single one. We have
\begin{eqnarray}\label{distr:gluons}
\langle p,S|\mathcal{T}^{\mu\nu\rho}_\pm(z_1,z_2,z_3)|p,S\rangle &=&-(p^+)^3M\int [dx]e^{-ip^+(x_1z_1+x_2z_2+x_3z_3)}
\\\nn &&
\times \Big(
\frac{\tilde s^\mu g_T^{\nu\rho}+\tilde s^\nu g_T^{\mu\rho}+\tilde s^\rho g_T^{\mu\nu}}{2(2-\epsilon)}G_\pm(x_1,x_2,x_3)
\\\nn &&+\frac{\tilde s^{\nu}g_T^{\mu\rho}Y_\pm(x_1,x_2,x_3) \mp \tilde s^\mu g_T^{\nu\rho}Y_{\pm}(x_2,x_1,x_3)
\mp \tilde s^\rho g_T^{\mu\nu}Y_\pm(x_1,x_3,x_2)}{1-2\epsilon}
\Big).
\end{eqnarray}
The overall minus sign is set in order to have a simple relation to the distributions defined in \cite{Braun:2009mi,Kang:2008ey}. The foundation for this parameterization is discussed in appendix \ref{app:tensor-decomposition}. Despite its  cumbersome appearance, this parameterization has some natural properties, that significantly simplify the  calculation. Time-reversal and hermiticity  imply  that
\begin{eqnarray}\label{distr:gluon-reverse}
G_\pm(x_1,x_2,x_3)=G_\pm(-x_3,-x_2,-x_1),\qquad Y_\pm(x_1,x_2,x_3)=Y_\pm(-x_3,-x_2,-x_1) ,
\end{eqnarray}
which reflects the fact that the gluon is a neutral particle and thus, ``anti-gluon'' distribution is equal to the ``gluon'' one. 
Due to the permutation properties of the operator, the distributions are highly symmetric. Namely, the distribution $G_-$ ($G_+$) is (anti-)symmetric with respect to permutation of any pair of arguments
\begin{eqnarray}\label{distr:gluon-anti}
G_\pm(x_1,x_2,x_3)=\mp G_\pm(x_2,x_1,x_3)=\mp G_\pm(x_1,x_3,x_2).
\end{eqnarray}
The distribution $Y_-$($Y_+$) is (anti-)symmetric with respect to to permutation of $x_1$ and $x_3$,
\begin{eqnarray}\label{distr:gluon-anti2}
Y_\pm(x_1,x_2,x_3)=\mp Y_\pm(x_3,x_2,x_1).
\end{eqnarray}
Additionally, the distributions $Y_\pm$ obey a cyclic rule
\begin{eqnarray}\label{distr:gluon-cyclic}
Y_\pm(x_1,x_2,x_3)+Y_\pm(x_2,x_3,x_1)+Y_\pm(x_3,x_1,x_2)=0.
\end{eqnarray}
The graphical representation of these transformation in barycentric coordinates is shown in fig.~\ref{fig:transformation}.

\begin{figure}[t]
\begin{center}
\includegraphics[width=0.7\textwidth]{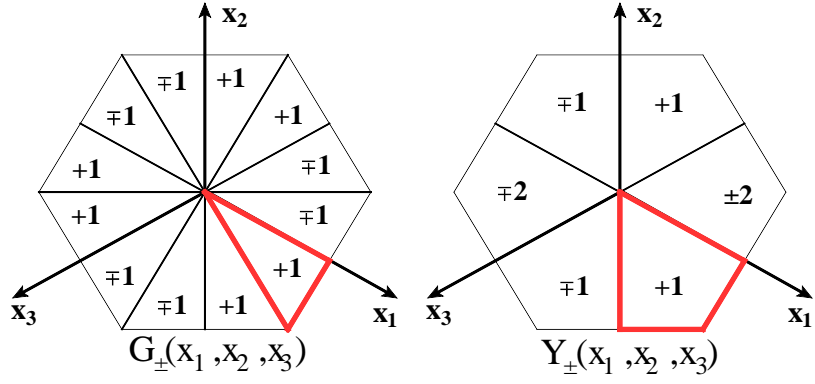}
\end{center}
\caption{\label{fig:GYplus} The value of functions $G_\pm$ and $Y_\pm$ in the whole domain is defined by values in the red segments. The values in other segments is obtained by turning/reflecting the values with respect to edges and multiplying by the factor shown within the segment.}
\end{figure}

The symmetry properties in eq.~(\ref{distr:gluon-reverse}-\ref{distr:gluon-cyclic}) significantly restrict the functional form of distributions. In particular, the functions $G_\pm$ are entirely defined by its values in the region $0<x_1/2<-x_2<x_1$. Whereas the functions $Y_\pm$ are defined by its values in the region $0<x_1/2<-x_2<2x_1$. Graphically these relations are demonstrated in fig.~\ref{fig:GYplus}.

The functions $G$ and $Y$  mix under  evolution. In many aspects they are similar to the functions $T$ and $\Delta T$ of the quark case. Nonetheless, the parametrization given here grants many simplification during calculation, because each of the structures in eq.~(\ref{distr:gluons}) belongs to an irreducible representation of the Lorentz group. For that reason these structures enter the dimensionally regularized expression with different $\epsilon$-dependent factors.

The relation of the functions $G_\pm$ and $Y_\pm$ to the functions used in \cite{Braun:2009mi} is
\begin{eqnarray}
T_{3F}^\pm(x_1,x_2,x_3)=G_\pm(x_1,x_2,x_3)+Y_\pm(x_1,x_2,x_3).
\end{eqnarray}
It is important to note that this comparison is made at $\epsilon=0$, because at $\epsilon\neq 0$ the comparison is impossible. The inverse relation is
\begin{eqnarray}\label{relationG-T}
G_\pm(x_1,x_2,x_3)&=&\frac{T_{3F}^\pm(x_1,x_2,x_3)-T_{3F}^\pm(x_2,x_1,x_3)-T_{3F}^\pm(x_1,x_3,x_2)}{3},
\\\label{relationY-T}
Y_\pm(x_1,x_2,x_3)&=&\frac{2T_{3F}^\pm(x_1,x_2,x_3)+T_{3F}^\pm(x_2,x_1,x_3)+T_{3F}^\pm(x_1,x_3,x_2)}{3}.
\end{eqnarray}
Therefore, our basis is equivalent to a decomposition of a general 3-variable function into antisymmetric and cyclic components. The reduction of three-variable notation used here and in \cite{Braun:2009mi} to the two-variable notation used in \cite{Kang:2008ey,Chen:2016dnp,Chen:2017lvx} is the same as for quarks in eq.~(\ref{def:T(3)->T(2)}). 
In~\cite{Beppu:2010qn,Dai:2014ala} a different notation is used, which again can be related to our functions at $\epsilon \to 0$. For a detailed comparison we refer to the discussion in~\cite{Beppu:2010qn}.

\section{Small-$b$ expansion for unpolarized and Sivers distributions}
\label{sec:TOdistr}

Having at hand the parametrization of the matrix elements we can obtain the matching coefficient for TMD distributions to collinear distributions. The standard protocol to achieve this  is  the following. We derive the TMD distribution using the operators $\mathcal{U}$ (compare eq.~(\ref{def:TMDPDF_Qop}) and eq.~(\ref{def:TMDop_DY})), 
\begin{eqnarray}\label{main-fourier}
\Phi^{[\gamma^+]}_{q\ot h}(x,\vec b)=\int \frac{dz}{2\pi}e^{-2ixzp^+}\langle p,S|\mathcal{U}^{\gamma^+}\(z,-z;\frac{\vec b}{2}\)|p,S\rangle.
\end{eqnarray}
Next, we substitute the expression for OPE eq.~(\ref{renorm:final}) into the matrix element and we evaluate the Fourier transform using the parameterization for collinear matrix elements. In this way we obtain the small-$b$ expansion for the TMD distribution $\Phi^{[\gamma^+]}$.  Collecting all terms with appropriate Lorentz structures, eq.~(\ref{param:TMDv}), we obtain the small-$b$ expansion for individual TMD distributions, in our case these are the unpolarized and Sivers distributions. The procedure is rather straightforward and it can be performed for each diagram independently. In sec.~\ref{sec:TOdistr1} we give several comments on the evaluation of it, while the final result is presented in sec.~\ref{sec:results}. The results for individual diagrams are presented in appendix \ref{app:diag-by-diag:matching}.

\subsection{From operators to distributions and tree level results}
\label{sec:TOdistr1}

The tree level order of OPE is given in eq.~(\ref{U_Taylor}). Applying the transformation in eq.~(\ref{main-fourier}) and using the definitions in eq.~(\ref{def:f1},~\ref{def:T}) we obtain\footnote{When evaluating matrix element one should also consider the matrix element of the first term in eq.~(\ref{U_DY_der}). For the unpolarized operator this matrix element is zero. The proof can be found in \cite{Scimemi:2018mmi}.}
\begin{eqnarray}
&&\Phi^{[\gamma^+]}_{q\ot h;\text{DY}}(x,\vec b)=f_1(x)
\\\nn && +i\tilde s_\mu b^\mu (p^+)^2M
\int \frac{dz}{2\pi}e^{-2ixzp^+} \(\int_{-\infty}^z+\int_{-\infty}^{-z}\)d\tau \int [dx]e^{-ip^+(x_1z+x_2\tau-x_3 z)}T(x_1,x_2,x_3).
\end{eqnarray}
To evaluate the second line we use the following trick. We consider the two integrals over $\tau$ separately and change the variables $x_{1,2,3}\to-x_{3,2,1}$, $\tau\to-\tau$ in the second one. The integrand is  invariant under such transformation, due to the property in eq.~(\ref{quark:symT}) while the limits of integration change to $(-z,+\infty)$. As a result the two integrals over $\tau$  can be combined into a single integral over $\tau$ from $-\infty$ to $+\infty$, 
\begin{eqnarray}\label{distr:11}
&&\Phi^{[\gamma^+]}_{q\ot h}(x,\vec b)=f_1(x)
\\\nn && +i\tilde s_\mu b^\mu (p^+)^2M
\int \frac{dz}{2\pi}e^{-2ixzp^+} \int_{-\infty}^\infty d\tau \int [dx]e^{-ip^+(x_1z+x_2\tau-x_3 z)}T(x_1,x_2,x_3).
\end{eqnarray}
Let us stress that \textit{the dependence on the intermediate gluon position $\tau$ disappears}. This property holds for all diagrams and allows to combine seemingly cumbersome expressions into simple ones.  It is the result of time-reversal symmetry. Therefore, to observe such cancellation, one should collect a diagram with its conjugated. I.e. the dependence on the intermediate point cancels in combination of diagrams $A$ and $A^*$, $C$ and $C^*$, $E$ and $E^*$, $D$ and $D^*$. The rest diagrams are self-conjugated.

The time-reversal symmetry  is also responsible of  the different relative sign in the matching of DY and SIDIS operators. Indeed, since the integrands are symmetric under time-reversal, the intermediate point cancels and the only thing that matters is a common global sign. This sign is necessarily different between DY and SIDIS expressions, due to different boundary conditions holding in two cases. In other words, all gluon fields in the DY case are connected to $-\infty$ and the corresponding integrals are $\int_{-\infty}$. Whereas for SIDIS they are connected to $+\infty$ and corresponding integrals are $\int_{+\infty}=-\int^{+\infty}$. In this way, we observe the well-known relation
\begin{eqnarray}
C_{1T;\text{DY}}^\perp(x_1,x_2,x_3,\vec b)=-C_{1T;\text{DIS}}^\perp(x_1,x_2,x_3,\vec b),
\end{eqnarray}
i.e. the matching (Wilson coefficient) of the Sivers function has a different  sign in DY and SIDIS. This observations agrees with the time-reversal property  of the Sivers distribution 
\begin{eqnarray}
f^\perp_{1T;\text{DY}}(x,\vec b)=-f^\perp_{1T;{DIS}}(x,\vec b),
\end{eqnarray}
observed a long ago \cite{Collins:2002kn}. 

Coming back to  eq.~(\ref{distr:11}), the integrals over $\tau$ and $z$ decouple and both produce a $\delta$-function. We obtain
\begin{eqnarray}
&&\Phi^{[\gamma^+]}_{q\ot h}(x,\vec b)=f_1(x) +i\pi\tilde s_\mu b^\mu M \int [dx]\delta(x_2)\delta(x-x_3)T(x_1,x_2,x_3).
\end{eqnarray}
Using the delta-function in the definition of $[dx]$ in eq.~(\ref{def:[dx]}), the integrals over $x$'s can be evaluated, 
\begin{eqnarray}
&&\Phi^{[\gamma^+]}_{q\ot h}(x,\vec b)=f_1(x) +i\pi\tilde s_\mu b^\mu M T(-x,0,x)+O(a_s)+O(\vec b^2).
\end{eqnarray}
This expression gives the leading order matching for unpolarized and Sivers TMD distributions in eq.~(\ref{param:TMDv})
\begin{eqnarray}
f_1(x,\vec b)&=&f_1(x)+O(a_s)+O(\vec b^2),
\\
f_{1T}^\perp(x,\vec b)&=&\pm\pi T(-x,0,x)+O(a_s)+O(\vec b^2),
\end{eqnarray}
where $+$ sign is for DY operator and $-$ sign is for SIDIS operator. The same procedure with minimal modifications can be done for each term of OPE also at higher orders. In appendix \ref{app:diag-by-diag:matching}, we present the expressions for each diagram at NLO  and the corresponding final result is given in the next section.

The $T$ and $\Delta T$ distributions defined on the line $x_2=0$ are generally known as Efremov-Teryaev-Qui-Sterman (ETQS) distributions \cite{Efremov:1983eb,Qiu:1991pp}. In the next section, we write explicitly the evolution equation for these functions in eq.~(\ref{result:tw3Evolution}). Here, we just  remind that \textit{the ETQS functions are not autonomous}, meaning that their evolution involves the values of these functions in a full domain of $x_{1,2,3}$. However, we have found that the finite part\footnote{\label{foot1}Following common terminology, we name $C(\mathbf{L}_\mu=0)$ as the finite part of the coefficient function $C(\mathbf{L}_\mu)$, whereas $C(\mathbf{L}_\mu)-C(\mathbf{L}_\mu=0)$ is named the logarithmic part.} of the small-$b$ matching coefficient involves only ETQS functions.

The line $x_2=0$ plays a special role in the matching of TMD distributions as  shown in red in fig.~\ref{fig:domain}. In the parton picture the distributions defined on this line can be interpreted as  ``gluonless''. Indeed, while the quarks are normally emitted and absorbed by a hadron (as in usual twist-2 distribution), here the gluon is in an ``intermediate state'' nor emitted, nor absorbed, but smoothly distributed all-over the space. This picture also supports the interpretation of variables $x$, as the parton momenta measured as the fraction of the hadron momentum. In such a momentum picture, the line $x_2=0$ corresponds to null-energy gluon.

The symmetry properties of the distributions allow some simplification along the line $x_2=0$. In particular, the $\Delta T$ function (which in principle appears when $x_2\neq 0$) does not explicitly contribute to the matching  due to eq.~(\ref{quark:symdT})
\begin{eqnarray}
\Delta T(-x,0,x)=0,
\end{eqnarray}
but  it will appear  in the evolution of the ETQS functions, as we show in the next section.

Due to the anti-symmetry property the function $G_\pm$  when one of their arguments in 0, they can be expressed as  ETQS distributions
\begin{eqnarray}
G_\pm(-x,0,x)=\mp G_\pm(x,0,-x)=\mp G_\pm(-x,x,0)=\mp G_\pm(0,-x,x).
\end{eqnarray}
The functions $Y_\pm$ at $x_i=0$ also can be expressed via ETQS distributions, but with a different rule
\begin{eqnarray}
Y_\pm(-x,x,0)=\mp Y_\pm(x,-x,0)=\mp Y_\pm(0,x,-x)=-\frac{Y_\pm(-x,0,x)}{2}.
\end{eqnarray}
The application of these rules significantly simplifies the calculation.

\subsection{Results at NLO}
\label{sec:results}

The NLO matching of Sivers TMD distribution at small-$b$ reads
\begin{eqnarray}\label{result:Sivers}
&&f_{1T;q\ot h;\text{DY}}^\perp(x,\vec b;\mu,\zeta)=\pi T(-x,0,x)+\pi a_s(\mu)\Big\{
\\\nn && \quad-2\mathbf{L}_\mu P \otimes T+C_F\(-\mathbf{L}_\mu^2+2\mathbf{l}_\zeta \mathbf{L}_\mu+3\mathbf{L}_\mu-\frac{\pi^2}{6}\)T(-x,0,x)
\\\nn &&\quad +
\int d\xi \int_0^1 dy \delta(x-y\xi)\Big[\(C_F-\frac{C_A}{2}\)2\bar y T(-\xi,0,\xi)+\frac{3 y \bar y }{2}\frac{G_+(-\xi,0,\xi)+G_-(-\xi,0,\xi)}{\xi}\Big]\Big\}
\\\nn && \hspace{0.7\textwidth} +O(a_s^2)+O(\vec b^2),
\end{eqnarray}
where on the right hand side all distributions are defined at the scale $\mu$, $\bar y=1-y$ and
\begin{eqnarray}
\mathbf{l}_\zeta=\ln\(\frac{\mu^2}{\zeta}\).
\end{eqnarray}
Eq.~(\ref{result:Sivers} ) is written for the DY definition of the TMD distribution. In the case of the SIDIS definition the factor $\pi$ in the first line should be replaced by $-\pi$.

The symbol $P\otimes T$ represents the  evolution kernel for the function $T(x_1,x_2,x_3)$ on the $x_2=0$ line. It reads 
\begin{eqnarray}\label{result:tw3Evolution}
&&\mu^2 \frac{d}{d\mu^2}T(-x,0,x)=2a_s(\mu)P\otimes T
=2a_s \int d\xi \int_0^1 dy \delta(x-y\xi)\Bigg\{
\\\nn &&\quad \(C_F-\frac{C_A}{2}\)\Big[\(\frac{1+y^2}{1-y}\)_+T(-\xi,0,\xi)+(2y-1)_+T(-x,\xi,x-\xi)-\Delta T(-x,\xi,x-\xi)\Big]
\\\nn &&\quad +\frac{C_A}{2}\Big[\(\frac{1+y}{1-y}\)_+T(-x,x-\xi,\xi)+\Delta T(-x,x-\xi,\xi)\Big]
\\\nn && \quad +\frac{1-2y\bar y}{4}\frac{G_+(-\xi,0,\xi)+Y_+(-\xi,0,\xi)+G_-(-\xi,0,\xi)+Y_-(-\xi,0,\xi)}{\xi}
\Bigg\},
\end{eqnarray}
where the plus-distribution is defined as usual
\begin{eqnarray}
\(f(y)\)_+=f(y)-\delta(\bar y)\int_0^1 dy' f(y').
\end{eqnarray}
Note that the gluon part is regular for $\xi\to 0$ since functions $G_\pm$ and $Y_\pm$ vanish at $x_{1,2,3}=0$. 

In  eq.~(\ref{result:Sivers},~\ref{result:tw3Evolution}) the integrals over $y$ and $\xi$ together with the $\delta(x-y\xi)$ reproduce the Mellin convolution. This convolution naturally appears during the calculation and it is defined for the whole range of $x$, $(-1<x<1)$ (and we recall that the anti-quark TMD distributions are given by values of $x<0$, see definition in eq.~(\ref{def:sivers_allX})). It should be understood literally
\begin{eqnarray}
\int d\xi\int_0^1 dy \delta(x-y\xi)f(y)g(\xi)=\left\{
\begin{array}{ll}
\Ds \int_{x}^1 \frac{d\xi}{\xi}f\(\frac{x}{\xi}\)g(\xi),&\qquad x>0,
\\
\Ds \int_{|x|}^1 \frac{d\xi}{\xi}f\(\frac{|x|}{\xi}\)g(-\xi),&\qquad x<0.
\end{array}\right.
\end{eqnarray}

\subsection{Discussion and comparison with earlier calculations}
\label{sec:discussion}

The evolution kernel in eq.~(\ref{result:tw3Evolution}) derived by us agrees with the known results in~\cite{Braun:2009mi,Kang:2012em}. Also, the matching of the twist-2 part coincides with earlier works exactly i.e. as the whole function of $\epsilon$. Altogether this provides a very strong check for the whole procedure and results derived by us.

It is instructive to compare eq.~(\ref{result:Sivers}) to the small-$b$ expansion of the unpolarized TMD distribution, which we have also reevaluated in this work to provide an additional cross-check.
 Following the notation of this work, it reads \cite{Collins:2011zzd,GarciaEchevarria:2011rb,Echevarria:2016scs,Vladimirov:2014aja}
\begin{eqnarray}\label{result:unpol}
&&f_{1}(x,\vec b;\mu,\zeta)=f_1(x)+a_s(\mu)\Big\{
-2\mathbf{L}_\mu P \otimes f_1+C_F\(-\mathbf{L}_\mu^2+2\mathbf{l}_\zeta \mathbf{L}_\mu+3\mathbf{L}_\mu-\frac{\pi^2}{6}\)f_1(x)
\\\nn &&\quad +
\int d\xi \int_0^1 dy \delta(x-y\xi)\Big[C_F 2\bar y f_1(\xi)+2 y \bar yg(\xi)\Big]\Big\}+O(a_s^2)+O(\vec b^2),
\end{eqnarray}
where the evolution kernel is
\begin{eqnarray}\label{result:unpol_EVOL}
\mu^2 \frac{d}{d\mu^2}f_1(x)&=&2a_s(\mu)P\otimes f_1
\\\nn &=&2a_s \int d\xi \int_0^1 dy \delta(x-y\xi)\Big\{C_F\(\frac{1+y^2}{1-y}\)_+f_1(\xi)+\frac{1-2y\bar y}{2}g(\xi)\Big\}.
\end{eqnarray}
One can see that eq.~(\ref{result:Sivers}) and eq.~(\ref{result:unpol}) have a very similar structure and, more precisely, \textit{the finite parts\footnoteref{foot1} of these expressions have the same $y$-behavior}. It is possible that this fact indicates some hidden correspondence which is to be understood in the future. 

Let us note that our calculation scheme (namely, the definition of distributions in $d$-dimensions, as it is discussed in sec.~\ref{sec:def-collinear}) affects only the quark-from-gluon  terms. In appendix \ref{app:diag-by-diag:matching} we present these mixing diagrams with the explicit designation of $\epsilon$'s from different sources. We have found that the scheme dependence enters the expressions via factors $\sim \epsilon/(1-\tilde \epsilon)$, where $\epsilon$ is the parameter of dimension regularization and $\tilde \epsilon$ is the parameter of $d$-dimensional definition of distributions. Therefore, the current choice of scheme influences only the $\epsilon$-suppressed terms of the final expression and thus it can contribute only from NNLO. Let us mention, that the same observation (namely, the suppression of the details of the $d$-dimensional definition in the  NLO coefficient function) is valid also in the case of the helicity distribution, which contains $\gamma^5$-matrix, see ref.~\cite{Gutierrez-Reyes:2017glx}.

The expressions for coefficient functions in eq.~(\ref{result:Sivers}-\ref{result:unpol}) are given for a general scale setting $(\mu,\zeta)$. For practical applications, it is convenient to use the $\zeta$-prescription \cite{Scimemi:2017etj,Scimemi:2018xaf}, where a TMD distribution is defined at the line $\zeta=\zeta(\mu)$. This line depends on certain boundary conditions that can be uniquely fixed and which define the so-called \textit{optimal TMD distribution}, see a detailed discussion in \cite{Scimemi:2018xaf}. The line $\zeta_\mu$ is universal for all TMD distributions and on this line the expression for the coefficient function simplifies. Namely, in eq.~(\ref{result:Sivers},~\ref{result:unpol}) one should set
\begin{eqnarray}
\text{in $\zeta$-prescription:}\qquad -\mathbf{L}_\mu^2+2\mathbf{l}_\zeta \mathbf{L}_\mu+3\mathbf{L}_\mu \to 0.
\end{eqnarray}
It is easy to see that in $\zeta$-prescription the TMD distribution is (naively-)independent on the scale $\mu$.

The matching coefficient for Sivers function can be found in the literature scattered in  different works: the quark-to-quark part has been deduced in~\cite{Sun:2013hua} and the quark-to-gluon part has been evaluated in~\cite{Dai:2014ala}. In both references the derivation of the matching coefficient has been made indirectly, refactorizing the factorized cross-section for SSA with the help of known matching for unpolarized TMD distribution. In our approach we evaluate the Sivers function directly, which grants us a better control over factors and schemes. Let us compare and comment on these works one-by-one.

In~\cite{Sun:2013hua} the quark-from-quark part of the matching  (the first term in square brackets in eq.~(\ref{result:Sivers})) is derived. 
A comparison with this work shows a disagreement in the  logarithmic part\footnoteref{foot1}, but an agreement in the finite part (i.e. compare eq.~(\ref{result:tw3Evolution}) with eq.~(12) of ~\cite{Sun:2013hua}). The origin of  this difference is clear. The calculation of ref.~\cite{Sun:2013hua} is based on the fixed-order calculation of SSA made in~\cite{Ji:2006ub,Koike:2007dg}. The latter considers only gluon-pole contributions and misses a quark-pole contribution, which roughly corresponds to our diagrams D (see detailed discussion in~\cite{Schafer:2012ra,Braun:2009mi,Kang:2012em}), which in turn, contributes only to the logarithmic part of matching coefficient, i.e. second line of eq.~ (\ref{result:Sivers})).

In~\cite{Dai:2014ala} the quark-to-gluon matching has been calculated. The result is presented  using the functions $N(x_1,x_2)$ and $O(x_1,x_2)$ which can be related to a combination of the functions $G$ and $Y$, similar to eq.~(\ref{relationG-T},~\ref{relationY-T}) (for a comparison of the definitions of these functions see~\cite{Beppu:2010qn}). In particular, $G_+(-x,0,x)+Y_+(-x,0,x)\simeq N(x,x)-N(x,0)$ and $G_-(-x,0,x)+Y_-(-x,0,x)\simeq O(x,x)-O(x,0)$.  Using these relations and  comparing with eq.~(44) of~\cite{Dai:2014ala}  we find a complete agreement with the logarithmic part (which is expected since it is given by the evolution kernel), but disagreement in the finite part. We claim that this disagreement is the result of a different parametrization of the gluon PDF used in~\cite{Dai:2014ala}. Indeed, according to eq.~(39) of \cite{Dai:2014ala}, the authors of \cite{Dai:2014ala} define PDF in $d$-dimensions, but  they do not decompose the tensors to irreducible representations and therefore $\epsilon$-dependent pre-factors of PDFs are different. 

In fact, the method of ref.~\cite{Dai:2014ala} could be inconsistent beyond LO. Indeed, the parameterization of the twist-3 matrix element used by~\cite{Dai:2014ala} is based on the 4-dimensional relation (see also~\cite{Beppu:2010qn})
\begin{eqnarray}\label{gepsilon=0}
g^{\mu\nu}\epsilon^{\alpha\beta\rho\delta}=
g^{\mu\alpha}\epsilon^{\nu\beta\rho\delta}+
g^{\mu\beta}\epsilon^{\alpha\nu\rho\delta}+
g^{\mu\rho}\epsilon^{\alpha\beta\nu\delta}+
g^{\mu\delta}\epsilon^{\alpha\beta\rho\nu},
\end{eqnarray}
which is used to reduce the number of degrees of freedom. In $d$-dimensions the relation in eq.~(\ref{gepsilon=0}) is not valid. Instead one has to use the decomposition to irreducible components (see discussion in appendix~\ref{app:tensor-decomposition}), as it is made in this work. In order to consistently use the parameterization based on eq.~(\ref{gepsilon=0}), the limit $\epsilon\to 0$ must be taken prior to the application of the parameterization, i.e. the approach one, as it is discussed in the introduction to the sec.~\ref{sec:def-collinear}. Contrary, the authors of~\cite{Dai:2014ala} have used a 4-dimensional parametrization within the $d$-dimensional calculation. There is no apparent contradiction at one-loop level, however, it can appear at higher perturbative orders.

\section{Conclusion}

We have derived the matching of the Sivers function to collinear distributions at NLO. The final result is given in eq.~(\ref{result:Sivers}) both for quark-to-quark and quark-to-gluon channels. The final result can be compared to the known calculations piece by piece: the logarithmic part agrees with the evolution kernel derived in~\cite{Braun:2009mi,Kang:2012em}, the finite quark-to-quark part agrees with the one derived in~\cite{Sun:2013hua} and the finite quark-to-gluon part is in disagreement with~\cite{Dai:2014ala}.
 In sec.~\ref{sec:discussion} we argue that the disagreement between our calculation and the calculation made in~\cite{Dai:2014ala} is due to the difference in  calculation schemes. 
The peculiarities of our calculation scheme are given in beginnings of sec.~\ref{sec:eval} and sec.~\ref{sec:def-collinear}. We also argue that our calculation scheme is equivalent to the scheme commonly used for twist-2 TMD matching, which we also confirm by comparing the twist-2 part of our calculation, eq.~(\ref{result:unpol}).

In contrast to all previous evaluations of Sivers function we do not consider any process but derive it directly from the definition of the TMD operator. The evaluation presented here is in many aspects novel, especially for the TMD community. Our calculation is made at the level of operators within the background field method which provides the most complete type of calculation and in the text we have described many details. In particular, for the first time, we explicitly demonstrate the appearance of rapidity divergences at the operator level, sec.~\ref{sec:rapidity_div} and explicitly demonstrate its renormalization at all twists of collinear OPE (sec.~\ref{sec:backrenormalization}). We also demonstrate the appearance of the famous sign flip for Sivers functions defined for DY and SIDIS, eq.~(\ref{SF:DY<->DIS}).

The method outlined in this work can be used also for the evaluation of the other leading order distributions which match on collinear twist-3 operators. All intermediate results of the calculation are presented in the appendix. Since the calculation is made at the level of operators, it contains the complete information on small-b OPE. In particular, it can be used to write down the matching of GTMD distributions to GPDs. Also, many diagrams can be used without recalculation for other polarizations. We expect 
 that this line of research will give new results in the near future and before the advent of the Electron Ion Collider (EIC).

\acknowledgments A.V. gratefully acknowledges V.~Braun and A.~Manashov for numerous stimulating discussions and help in  clarifying several aspects of higher twist calculus.
I.S. is supported by the Spanish MECD grant FPA2016-75654-C2-2-P. A.T. is grateful to J.W. Qiu and W. Vogelsang for valuable discussions and is supported by the U.S. Department of Energy, Office of Science, Office of Nuclear Physics under contract DE-AC02-98CH10886 and in part by the US DOE Transverse Momentum Dependent (TMD) Topical Theory Collaboration.

\appendix

\section{Parametrization of twist-3 operators and decomposition of 3-tensors}
\label{app:tensor-decomposition}

The light-cone gluon operators that enter our calculation are
\begin{eqnarray}\label{app:T+}
\mathcal{T}^{\mu\nu\rho}_+(z_1,z_2,z_3)&=&igf^{ABC}F^{A;\mu+}(z_1n)F^{B;\nu+}(z_2n)F^{C;\rho+}(z_3n),
\\\label{app:T-}
\mathcal{T}^{\mu\nu\rho}_-(z_1,z_2,z_3)&=&gd^{ABC}F^{A;\mu+}(z_1n)F^{B;\nu+}(z_2n)F^{C;\rho+}(z_3n),
\end{eqnarray}
where $f^{ABC}$ and $d^{ABC}$ are structure constants of the gauge-group. Here we omit the Wilson lines, for simplicity. To find an appropriate parametrization of these operators in  dimensional regularization, we proceed as the following. First of all, we decompose the $V\times V\times V$-tensor (with $V$ being a $2-2\epsilon$ dimensional vector) into irreducible components. There are 7 irreducible components, which can be selected by appropriate projectors. Explicitly the projectors read \cite{Cvitanovic:2008zz},
\begin{eqnarray}
\text{symmetric-traceless}\qquad P_1^{\mu\lambda\nu;\mu'\lambda'\nu'}&=& S^{\mu\nu\lambda;\mu'\nu'\lambda'}-P_2^{\mu\lambda\nu;\mu'\lambda'\nu'},
\\
\text{symmetric}\qquad P_2^{\mu\lambda\nu;\mu'\lambda'\nu'}&=& \frac{3}{4-2\epsilon} S^{\mu\nu\lambda;\alpha \beta \beta}S^{\alpha \gamma \gamma;\mu'\nu'\lambda'},
\\
\text{$\mu\nu$-symmetric-traceless}\qquad P_3^{\mu\lambda\nu;\mu'\lambda'\nu'}&=& \frac{4}{3}S^{\mu\nu;\alpha\beta}A^{\beta\lambda;\gamma\lambda'}S^{\alpha\gamma;\mu'\nu'}-
P_4^{\mu\lambda\nu;\mu'\lambda'\nu'},
\\
\text{$\mu\nu$-symmetric}\qquad P_4^{\mu\lambda\nu;\mu'\lambda'\nu'}&=& \frac{2}{1-2\epsilon} S^{\mu\nu;\alpha \beta}A^{\beta \lambda;\alpha \gamma}A^{\rho\gamma;\sigma \lambda'} S^{\rho \sigma ;\mu'\nu'},
\\
\text{$\mu\nu$-antisymmetric-traceless}\qquad P_5^{\mu\lambda\nu;\mu'\lambda'\nu'}&=& \frac{4}{3}A^{\mu\nu;\alpha\beta}S^{\beta\lambda;\gamma\lambda'}A^{\alpha\gamma;\mu'\nu'}-
P_6^{\mu\lambda\nu;\mu'\lambda'\nu'},
\\
\text{$\mu\nu$-antisymmetric}\qquad P_6^{\mu\lambda\nu;\mu'\lambda'\nu'}&=& \frac{2}{1-2\epsilon} A^{\mu\nu;\alpha \lambda}A^{\alpha\lambda';\mu'\nu'},
\\
\text{anti-symmetric}\qquad P_7^{\mu\lambda\nu;\mu'\lambda'\nu'}&=& A^{\mu\nu\lambda;\mu'\nu'\lambda'},
\end{eqnarray}
where $S^{\mu_1..\mu_n;\nu_1..\nu_n}$ ($A^{\mu_1..\mu_n;\nu_1..\nu_n}$) are (anti)symmetric products of $n$ $g^{\mu_i\nu_j}_T$'s, with normalization factor $1/n!$. These projectors satisfy
\begin{eqnarray}
g_T^{\mu\mu'}g_T^{\nu\nu'}g_T^{\lambda\lambda'}&=&\sum_{n=1}^7  P_i^{\mu\nu\lambda;\mu'\nu'\lambda'},
\qquad
 P_i^{\mu\nu\lambda;\alpha\beta\gamma}P_j^{\alpha\beta\gamma;\mu'\nu'\lambda'}=\delta_{ij}P_i^{\mu\nu\lambda;\mu'\nu'\lambda'}.
\end{eqnarray}
The dimension of corresponding irreducible sub-spaces are
\begin{eqnarray}
\text{dim}_i=P_i^{\mu\nu\lambda;\mu\nu\lambda}=\Big\{\frac{(\tilde d-1)\tilde d(\tilde d+4)}{6},\tilde d, \frac{\tilde d(\tilde d^2-4)}{3},\tilde d,\frac{\tilde d(\tilde d^2-4)}{3},\tilde d,\frac{\tilde d(\tilde d-1)(\tilde d-2)}{6}\Big\},
\end{eqnarray}
here $\tilde d=2(1-\epsilon)$. So, one can see that $3$'d, $5$'th and $7$'th subspaces vanishes at $\epsilon\to 0$. These subspaces represent evanescent components of operator.

In the next step we construct tensors that belong to particular subspaces,
\begin{eqnarray}
P^{\mu\nu\lambda;\alpha\beta\gamma}t_j^{\alpha\beta\gamma}=\delta_{ij}t_i^{\mu\nu\alpha}.
\end{eqnarray}
These tensors parameterize the forward matrix element and thus can be built out of single $s^\mu$, $a^{\mu\nu}$ (a $d$-dimensional analog of $\epsilon_T^{\mu\nu}$) and $g_T^{\mu\nu}$. We found
\begin{eqnarray}
t_2^{\mu\nu\lambda}&=&s^\alpha a^{\mu\alpha}g_T^{\nu\lambda}+s^\alpha a^{\nu \alpha}g_T^{\lambda \mu}+s^\alpha a^{\lambda \alpha}g_T^{\mu\nu},
\\
t_3^{\mu\nu\lambda}&=&s^\alpha a^{\mu\alpha}g_T^{\nu\lambda}-2s^\alpha a^{\nu \alpha}g_T^{\lambda \mu}+s^\alpha a^{\lambda \alpha}g_T^{\mu\nu}
+(1-2\epsilon)(s^\mu a^{\nu\lambda}-s^\lambda a^{\mu\nu}),
\\
t_4^{\mu\nu\lambda}&=&-s^\alpha a^{\mu\alpha}g_T^{\nu\lambda}+2s^\alpha a^{\nu \alpha}g_T^{\lambda \mu}-s^\alpha a^{\lambda \alpha}g_T^{\mu\nu},
\\
t_5^{\mu\nu\lambda}&=&3s^\alpha a^{\mu\alpha}g_T^{\nu\lambda}-3s^\alpha a^{\lambda \alpha}g_T^{\mu\nu}
+(1-2\epsilon)(-s^\mu a^{\nu\lambda}+2s^\nu a^{\lambda \mu}-s^\lambda a^{\mu\nu}),
\\
t_6^{\mu\nu\lambda}&=&s^\alpha a^{\mu\alpha}g_T^{\nu\lambda}-s^\alpha a^{\lambda \alpha}g_T^{\mu\nu},
\\
t_7^{\mu\nu\lambda}&=&-s^\mu a^{\nu\lambda}-s^\nu a^{\lambda \mu}-s^\lambda a^{\mu\nu}.
\end{eqnarray}
Note, that $s^\mu a^{\nu\mu}=\tilde s^\nu$. The tensor $t_1^{\mu\nu\lambda}=0$, since it is not possible to build completely traceless tensor with a single entry of a vector.

Finally, we parametrize the matrix element as
\begin{eqnarray}
\langle p,S|\mathcal{T}^{\mu\nu\lambda}_\pm(z_1,z_2,z_3)|p,S\rangle =(p^+)^3M\int [dx] e^{-ip^+\sum x_i z_i}\sum_{i=2}^7t_i^{\mu\nu\lambda}F^\pm_i(x_1,x_2,x_3),
\end{eqnarray}
where the integral measure is defined in eq.~(\ref{def:[dx]}). 
The distributions $F_{3,5,7}$ do not mix with other distributions at the perturbative order that we discuss here. Therefore, they could be safely set to zero. Therefore, we have three functions $F_{2,4,6}$ that survive in the limit $\epsilon\to 0$.

The operators $\mathcal{T}_\pm$ have the following property under  permutation of arguments
\begin{eqnarray}\label{app:T=T=T}
\mathcal{T}_\pm^{\mu\nu\lambda}(z_1,z_2,z_3)=\mp \mathcal{T}_\pm^{\mu\lambda\nu}(z_1,z_3,z_2)
=\mp \mathcal{T}_\pm^{\nu\mu\lambda}(z_2,z_1,z_3),
\end{eqnarray}
which  put some constraints on the functions $F_{2,4,6}$. Consequently, the function $F^-_2$($F_2^+$) is completely (anti)symmetric,
\begin{eqnarray}
F_2^\pm(x_1,x_2,x_3)=\mp F_2^\pm(x_2,x_1,x_3)=\mp F_2^\pm(x_1,x_3,x_2).
\end{eqnarray}
Another consequence of relation (\ref{app:T=T=T}) is that functions $F_4$ and $F_6$ are related to each other. We find it convenient to use $F_4$ as independent, setting
\begin{eqnarray}
F_6^\pm(x_1,x_2,x_3)=\pm\(F_4^\pm(x_1,x_3,x_2)-F_4^\pm(x_2,x_1,x_3)\).
\end{eqnarray}
The function $F_4^\pm$ has the following symmetry properties
\begin{eqnarray}
&&F_4^\pm(x_1,x_2,x_3)=\mp F_4^\pm(x_3,x_2,x_1),
\\\nn 
&&F_4^\pm(x_1,x_2,x_3)+F_4^\pm(x_2,x_3,x_1)+F_4^\pm(x_3,x_1,x_2)=0.
\end{eqnarray}

For convenience of comparison we introduce additional $\epsilon$-dependent factors and denote
\begin{eqnarray}
F_2^\pm(x_1,x_2,x_3)=-\frac{G_\pm(x_1,x_2,x_3)}{2(2-\epsilon)},\qquad
F_4^\pm(x_1,x_2,x_3)=-\frac{Y_\pm(x_1,x_2,x_3)}{2(1-2\epsilon)},
\end{eqnarray}
and we obtain the parametrization in eq.~(\ref{distr:gluons}).

Let us also make an analogy with the parameterization of quark operator. The general quark operator with positive parity has three indices (if we omit evanescent operators with anti-symmetric products of 4, 6, etc. indices). It reads
\begin{eqnarray}
\mathcal{T}^\nu_{\gamma^+\gamma^\mu \gamma^\lambda}(z_1,z_2,z_3)&=&g\bar q(z_1n)\gamma^+ \gamma^\mu F^{\nu+}(z_2n)\gamma^\lambda q(z_3n),
\end{eqnarray}
where all indices are transverse. Here, we omit the Wilson lines, for simplicity. Therefore, it is parameterized by the same set of tensors,
\begin{eqnarray}
\langle p,S|\mathcal{T}^{\nu}_{\gamma^+\gamma^\mu\gamma^\lambda}(z_1,z_2,z_3)|p,S\rangle=(p^+)^2M\int [dx] e^{-ip^+\sum x_i z_i}\sum_{i=2}^7t_i^{\mu\nu\lambda}Q_i(x_1,x_2,x_3).
\end{eqnarray}
For the same reasons as for the gluon operator we drop all functions $Q_{3,5,7}$. The remaining functions $Q_{2,4,6}$ are not independent, but can be related by time-reversal symmetry. In particular we get $Q_2=Q_4$. Comparing to the parameterizations in eq.~(\ref{def:T},~\ref{def:DeltaT}) we get
\begin{eqnarray}
T(x_1,x_2,x_3)=\frac{Q_2(x_1,x_2,x_3)}{3(1-\epsilon)},\qquad \Delta T(x_1,x_2,x_3)=-\frac{Q_6(x_1,x_2,x_3)}{2}.
\end{eqnarray}
Therefore, we can conclude that the function $\Delta T$ is the quark analog of $F_6$ gluon distribution.

\section{Example of evaluation: diagram E}
\label{app:exampleDiag}

In this appendix we give a detailed technical description of the evaluation of a diagram. For demonstration purposes we have selected the diagram E (see fig.~\ref{fig:3point}) since it is the most involved diagram, which allows to demonstrate all particularities of the calculation. The remaining diagrams are obtained in a similar manner, albeit the evaluation is typically shorter.

\subsection{Evaluation of contribution to OPE}

The diagram reads
\begin{eqnarray}
\widetilde{\mathcal{U}}_{\mathbf{E}}=
\contraction[6pt]{\Big(ig\int d^du \,\bar q\fnot B}{\psi}{(u)\Big)\Big\{}{\bar \psi}
\contraction[6pt]{\Big(ig\int d^du \,\bar q\fnot B\psi(u)\Big)\Big\{
\bar \psi(z_1+\vec b)\gamma^+ }{\psi}{(z_2-\vec b)\Big\}
\Big(ig\int d^dx }{\bar \psi}
\contraction[6pt]{\Big(ig\int d^du \,\bar q\fnot B\psi(u)\Big)\Big\{
\bar \psi(z_1+\vec b)\gamma^+ \psi(z_2-\vec b)\Big\}
\Big(ig\int d^dx \bar \psi\fnot A}{\psi}{(x)\Big)\Big(ig\int d^dy }{\bar \psi}
\contraction[8pt]{~\Big(ig\int d^du \,\bar q}{\fnot B}{\psi(u)\Big)\Big\{
\bar \psi(z_1+\vec b)\gamma^+ \psi(z_2-\vec b)\Big\}
\Big(ig\int d^dx \bar \psi\fnot A\psi(x)\Big)
\Big(ig\int d^dy \bar \psi}{\fnot B}
\Big(ig\int d^du \,\bar q\fnot B\psi(u)\Big)\Big\{
\bar \psi(z_1+\vec b)\gamma^+ \psi(z_2-\vec b)\Big\}
\Big(ig\int d^dx \bar \psi\fnot A\psi(x)\Big)
\Big(ig\int d^dy \bar \psi\fnot Bq(y)\Big),\nn
\end{eqnarray}
where the factors in the round brackets come from the expansion of the action exponent. Using the expressions for propagators in dimension regularization (with $d=4-2\epsilon$)
\begin{eqnarray}
\contraction{}{\psi}{(x)}{\bar \psi}\psi(x)\bar \psi(y)&=&\frac{\Gamma(2-\epsilon)}{2\pi^{d/2}}\frac{i(\fnot x-\fnot y)}{(-(x-y)^2+i0)^{2-\epsilon}}
\\
\contraction{}{B}{\,(x)~}{B}
B_\mu^A(x)B_\nu^B(y)&=&\frac{\Gamma(1-\epsilon)}{4\pi^{d/2}}\frac{-g_{\mu\nu}\delta^{AB}}{(-(x-y)^2+i0)^{1-\epsilon}},
\end{eqnarray}
we obtain
\begin{eqnarray}\label{app:d2}
\widetilde{\mathcal{U}}_{\mathbf{E}}&=& -g^3\frac{\Gamma^3(2-\epsilon)\Gamma(1-\epsilon)}{32\pi^{2d}}\(C_F-\frac{C_A}{2}\)\int d^du d^dxd^dy 
\\\nn &&
 \frac{\bar q(u)A_\mu(x)\gamma_\nu(\fnot u-\fnot b)\gamma^+(\fnot x+\fnot b)\gamma^\mu(\fnot x-\fnot y)\gamma^\nu q(y)}{[-(u-z_1-b)^2+i0]^{2-\epsilon}[-(x-z_2+b)^2+i0]^{2-\epsilon}[-(x-y)^2+i0]^{2-\epsilon}[-(u-y)^2+i0]^{1-\epsilon}},
\end{eqnarray}
and we have used that $\gamma^+\gamma^+=0$.

The expression in eq.~(\ref{app:d2}) should be understood as a generating function that contributes to all orders of small-$b$ expansion. The typical task requires a consideration of terms with a particular counting only. For instance, in this work we need only the terms proportional to $b^\mu$. The most straightforward approach to extract particular contributions from such generating function is to  Taylor expand all fields around a point (say 0) and evaluating  the loop integral that decouples from the fields. In the resulting series, the desired contributions are to be sorted out and resummed back to the non-local form. However, this is a very algebraically heavy way. Here we use an equivalent, but much more efficient,  strategy that requires the evaluation of only several terms. It is described in the following.

First of all we decouple the expansion parameter (here the vector $b$) from the integration variables. The natural way to do so, is to join propagators by the Feynman variables and make the shift of variables. For this diagram we introduce four Feynman variables $\alpha$, $\beta$, $\gamma$ and $\rho$ for propagators from left to right in (\ref{app:d2}). Then we make a shift of variables
\begin{eqnarray*}
x&\to& x+r_x=x+\frac{\alpha\gamma\rho}{\lambda}z_1+\frac{\alpha\beta\gamma+\alpha\beta\rho+\beta\gamma\rho}{\lambda}z_2-\vec b\(1-2\frac{\alpha\gamma\rho}{\lambda}\)
\\
y&\to & y+r_y =y+\frac{\alpha\gamma\rho+\alpha\beta\rho}{\lambda}z_1+\frac{\alpha\beta\gamma+\beta\gamma\rho}{\lambda}z_2-\vec b\(1-2\frac{\alpha\gamma\rho+\alpha\beta\rho}{\lambda}\)
\\
u&\to& u+r_u=u+\frac{\alpha\gamma\rho+\alpha\beta\gamma+\alpha\beta\rho}{\lambda}z_1+\frac{\beta\gamma\rho}{\lambda}z_2+\vec b\(1-2\frac{\beta\gamma\rho}{\lambda}\)
\end{eqnarray*}
where
\begin{eqnarray}\label{app:lambda}
\lambda=\alpha\gamma\rho+\alpha\beta\gamma+\alpha\beta\rho+\beta\gamma\rho.
\end{eqnarray}
After these transformations the expression for the diagram is
\begin{eqnarray}\nn
&&\widetilde{\mathcal{U}}_{\mathbf{E}}= -g^3\frac{\Gamma(7-4\epsilon)}{32\pi^{2d}}\(C_F-\frac{C_A}{2}\)\int d^du d^dxd^dy  \int [d\alpha d\beta d\gamma d\rho]
(\alpha\beta\gamma)^{1-\epsilon}\rho^{-\epsilon} \\\nn &&
\frac{\bar q(u+r_u)A_\mu(x+r_x)\gamma_\nu(\fnot u-2\frac{\beta\gamma\rho}{\lambda}\fnot b)\gamma^+(\fnot x+2\frac{\alpha\gamma\rho}{\lambda}\fnot b)\gamma^\mu(\fnot x-\fnot y-\frac{\alpha\beta\rho}{\lambda}(2\fnot b+z_{12}\gamma^+))\gamma^\nu q(y+r_y)}{[-(\beta+\gamma)x^2-(\gamma+\rho)y^2-(\alpha+\rho)u^2+2\rho (uy)+2\gamma(xy)+4\frac{\alpha\beta\gamma\rho}{\lambda}\vec b^2+i0]^{7-4\epsilon}},
\end{eqnarray}
where $z_{12}=z_1-z_2$. The next step is to expand the fields around the points $r_i$. The resulting expression is a series of integrals with a given propagator and monomials built of $x^\mu$, $y^\mu$ and $u^\mu$. The open indices of such integral can result only into the metric tensors $g^{\mu\nu}$. The dimension of the loop-integral is carried entirely by $\vec b^2$ and can be easily computed. The loop-integral and the numerator is the only source of $\vec b$. It also enters the argument of the fields, but this source is independent from the loop-computation and can be considered later. Thus we sort all terms in the expressions in  powers of $\vec b$ and select the ones that are linearly proportional to $\vec b$.

Note, that the terms with the same dimension do not necessary have the same $\vec b$-counting. As so, all terms without $z_{12}$ has counting $n+1$ (where $n$ is the number of fields derivatives). Therefore, only terms without field derivative contribute in this case. The terms that contain factor $z_{12}$ has counting $n+0$ and require the expansion of the fields up to one derivative. Let us note, that the expansion of fields in $\vec b$ rises the counting even more and so it does not contribute at considered order. For that reason we can neglect $\vec b$ in the argument of fields (Such contributions can appear only in the diagrams that also contribute to twist-2, i.e. A, B and L).

The loop integration is straightforward. We have
\begin{eqnarray}\label{LoopInt:3}
&&\int d^dx d^dy d^d u\frac{1}{[\Delta+i0]^{7-4\epsilon}}
=\frac{-i\pi^{3d/2}\Gamma(-\epsilon)}{\Gamma(7-4\epsilon)}\frac{\epsilon\lambda^{\epsilon-2}}{X^{1-\epsilon}}
\\
&&\int d^dx d^dy d^d u\frac{\{x^\mu x^\nu,y^\mu y^\nu,u^\mu u^\nu,x^\mu y^\nu, x^\mu u^\nu,y^\mu u^\nu\}}{[\Delta+i0]^{7-4\epsilon}}
=\frac{-i\pi^{3d/2}\Gamma(-\epsilon)}{\Gamma(7-4\epsilon)}\frac{\lambda^{\epsilon-3}}{X^{-\epsilon}}\frac{g^{\mu\nu}}{2}
\\\nn&&\qquad\qquad\{\alpha\rho +\alpha \gamma+\rho \gamma,~(\alpha+\rho)(\gamma+\beta),~\beta\rho+\rho \gamma+\gamma\beta,~
(\alpha+\rho)\gamma,~\rho\gamma ,~\rho(\gamma+\beta)\},
\\ 
&&
\int d^dx d^dy d^du\frac{\{\overbrace{x^\mu .. y^\nu}^{\text{odd \#}}\}}{[\Delta+i0]^{7-4\epsilon}}
=0,
\end{eqnarray}
where $\Delta=-(\gamma+\beta)x^2-(\rho+\gamma)y^2-(\alpha+\rho)u^2+2\gamma (xy)+2\rho (yu)+X$, with $X=4\alpha\beta\gamma\rho\vec b^2/\lambda >0$ and $\lambda$ is defined in (\ref{app:lambda}). 
The obtained expression can be drastically simplified once we pass to dual Feynman variables. They are defined as
\begin{eqnarray*}
\alpha'=\frac{\beta\gamma\rho}{\lambda},
\qquad
\beta=\frac{\alpha\gamma\rho}{\lambda},
\qquad
\gamma=\frac{\alpha\beta\rho}{\lambda}
\qquad
\rho=\frac{\alpha\beta\gamma}{\lambda}.
\end{eqnarray*}
The integration domain of dual variables coincides with the integration domain of original variables and the Jacobian of transformation is
\begin{eqnarray}
\frac{[d\alpha' d\beta' d\gamma' d\rho']}{[d\alpha d\beta d\gamma d\rho]}=\frac{(\alpha\beta\gamma\rho)^2}{\lambda^4}.
\end{eqnarray}
In fact, the dual Feynman variables are the variables that appear if one calculates the loop-integration in  momentum space. The arguments of the fields $r_i$ in the terms of dual variables take a simple form
\begin{eqnarray}
r_x=z_{21}^\beta,\qquad r_y=z_{21}^{\beta+\gamma},\qquad r_u=z_{12}^\alpha,
\end{eqnarray}
where $z_{ij}^\alpha=z_i(1-\alpha)+z_j\alpha$.

After these transformations and minor algebraic simplifications, we obtain 
\begin{eqnarray}\label{app:diagE_final_OPE}
\widetilde{\mathcal{U}}_{\mathbf{E}}&&=-2iga_s\vec b^{2\epsilon}\Gamma(-\epsilon)\(C_F-\frac{C_A}{2}\) b_\mu\int [d\alpha d\beta d\gamma d\rho]\Big\{
\\\nn &&
(1-\epsilon)[1+z_{12}(\alpha \partial_1+\bar \beta \partial_2+(1-\beta-\gamma)\partial_3)]\mathcal{Q}^\mu_{\gamma^+}(z_{12}^\alpha,z_{21}^\beta,z_{21}^{\beta+\gamma})
\\\nn &&-(1+\epsilon)[3-z_{12}(\alpha \partial_1+\bar \beta \partial_2+(1-\beta-\gamma)\partial_3)]\mathcal{Q}^\nu_{\gamma^+\gamma^{\nu\mu}}(z_{12}^\alpha,z_{21}^\beta,z_{21}^{\beta+\gamma})\Big\},
\end{eqnarray}
where the definition of $\mathcal{Q}$ is given in (\ref{app:Q}) and $\partial_{1,2,3}$ is the $\partial_+$ that acts on $\bar q$, $A$, $q$ in $\mathcal{Q}$. The expression for the diagram $\mathbf{E^*}$ could be obtained from this one by inversion of order of $\gamma$-matrices and field order and  $z_1\leftrightarrow z_2$.  The analogous expressions for other diagrams are given in appendix \ref{app:diag-by-diag_OP}.

\subsection{Evaluation of matrix element}
The contribution of the diagram $\mathbf{E}$ to the matching expression is calculated by
\begin{eqnarray}\label{app:toE}
f_{\mathbf{E}}=\int \frac{dz}{2\pi} e^{-2ixp^+z}\langle p,S|\widetilde{\mathcal{U}}_{\mathbf{E}}\(z_1=-z_2=z,\frac{\vec b}{2}\)|p,S\rangle.
\end{eqnarray}
In order to illustrate this calculation  we consider, for definiteness, DY induced operator.

As a first step, we rewrite the operators $\mathcal{Q}^\mu_{\Gamma}$  in terms of operators $\mathcal{T}^\mu_\Gamma$ whose matrix elements define the twist-3 collinear distributions eq.~(\ref{distr:T1},~\ref{distr:T2}). To do so, we eliminate light-cone derivatives in eq.~(\ref{app:diagE_final_OPE}) using integration by parts over the Feynman parameters. For example,
\begin{eqnarray}
&&\int [d\alpha d\beta d\gamma d\rho] z_{12}\bar \beta (\partial_2+\partial_3) \mathcal{Q}^\mu_{\gamma^+}(z_{12}^\alpha,z_{21}^\beta,z_{21}^{\beta+\gamma})=
\int [d\alpha d\beta d\gamma d\rho] \bar \beta \partial_\beta \mathcal{Q}^\mu_{\gamma^+}(z_{12}^\alpha,z_{21}^\beta,z_{21}^{\beta+\gamma})
\\\nn &&=\int [d\alpha d\beta d\gamma] \(\bar \beta \mathcal{Q}^\mu_{\gamma^+}(z_{12}^\alpha,z_{21}^\beta,z_{21}^{\beta+\gamma})-
\mathcal{Q}^\mu_{\gamma^+}(z_{12}^\alpha,z_2,z_{21}^{\gamma})\)+\int [d\alpha d\beta d\gamma d\rho]\mathcal{Q}^\mu_{\gamma^+}(z_{12}^\alpha,z_{21}^\beta,z_{21}^{\beta+\gamma})
\\\nn &&=\int [d\alpha d\beta d\gamma d\rho] \(1+\bar \beta \delta(\rho)-\delta(\beta)\)\mathcal{Q}^\mu_{\gamma^+}(z_{12}^\alpha,z_{21}^\beta,z_{21}^{\beta+\gamma}),
\end{eqnarray}
and similarly for other derivatives. As a result of this procedure we get
\begin{eqnarray}\label{app:diagE_0}
\widetilde{\mathcal{U}}_{\mathbf{E}}&&=-2iga_s\vec b^{2\epsilon}\Gamma(-\epsilon)\(C_F-\frac{C_A}{2}\) b_\mu\int [d\alpha d\beta d\gamma d\rho]\Big\{
\\\nn &&
(1-\epsilon)[4 -\delta(\beta)]\mathcal{Q}^\mu_{\gamma^+}(z_{12}^\alpha,z_{21}^\beta,z_{21}^{\beta+\gamma})
-(1+\epsilon)[-1+\delta(\beta)]\mathcal{Q}^\nu_{\gamma^+\gamma^{\nu\mu}}(z_{12}^\alpha,z_{21}^\beta,z_{21}^{\beta+\gamma})\Big\}.
\end{eqnarray}
We also replace $A^\mu$ by $F^{\mu+}$ using the identity valid in the light-cone gauge
\begin{eqnarray}
A^\mu(z n)&=&-\int_{-\infty}^{z} d\sigma~F^{\mu+}(\sigma n).
\end{eqnarray}
This is valid for the operator in the DY kinematics while in SIDIS kinematics the identity eq.~(\ref{gauge:A=F_adv}) should be used instead. The result of these operations reads
\begin{eqnarray}\label{app:diagE_1}
\widetilde{\mathcal{U}}_{\mathbf{E}}&&=2ia_s\vec b^{2\epsilon}\Gamma(-\epsilon)\(C_F-\frac{C_A}{2}\) b_\mu\int [d\alpha d\beta d\gamma d\rho]\Big\{
\\\nn &&
(1-\epsilon)[4-\delta(\beta)]\int_{-\infty}^{z_{21}^\beta}d\sigma\mathcal{T}^\mu_{\gamma^+}(z_{12}^\alpha,\sigma,z_{21}^{\beta+\gamma})
-(1+\epsilon)[-1+\delta(\beta)]\int_{-\infty}^{z_{21}^\beta}d\sigma\mathcal{T}^\nu_{\gamma^+\gamma^{\nu\mu}}(z_{12}^\alpha,\sigma,z_{21}^{\beta+\gamma})\Big\} .
\end{eqnarray}

Next, we evaluate the matrix element of eq.~(\ref{app:diagE_1}) by applying the definitions in eq.~(\ref{def:T},~\ref{def:DeltaT}):
\begin{eqnarray}\label{app:diagE_11}
&&\langle p,S|\widetilde{\mathcal{U}}_{\mathbf{E}}|p,S\rangle=4ia_s M (p^+)^2 \vec b^{2\epsilon}\Gamma(-\epsilon)\(C_F-\frac{C_A}{2}\) \tilde s^\mu b_\mu\int [d\alpha d\beta d\gamma d\rho]\int[dx]\Big\{
\\\nn &&\qquad
(1-\epsilon)[4-\delta(\beta)]\int_{-\infty}^{z_{21}^\beta}d\sigma e^{-ip^+(x_1 z_{12}^\alpha+x_2 \sigma+x_3 z_{21}^{\beta+\gamma})} T(x_1,x_2,x_3)
\\\nn &&\qquad+(1+\epsilon)[-1+\delta(\beta)]\int_{-\infty}^{z_{21}^\beta}d\sigma e^{-ip^+(x_1 z_{12}^\alpha+x_2 \sigma+x_3 z_{21}^{\beta+\gamma})} \Delta T(x_1,x_2,x_3)\Big\},
\end{eqnarray}
where $[dx]=dx_1dx_2dx_3\delta(x_1+x_2+x_3)$. In the case of forward matrix element, the further evaluation can be essentially simplified by adding the conjugated diagram $\mathbf{E^*}$. After the same manipulations, diagram $\mathbf{E^*}$ is
\begin{eqnarray}\label{app:diagE_2}
&&\langle p,S|\widetilde{\mathcal{U}}_{\mathbf{E^*}}|p,S\rangle=4ia_s M (p^+)^2 \vec b^{2\epsilon}\Gamma(-\epsilon)\(C_F-\frac{C_A}{2}\) \tilde s^\mu b_\mu\int [d\alpha d\beta d\gamma d\rho]\int[dx]\Big\{
\\\nn &&\qquad
(1-\epsilon)[4-\delta(\beta)]\int_{-\infty}^{z_{12}^\beta}d\sigma e^{-ip^+(x_1 z_{12}^{\beta+\gamma}+x_2 \sigma+x_3 z_{21}^\alpha)} T(x_1,x_2,x_3)
\\\nn &&\qquad-(1+\epsilon)[-1+\delta(\beta)]\int_{-\infty}^{z_{12}^\beta}d\sigma e^{-ip^+(x_1 z_{12}^{\beta+\gamma}+x_2 \sigma+x_3 z_{21}^\alpha)} \Delta T(x_1,x_2,x_3)\Big\}.
\end{eqnarray}
The sum of conjugated diagrams can be simplified with the help of symmetry relations see eq.~(\ref{quark:symT},~\ref{quark:symdT}). Let us show this procedure taking as an 
 example  the first term in the curly brackets of eq.~(\ref{app:diagE_11},~\ref{app:diagE_2}). We have
\begin{eqnarray}
&&\nn
\int [dx]T(x_1,x_2,x_3)\(
\int_{-\infty}^{z_{21}^\beta}d\sigma e^{-ip^+(x_1 z_{12}^\alpha+x_2 \sigma+x_3 z_{21}^{\beta+\gamma})} +\int_{-\infty}^{z_{12}^\beta}d\sigma e^{-ip^+(x_1 z_{12}^{\beta+\gamma}+x_2 \sigma+x_3 z_{21}^\alpha)} \)
\\\nn&&=\int [dx]T(x_1,x_2,x_3)\(
\int_{-\infty}^{z_{21}^\beta}d\sigma e^{-ip^+(x_1 z_{12}^\alpha+x_2 \sigma+x_3 z_{21}^{\beta+\gamma})} +\int_{-\infty}^{z_{12}^\beta}d\sigma e^{-ip^+(-x_3 z_{12}^{\beta+\gamma}-x_2 \sigma-x_1 z_{21}^\alpha)} \)
\\&&=\int [dx]\int_{-\infty}^{\infty}d\sigma T(x_1,x_2,x_3)e^{-ip^+(x_1 z_{12}^\alpha+x_2 \sigma+x_3 z_{21}^{\beta+\gamma})},
\end{eqnarray}
where in the second line we have changed $x_{1,2,3}\to -x_{3,2,1}$, and in the third line we have changed $\sigma \to -\sigma+z_1+z_2$ for the contribution of the diagram $\mathbf{E^*}$. 

The integral over $\sigma$ is equal to  $2\pi\delta(x_2)$ and we obtain for the full diagram
\begin{eqnarray}\label{app:diagE_4}
&&\langle p,S|\widetilde{\mathcal{U}}_{\mathbf{E}+\mathbf{E^*}}|p,S\rangle=8\pi ia_s M p^+\vec b^{2\epsilon}\Gamma(-\epsilon)\(C_F-\frac{C_A}{2}\) \tilde s^\mu b_\mu\int [d\alpha d\beta d\gamma d\rho]\int[dx]\Big\{
\\\nn &&\qquad
(1-\epsilon)[4-\delta(\beta)]\delta(x_2)e^{-ip^+(x_1 z_{12}^{\alpha}+x_3 z_{21}^{\beta+\gamma})} T(x_1,x_2,x_3)
\\\nn &&\qquad+(1+\epsilon)[-1+\delta(\beta)]\delta(x_2) e^{-ip^+(x_1 z_{12}^{\alpha}+x_3 z_{21}^{\beta+\gamma})} \Delta T(x_1,x_2,x_3)\Big\}.
\end{eqnarray}
The last line of eq.~(\ref{app:diagE_4}) is zero since $\Delta T(x,0,-x)=0$. At the point $z_1=-z_2=z$ the expression simplify further
\begin{eqnarray}\label{app:diagE_5}
&&\langle p,S|\widetilde{\mathcal{U}}_{\mathbf{E}+\mathbf{E^*}}(z_1=-z_2=z)|p,S\rangle=8\pi ia_s M p^+\vec b^{2\epsilon}\Gamma(-\epsilon)\(C_F-\frac{C_A}{2}\) \tilde s^\mu b_\mu
\\\nn &&\qquad \int [d\alpha d\beta d\gamma d\rho]\int_{-1}^1 dx_1
(1-\epsilon)[4-\delta(\beta)]e^{-ip^+x_1 z \rho} T(x_1,0,-x_1)
\\\nn &&=8\pi ia_s M p^+\vec b^{2\epsilon}\Gamma(-\epsilon)\(C_F-\frac{C_A}{2}\) \tilde s^\mu b_\mu
\int_0^1 d\rho \int_{-1}^1 dx_1
(1-\epsilon)\bar \rho(1-2\rho)e^{-ip^+x_1 z \rho} T(x_1,0,-x_1).
\end{eqnarray}
Finally, making Fourier transformation to momentum faction $x$ as in eq.~(\ref{app:toE}) we get
\begin{eqnarray}
f_{\mathbf{E}+\mathbf{E^*}}&=&2 \pi i M a_s  (1-\epsilon)\(C_F-\frac{C_A}{2}\)\Gamma(-\epsilon)\(\frac{\vec b^2}{4}\)^{\epsilon} (\tilde s \cdot b)
\\\nn &&\int d\xi \int_0^1 dy \delta(x-y\xi)\bar y(1-2y) T(-,\xi,0,\xi),
\end{eqnarray}
where we rename $\rho\to y$ and $x_1\to \xi$, and rescale $b\to b/2$.

All other diagrams are evaluated in the same manner, with the only difference that self-conjugated diagrams are already symmetric with respect to $x_{1,2,3}\to-x_{3,2,1}$. The diagram-by-diagram expressions are given in appendix \ref{app:diag-by-diag:matching}.

\section{Diagram-by-diagram expressions}
\label{app:diag-by-diag}

In this appendix we collect the expressions for diagrams presented in figs.~\ref{fig:2point},~\ref{fig:3point} and~\ref{fig:QG}.

\subsection{Expressions for OPE}
\label{app:diag-by-diag_OP}

In this appendix we provide the full set of expressions obtained from the evaluation of diagrams in  background field. The expressions are given in  light-cone gauge for the Drell-Yan operator eq.~(\ref{def:TMDop_DY}) (i.e. with retarded eq.~(\ref{gauge:ret}) boundary conditions). The analogous expressions for the SIDIS operator, eq.~(\ref{def:TMDop_DIS}), are obtained by replacing  $-\infty$ with $+\infty$ in the integration limits, as it is discussed in sec.~\ref{sec:DY-SIDIS-difference}. We stress that the calculation has been done for an operator with unrelated light cone positions of fields $z_1$ and $z_2$. Therefore, the OPE presented here  is also suitable for evaluating the matching of the GTMD distributions.

We use the following shorthand notation 
\begin{eqnarray}
&&\bar \alpha =1-\alpha,\qquad z_{ij}^\alpha = z_i \bar \alpha+z_j \alpha,\qquad z_{ij}=z_i-z_j,
\\
&&\vec b^2=-b^2>0,\qquad a_s=\frac{g^2}{(4\pi)^2}.
\end{eqnarray}
The combination $z_{i\sigma}^\alpha$ is a shorthand notation for $z_{ij}^\alpha$ with $z_j=\sigma$ and analogously for $z_{\sigma i}^\alpha$. The variables $\alpha$, $\beta$, $\gamma$ and $\rho$ are usual Feynman variables, which satisfy $(\alpha+\beta+\gamma+\rho=1)$. For convenience we put this restriction into the definition of the integration measure $[d\alpha d\beta ...]$ (here the dots indicate the number of Feynman variables participating in a diagram). For example for three variables we define
\begin{eqnarray}
\int [d\alpha d\beta d\gamma]f(\alpha,\beta,\gamma)\equiv \int d\alpha d\beta d\gamma \delta(1-\alpha-\beta-\gamma)f(\alpha,\beta,\gamma).
\end{eqnarray}

Here are the expressions for individual diagram contributions into the OPE: 
\begin{align}
\widetilde{\mathcal{U}}_{\mathbf{A}}&=2a_sC_F\Gamma(-\epsilon) \vec b^{2\epsilon}
 \int_{-\infty}^{z_1} d\sigma  
\int_0^1 d\alpha \,\bar \alpha~\bar q( z_1n +\vec b)\gamma^+ \,\overrightarrow{\partial_+}q( z^\alpha_{2\sigma}n - (1-2\alpha)\vec b),
\\
\widetilde{\mathcal{U}}_{\mathbf{A^*}}&=2a_sC_F \Gamma(-\epsilon)\vec b^{2\epsilon}
 \int_{-\infty}^{z_2} d\sigma  
\int_0^1 d\alpha\,\bar\alpha ~\bar q( z_{1\sigma}^\alpha n +(1-2\alpha)\vec b)~\overleftarrow{\partial_+}\gamma^+ q(z_2 n -\vec b),
\\ 
\widetilde{\mathcal{U}}_{\mathbf{B}}&=
2a_sC_F\Gamma(-\epsilon)\vec b^{2\epsilon}\int[d\alpha d\beta d\gamma]\Big\{(1-\epsilon)
~
\bar q(z_{12}^\alpha n+\vec b \(1-2\alpha\))
\gamma^+\,
q(z_{21}^\beta n-\vec b \(1-2\beta\))
\\&\nn +b_\mu  \bar q(z_{12}^\alpha n)\gamma^+\Big[(1-\epsilon)\((1-2\alpha)\overleftarrow{\partial^\mu}-(1-2\beta)\overrightarrow{\partial^\mu}\)
-(1+\epsilon)(\overleftarrow{\partial_{\nu}}+\overrightarrow{\partial_{\nu}})
\gamma^{\nu\mu}\,
\Big]q(z_{21}^\beta n)\Big\},
\\ 
\widetilde{\mathcal{U}}_{\mathbf{C}}&=
-2iga_s\Gamma(-\epsilon)\vec b^{2\epsilon}\(C_F-\frac{C_A}{2}\)b_\mu\int_{-\infty}^{z_1} d\sigma \int[d\alpha d\beta d\gamma]\Big\{
\\ &\nn ((1-2\beta) \partial_2+2\alpha \partial_3)\mathcal{Q}^\mu_{\gamma^+}(z_1,z_{2\sigma}^\beta,z_{\sigma2}^\alpha)
+\partial_2 \mathcal{Q}^\nu_{\gamma^+ \gamma^{\nu\mu}}(z_1,z_{2\sigma}^\beta,z_{\sigma2}^\alpha)\Big\},
\\
\widetilde{\mathcal{U}}_{\mathbf{C^*}}&=
-2iga_s\Gamma(-\epsilon)\vec b^{2\epsilon}\(C_F-\frac{C_A}{2}\)b_\mu\int_{-\infty}^{z_1} d\sigma \int[d\alpha d\beta d\gamma]\Big\{
\\ &\nn ((1-2\beta) \partial_2+2\alpha \partial_1)\mathcal{Q}^\mu_{\gamma^+}(z_{\sigma 1}^\alpha,z_{1\sigma}^\beta,z_2)
-\partial_2 \mathcal{Q}^\nu_{\gamma^+ \gamma^{\nu\mu}}(z_{\sigma 1}^\alpha,z_{1\sigma}^\beta,z_2)\Big\},
\\
\widetilde{\mathcal{U}}_{\mathbf{D}}&=
-2iga_s\Gamma(-\epsilon)\vec b^{2\epsilon}\frac{C_A}{2}b_\mu \int_{-\infty}^{z_1} d\sigma \int[d\alpha d\beta d\gamma]\Big\{
\\\nn&
((1-2\alpha) \partial_2-2\bar \beta \partial_3)\mathcal{Q}^\mu_{\gamma^+}(z_1,z_{\sigma 2}^\alpha,z_{2\sigma}^\beta)-\partial_2\mathcal{Q}^\nu_{\gamma^+\gamma^{\nu\mu}}(z_1,z_{\sigma 2}^\alpha,z_{2\sigma}^\beta)
\Big\},
\\
\widetilde{\mathcal{U}}_{\mathbf{D^*}}&=
-2iga_s\Gamma(-\epsilon)\vec b^{2\epsilon}\frac{C_A}{2}b_\mu\int_{-\infty}^{z_2} d\sigma \int[d\alpha d\beta d\gamma] \Big\{
\\\nn&
((1-2\alpha)\partial_2-2\bar \beta \partial_1)\mathcal{Q}^\mu_{\gamma^+}(z_{1\sigma}^\beta,z_{\sigma 1}^\alpha,z_2)+\partial_2\mathcal{Q}^\nu_{\gamma^+\gamma^{\nu\mu}}(z_{1\sigma}^\beta,z_{\sigma 1}^\alpha,z_2)\Big\},
\end{align}
\begin{align}
\widetilde{\mathcal{U}}_{\mathbf{E}}&=-2iga_s\vec b^{2\epsilon}\Gamma(-\epsilon)\(C_F-\frac{C_A}{2}\) b_\mu\int [d\alpha d\beta d\gamma d\rho]\Big\{
\\\nn &
(1-\epsilon)[1+z_{12}(\alpha \partial_1+\bar \beta \partial_2+(1-\beta-\gamma)\partial_3)]\mathcal{Q}^\mu_{\gamma^+}(z_{12}^\alpha,z_{21}^\beta,z_{21}^{\beta+\gamma})
\\\nn &-(1+\epsilon)[3-z_{12}(\alpha \partial_1+\bar \beta \partial_2+(1-\beta-\gamma)\partial_3)]\mathcal{Q}^\nu_{\gamma^+\gamma^{\nu\mu}}(z_{12}^\alpha,z_{21}^\beta,z_{21}^{\beta+\gamma})\Big\},
\\
\widetilde{\mathcal{U}}_{\mathbf{E^*}}&=-2iga_s\vec b^{2\epsilon}\Gamma(-\epsilon)\(C_F-\frac{C_A}{2}\) b_\mu\int [d\alpha d\beta d\gamma d\rho]\Big\{
\\\nn &
(1-\epsilon)[1+z_{21}((1-\beta-\gamma)\partial_1+\bar \beta \partial_2+\alpha \partial_3)]\mathcal{Q}^\mu_{\gamma^+}(z_{12}^{\beta+\gamma},z_{12}^\beta,z_{21}^{\alpha})
\\\nn & -(1+\epsilon)[3-z_{21}((1-\beta-\gamma)\partial_1+\bar \beta \partial_2+\alpha \partial_3)]\mathcal{Q}^\nu_{\gamma^+\gamma^{\nu\mu}}(z_{12}^{\beta+\gamma},z_{12}^\beta,z_{21}^{\alpha})\Big\},
\\
\widetilde{\mathcal{U}}_{\mathbf{F}}&=8iga_s\vec b^{2\epsilon}\Gamma(-\epsilon)\frac{C_A}{2}(1-\epsilon) b_\mu\int [d\alpha d\beta d\gamma d\rho]\mathcal{Q}^\mu_{\gamma^+}(z_{12}^\alpha,z_{12}^{\alpha+\gamma},z_{21}^\beta),
\end{align}
and  we have used the notation
\begin{eqnarray}\label{app:Q}
\mathcal{Q}^\mu_\Gamma(z_1,z_2,z_3)=\bar q(z_1 n)A^\mu(z_2n) \Gamma q(z_3n).
\end{eqnarray}
The symbols $\partial_{1,2,3}$ denote the $\partial_+$ that acts on field $\bar q$, $A$, $q$, correspondingly. In the diagrams A and B we have left the fields unexpanded in $\vec b$. It should be understood as a generating function for higher twist-operators. Note, that the diagrams A contains rapidity divergences, as it is discussed in sec.~\ref{sec:rapidity_div}. The expressions for SIDIS kinematics are obtained by replacement $-\infty$ by $+\infty$ in diagrams A, C and D.

The expressions for diagrams that mix the gluon and quark operators are
\begin{align}
\widetilde{\mathcal{U}}_{\mathbf{L}}&= 2ia_s\Gamma(-\epsilon)\vec b^{2\epsilon}\int[d\alpha d\beta d\gamma]
\Bigg\{
\\\nn & A^A_\mu(z_{12}^\alpha n+(1-2\alpha)\vec b)\Big[
g^{\mu\nu}\(\bar \alpha \overleftarrow{\partial_+}-\bar \beta \overrightarrow{\partial_+}\)
+2\epsilon\frac{b^\mu b^\nu}{\vec b^{2}}\((1-2\alpha) \overleftarrow{\partial_+}-(1-2\beta) \overrightarrow{\partial_+}\)
\\\nn &
\hskip0.2\textwidth -z_{12}g^{\mu\nu}\(\alpha \overleftarrow{\partial_+}+\bar \beta \overrightarrow{\partial_+}\)\(\bar \alpha \overleftarrow{\partial_+}+\beta \overrightarrow{\partial_+}\)\Big]
A^A_\nu(z_{21}^\beta n -(1-2\beta)\vec b)
\\&\nn
+b_\rho A^A_\mu(z_{12}^\alpha n)\Bigg[
g^{\mu\nu}\Big\{-\overleftarrow{\partial^\rho}(2\alpha\bar \alpha \overleftarrow{\partial_+}+(1-2\alpha \bar \beta)\overrightarrow{\partial_+})-\overrightarrow{\partial^\rho}((1-2\bar \alpha \beta)\overleftarrow{\partial_+}+2\beta \bar \beta \overrightarrow{\partial_+})\Big\}
\\\nn &
+g^{\mu\rho} \overleftarrow{\partial^\nu}\Big\{2\alpha(1-2\alpha)\overleftarrow{\partial_+}+(1-2\alpha(1-2\beta))\overrightarrow{\partial_+}\Big\}
+g^{\mu\rho} \overrightarrow{\partial^\nu}\Big\{(1-2\alpha)(1-2\beta)\overleftarrow{\partial_+}+4\beta \bar \beta \overrightarrow{\partial_+}\Big\}
\\\nn &
+g^{\nu\rho}\overrightarrow{\partial^\mu}\Big\{(1-2\beta(1-2\alpha))\overleftarrow{\partial_+}+2\beta(1-2\beta)\overrightarrow{\partial_+}\Big\}
+g^{\nu\rho} \overleftarrow{\partial^\mu}\Big\{4\alpha \bar \alpha \overleftarrow{\partial_+}+ (1-2\alpha)(1-2\beta)\overrightarrow{\partial_+}\Big\}
\\&\nn 
\hskip0.7\textwidth \Bigg]A^A_\nu(z_{21}^\beta n )\Bigg\},
\end{align}
\begin{align}
\widetilde{\mathcal{U}}_{\mathbf{M}}&=-ga_s\Gamma(-\epsilon)\vec b^{2\epsilon} \int [d\alpha d\beta d\gamma d\rho] A_\mu^A(z_{21}^\beta n)A_\sigma^B(z_{21}^{\beta+\gamma} n) A_\nu^C(z_{12}^\alpha n)(d^{ABC}+if^{ABC})\Bigg\{
\\\nn &
g^{\mu\nu}b^\sigma \((1+4\beta)\partial_1-2(1-2(\beta+\gamma))\partial_2-(1+4\alpha)\partial_3\)
\\&\nn
+g^{\mu\sigma}b^\nu \((1-4\beta)\partial_1-4(\beta+\gamma)\partial_2-(1-4\alpha)\partial_3\)
\\&\nn
+g^{\sigma\nu}b^\mu \((1-4\beta)\partial_1+4(1-\beta-\gamma)\partial_2-(1-4\alpha)\partial_3\)
\\\nn &
+z_{12}g^{\mu\nu}b^\sigma \(
\beta \partial_1^2+\alpha \partial_3^2+(1+\gamma+\rho)\partial_1\partial_3+(1-\beta-\gamma)\partial_2\partial_3+(\beta+\gamma)\partial_1\partial_2
\)
\\&\nn
+z_{12}(g^{\mu\sigma}b^\nu+g^{\sigma\nu}b^\mu) \Big[
\beta(1-2\beta)\partial_1^2+\alpha(1-2\alpha)\partial_3^2+2(\alpha+\rho)(\beta+\gamma)\partial_2^2
\\\nn &
+(2\alpha(\beta+\gamma)+(1-2\alpha)(\alpha+\rho))\partial_2\partial_3
+(2\beta(\alpha+\rho)+(1-2\beta)(\beta+\gamma))\partial_1\partial_2
\\\nn &-(\alpha+\beta-4\alpha\beta)\partial_1\partial_3
\Big] +4\epsilon \frac{b^\mu b^\sigma b^\nu}{\vec b^{2}}\((1-2\beta)\partial_1+(1-2(\beta+\gamma))\partial_2-(1-2\alpha)\partial_3\)
\Bigg\},
\end{align}
where we explicitly show the color indices. In the expression for the diagram L , the fields are left unexpanded in $\vec b$. In the expression for diagram M $\partial_{1,2,3}$ is $\partial_+$ that acts on $A_\mu$, $A_\sigma$ and $A_\nu$ correspondingly.

\subsection{Expressions for TMD distributions}
\label{app:diag-by-diag:matching}

In this section, we present the results for the matrix element in eq.~(\ref{main-fourier}) of the OPE contributions,
\begin{eqnarray}
f_{\text{diag}}=\int \frac{dz}{2\pi} e^{-2ixp^+z}\langle p,S|\widetilde{\mathcal{U}}_{\text{diag}}\(z_1=-z_2=z,\frac{\vec b}{2}\)|p,S\rangle.
\end{eqnarray}
 We  collect  all diagrams with their corresponding time-reversal  
and
we have
\begin{eqnarray}
f_{\mathbf{A}+\mathbf{A^*}}&=&2a_sC_F\Gamma(-\epsilon)\mathbf{B}^\epsilon \int d\xi \int_0^1 dy \delta(x-y\xi)
\Bigg\{
\\\nn && \Bigg[\(\frac{2y}{1-y}\)_+-2\delta(\bar y)\(1+\ln\(\frac{\delta}{p^+}\)\)\Bigg](f_1(\xi)+\mathbf{s}T(-\xi,0,\xi))
-2y \mathbf{s}T(-\xi,0,\xi)\Bigg\},
\\
f_{\mathbf{B}}&=&2a_sC_F(1-\epsilon)\Gamma(-\epsilon)\mathbf{B}^\epsilon \int d\xi \int_0^1 dy \delta(x-y\xi)
\Bigg\{\bar yf_1(\xi)+2y\bar y\mathbf{s}T(-\xi,0,\xi)\Bigg\},
\end{eqnarray}
\begin{eqnarray}
f_{\mathbf{C}+\mathbf{C^*}}&=&2a_s\(C_F-\frac{C_A}{2}\)\Gamma(-\epsilon)\mathbf{B}^\epsilon \mathbf{s} \int d\xi \int_0^1 dy \delta(x-y\xi)
\Bigg\{
\\\nn && 2y T(-\xi,0,\xi)-(1-2y)T(-x,\xi,x-\xi)-\Delta T(-x,\xi,x-\xi)\Bigg\},
\\
f_{\mathbf{D}+\mathbf{D^*}}&=&2a_s\frac{C_A}{2}\Gamma(-\epsilon)\mathbf{B}^\epsilon \mathbf{s} \int d\xi \int_0^1 dy \delta(x-y\xi)
\Bigg\{
\\\nn && \(\frac{-2y^2}{1-y}-2\delta(\bar y)\) T(-\xi,0,\xi)+\frac{1+y}{1-y}T(-x,x-\xi,\xi)+\Delta T(-x,x-\xi,\xi)\Bigg\},
\end{eqnarray}
\begin{eqnarray}
f_{\mathbf{E}+\mathbf{E^*}}&=&2a_s(1-\epsilon)\(C_F-\frac{C_A}{2}\)\Gamma(-\epsilon)\mathbf{B}^\epsilon \mathbf{s} \int d\xi \int_0^1 dy \delta(x-y\xi)
\,\bar y(1-2y) T(-\xi,0,\xi),
\\
f_{\mathbf{F}}&=&-4a_s(1-\epsilon)\frac{C_A}{2}\Gamma(-\epsilon)\mathbf{B}^\epsilon \mathbf{s} \int d\xi \int_0^1 dy \delta(x-y\xi)
\,y\bar y T(-\xi,0,\xi),
\end{eqnarray}
where
\begin{eqnarray}
\mathbf{s}=i\pi \tilde s_\mu b^\mu M,\qquad \mathbf{B}=\frac{\vec b^2}{4}>0.
\end{eqnarray}
Let us note that all diagrams with ladder-like topologies enter with a factor $(1-\epsilon)$.

The expression for the diagrams with quark-gluon mixing are
\begin{eqnarray}
f_{\mathbf{L}}&=&a_s\Gamma(-\epsilon)\mathbf{B}^\epsilon \int d\xi \int_0^1 dy \delta(x-y\xi)
\Bigg\{ 2\Bigg[\frac{1-2y\bar y}{2}-\frac{\epsilon \, y\bar y}{1-\tilde \epsilon}\Bigg]g(\xi)
\\\nn && 
+\mathbf{s}\Bigg[y(3-8y+6y^2)\frac{G_+(-\xi,0,\xi)}{\xi}+y^2 \frac{Y_+(-\xi,0,\xi)}{\xi}-\frac{6\epsilon\, y^2\bar y}{2-\tilde \epsilon}\frac{G_+(-\xi,0,\xi)}{\xi}\Big]\Bigg\},
\end{eqnarray}
\begin{eqnarray}
&&f_{\mathbf{M}}=2a_s\Gamma(-\epsilon)\mathbf{B}^\epsilon\mathbf{s} \int d\xi \int_0^1 dy \delta(x-y\xi)
\Bigg\{
\\\nn &&
(1-2y)(1-6y\bar y)\frac{G_+(-\xi,0,\xi)}{\xi}+(1-2y)\frac{Y_+(-\xi,0,\xi)}{\xi}+(1-2y\bar y)\frac{G_-(-\xi,0,\xi)+Y_-(-\xi,0,\xi)}{\xi}
\\\nn &&
\qquad\qquad\qquad-\frac{\epsilon}{2-\tilde \epsilon}\Big[6y\bar y (1-2y)\frac{G_+(-\xi,0,\xi)}{\xi}+6y\bar y \frac{G_-(-\xi,0,\xi)}{\xi}\Big]
\Bigg\}.
\end{eqnarray}
In these expressions we distinguish the parameter $\epsilon$ that comes from the dimensional regularization (i.e. from the loop integral measure $d^{4-2\epsilon}x$) and the parameter $\tilde \epsilon$ that comes from the definition of distributions in $4-2\tilde \epsilon-$dimensions, their normalization and tensor convolutions. The parameters $\epsilon$ and $\tilde \epsilon$ enter only as a universal composition $\epsilon/(1-\tilde \epsilon)$ and thus at this order of perturbative expressions the difference between schemes is absent.

Combining these expressions with the renormalization constants and taking the limit $\epsilon\to 0$, as it is discussed in eq.~(\ref{renorm:final-final}) we find  eq.~(\ref{result:Sivers},~\ref{result:tw3Evolution},~\ref{result:unpol},~\ref{result:unpol_EVOL}).

\bibliographystyle{JHEP}
\bibliography{TMD_ref}

\end{document}